\documentclass[review]{elsarticle}
 \biboptions{comma,sort&compress}
\usepackage{graphicx}
\usepackage{amsmath}
\usepackage{here}
\usepackage{cuted}
\usepackage{xcolor}

\def\beq{\begin{equation}}
\def\eeq{\end{equation}}
\def\beqn{\begin{equation*}}
\def\eeqn{\end{equation*}}
\def\bea{\begin{eqnarray}}
\def\eea{\end{eqnarray}}
\def\bq{\begin{quote}}
\def\eq{\end{quote}}
\def\ve{\vert}
\def\nnb{\nonumber}
\def\ga{\left(}
\def\dr{\right)}

\def\bma{\boldmath}															
\def\nnb{\nonumber}

\def\la{\langle}
\def\ra{\rangle}
\def\nin{\noindent}
\def\ba{\vspace*{-0.2cm}\begin{array}}
\def\ea{\end{array}\vspace*{-0.2cm}}

\def\b{$\bullet~$}
\def\d{$\diamond~$}

\def\als{\alpha_s}

\def\gg2{ \la\alpha_s G^2 \ra}
\def\gg3{g^3f_{abc}\la G^aG^bG^c \ra}
\def\ggg4{\la\als^2G^4\ra}

\usepackage{amsmath}
\usepackage{slashed}
\usepackage{color}

\begin{document}

\begin{frontmatter}

\title{Pseudoscalar and Vector $T_{QQ\bar q\bar q'}$ Spectra and Couplings from LSR at NLO}
\author{R. Albuquerque}
\address{Faculty of Technology,Rio de Janeiro State University (FAT,UERJ), Brazil}
\ead{raphael.albuquerque@uerj.br}
\author{S. Narison\corref{cor1}
}
\address{Laboratoire
Univers et Particules de Montpellier (LUPM), CNRS-IN2P3, \\
Case 070, Place Eug\`ene
Bataillon, 34095 - Montpellier, France\\
and\\
Institute of High-Energy Physics of Madagascar (iHEPMAD)\\
University of Ankatso, Antananarivo 101, Madagascar}
\ead{snarison@yahoo.fr}
\author{D. Rabetiarivony}
\ead{rd.bidds@gmail.com}

\address{Institute of High-Energy Physics of Madagascar (iHEPMAD)\\
University of Ankatso, Antananarivo 101, Madagascar}

\begin{abstract}
\noindent
We present systematic and improved estimates of the masses  and couplings of the  Pseudoscalar $(0^-)$ and  Vector  $(1^-)$ $T_{QQ\bar q\bar q'}$ states ($Q\equiv c,b~;~q,q'\equiv u,d,s$) using (inverse) QCD Laplace sum rules (LSR) and their ratios ${\cal R}$ within $\tau$ and $t_c$-stabilities criteria complemented by the (rigorous) condition: $R_{P/C} \equiv$ { Pole (Resonance)  over the QCD continuum contributions} $ \geq 1$.  Next-to-leading order (NLO) factorized perturbative (PT) QCD corrections are included for giving a meaning on the choice of the used running $\overline {MS}$ heavy quark mass, while the operator product expansion (OPE) is truncated at the under-controlled $d=6$ dimension condensates.  
Our results are compiled in Tables\,\ref{tab:resc}, \,\ref{tab:resb}, \,\ref{tab:res-rad-c}, \ref{tab:res-rad-b} and compared with some LO existing ones.  
We observe that the interpolating currents lead to two classes : Class H (Heavy) states with masses around 6 (resp. 13) GeV for charm (resp. bottom) channels. 
Class L (Light) states  $T_{cc\bar q\bar q'}(3.8\sim 4.4)$ where the pseudoscalar (resp. all
vector states)  are below the 
open charm thresholds and $T_{bb\bar q\bar q'}(\simeq 10.4)$ 
where all of them are below the open beauty thresholds. Mass-splittings due to SU3 breakings are tiny ($\leq$ 50 MeV). Though more accessible experimentally, Class L states have weaker couplings to the  currents than the Class H ones and may be difficult to observe.  The mass-splittings between the 1st radial excitation and the ground state of $T_{QQ\bar q\bar d}$ ($q\equiv,u,d)$  are about 2 GeV which are (almost) heavy flavour and current-type independent while their couplings are large signaling new dynamics of these exotic states. Quark masses behaviours of the masses and couplings based on empirical observation are discussed. 
The eventual findings of the $T_{cc\bar u\bar d}$(6.3) pseudoscalar ground state which may not be obscured by the Class L 1st radial excitations 
can be an alternative way to test the vacuum saturation violation  of the four-quark $d=6$ condensates. 
\end{abstract}
\end{frontmatter}
\pagestyle{plain}
 \section{Introduction}
 In this paper, we extend our work in Ref.\,\cite{Tcc} on the scalar and axial-vector $T_{QQ}$-like states to their chiral multiplet partners pseudoscalar and vector states. Some relevant references (original papers, reviews) on the multiquark states have been already quoted in this previous paper and will not be repeated here.


In a series of papers\,\cite{MOLE12,MOLE16,MOLE16X,SU3,4Q,DK,Zc,Zb}, we have used QCD spectral sum rules (QSSR) \`a la Shifman-Vainshtein-Zakharov (SVZ)\,\cite{SVZa,ZAKA}\,\footnote{For reviews, see e.g.  \cite{SNB1,SNB2,SNB3,IOFFEb,RRY,DERAF,BERTa,YNDB,PASC,DOSCH,COL}.}  within stability criteria to estimate the masses and couplings of different exotic XYZ states. Compared to the existing papers in the literature, we have emphasized that the inclusion of PT radiative corrections is important for justifying the choice of the input value of the heavy quark mass which plays a capital role in the analysis. In so doing, we have observed that, in the $\overline {MS}$ scheme, this correction is tiny which {\it a posteriori} explains the success of these lowest order (LO) results using the unjustified choice of the quark mass value in this $\overline {MS}$ scheme as to LO the definition of the heavy quark mass is ill-defined. 

 More recently, we have applied the LSR\,\cite{SVZa,BELLa,SNR} for interpreting the new states around (6.2-6.9) GeV found by the LHCb-group\,\cite{LHCb1} to be a doubly/fully hidden-charm molecules $(\bar QQ) (Q\bar Q)$ and $( \bar Q \bar Q)(QQ)$ tetraquarks states\,\cite{4Q}, while the new states found by the same group from the $DK$ invariant mass\,\cite{LHCb3} have been interpreted by a $0^+$ and $1^-$ tetramoles (superposition of almost degenerate molecules and tetraquark states having the same quantum numbers and couplings) slightly mixed with their radial excitations\,\cite{DK}. 
 We have also systematically studied the $Z_c$-like spectra and interpreted the $Z_c(3900)$ and the $Z_{cs}(3983)$ state found by BESIII\,\cite{BES3} as good candidates for $(1^+)$ tetramole states\,\cite{Zc}. 

Motivated by the recent LHCb discovery of a $1^{+}$ state at 3878 MeV\,\cite{LHCb4}, just below the $D^*D$ threshold, which is a good isoscalar ($I=0$) $T_{cc\bar u\bar d}$ axial vector $(J^P=1^+)$ candidate, we have also improved\,\cite{Tcc} the existing QSSR results for the masses and couplings by combining the direct mass determinations from the ratios ${\cal R}$ of Inverse Laplace sum rule (LSR)  with the ratio of masses from the double ratio of sum rules (DRSR)\,\cite{DRSR88}.

\section{The Inverse Laplace sum rules (LSR) approach}
We shall be concerned with the two-point correlator :
 \bea
\hspace*{-0.6cm} \Pi^{\mu\nu}_{\cal H}(q^2)&=&i\int \hspace*{-0.15cm}d^4x ~e^{-iqx}\la 0\vert {\cal T} {\cal O}^\mu_{\cal H}(x)\ga {\cal O}^\nu_{\cal H}(0)\dr^\dagger \vert 0\ra \nnb\\
&\equiv& -\ga g^{\mu\nu}-\frac{q^\mu q^\nu}{q^2}\dr\Pi^{(1)}_{\cal H}(q^2)+\frac{q^\mu q^\nu}{q^2} \Pi^{(0)}_{\cal H}(q^2)
 \label{eq:2point}
 \eea
built from the local hadronic operators ${\cal O}^\mu_{\cal H}(x)$  (see Table\,\ref{tab:current}).
It obeys the Finite Energy Inverse Laplace Transform Sum Rule (LSR) and their ratios:
\beq
\hspace*{-0.2cm} {\cal L}^c_n\vert_{\cal H}(\tau,\mu)=\int_{(2M_c+m_q+m_{q'})^2}^{t_c}\hspace*{-0.5cm}dt~t^n~e^{-t\tau}\frac{1}{\pi} \mbox{Im}~\Pi_{\cal H}(t,\mu)~: ~n=0,1~;~~~~~~~~~~~
 {\cal R}^c_{\cal H}(\tau)=\frac{{\cal L}^c_{1}\vert_{\cal H}} {{\cal L}^c_0\vert_{\cal H}},
\label{eq:lsr}
\eeq
 where $M_c$ and $m_s$ (we shall neglect $u,d$ quark masses) are the on-shell / pole charm and running strange quark masses, $\tau$ is the LSR variable, $t_c$ is  the threshold of the ``QCD continuum" which parametrizes, from the discontinuity of the Feynman diagrams, the spectral function  ${\rm Im}\,\Pi_{\cal H}(t,m_c^2,m_s^2,\mu^2)$.  In the minimal duality ansatz :
 \beq
 \frac{1}{\pi}{\rm Im}\,\Pi^{(1,0)}_{\cal H}(t) =  2f_{\cal H}^2M^8_{\cal H}\,\delta(t-M_{\cal H}^2)+ \frac{1}{\pi}{\rm Im}\,\Pi^{(1,0)}_{\cal H}(t)\vert_{\rm QCD}\,\theta(t-t_c),
\eeq
one can deduce the mass squared from the ratio of LSR at the optimization point $\tau_0$\,:
\beq
 {\cal R}^c_{\cal H}(\tau_0)= M_{\cal H}^2.
\eeq
We shall use this quantity for directly determining the mass of the $T_{QQ\bar q\bar q'}$ states. 

We also shall use  the double ratio of sum rule (DRSR)\,\cite{DRSR88}\,:
\beq
r_{{\cal H'}/{\cal H}}(\tau_0)\equiv \sqrt{\frac{{\cal R}^c_{\cal H'}}{{\cal R}^c_{\cal H}}}=\frac{M_{\cal H'}}{M_{\cal H}},
\eeq
which can be free from systematics provided that ${\cal R}^c_{\cal H}$ and ${\cal R}^c_{\cal H'}$ optimize at the same values of $\tau$ and of $t_c$:
\beq
 \tau_0\vert_{\cal H}\simeq \tau_0\vert_{\cal H'}~,~~~~~~~~~~~~t_c\vert_{\cal H}\simeq t_c\vert_{\cal H'}~.
\eeq
for extracting the $SU3$ mass-spilttings between the $T_{QQ\bar q\bar q'}$-like states.
This DRSR has been used in different channels for predicting successfully the few MeV mass-splittings (SU3-breakings, parity splittings,...) between different hadrons\,\cite{DRSR88,DRSR94,DRSR96,DRSR07,HBARYON1,DRSR11,DRSR11a}. 
\section{The interpolating operators}
In this paper, we choose to work with the $\bar 3_c3_c$ lowest dimension interpolating currents of the four-quark states  given  in Table\,\ref{tab:current}. 
\vspace*{-1cm} 
\begin{center}
   {\scriptsize
\begin{table}[hbt]
\setlength{\tabcolsep}{3pc}
    {\small
  \begin{tabular}{ll}
&\\
\hline
\hline
States& $\bar 3_c3_c$ Four-quark  Currents \\ 
 \hline
 {\it  Pseudoscalar $0^-$} \\
$T_{cc\bar u\bar d}$
 &  $ {\cal O}_{T_{ud}^{0^-} }= \frac{1}{\sqrt{2}}\epsilon_{i j k} \:\epsilon_{m n k} \left(
 c_i^T\, C  \gamma^\mu \,c_j \right) \big{[} \left( \bar{u}_m\,\gamma_5\, \gamma_\mu
  C \,\bar{d}_n^T\right) - \left( \bar{d}_m\,\gamma_5\, \gamma_\mu
  C \,\bar{u}_n^T\right)\big{]}$\\
  
    $T_{cc\bar u\bar s}$
 &  $ {\cal O}_{T^{0^-}_{us}}   = \epsilon_{i j k} \:\epsilon_{m n k}
    \left( c_i \, C \gamma_{\mu } c_j^T \right) 
    \left( \bar{u}_m\,\gamma_5 \,\gamma^{\mu } C \bar{s}_n^T \right)$\\

  $T_{cc\bar s\bar s}$
 &   $ {\cal O}_{T^{0^-}_{ss} }  = \epsilon_{i j k} \:\epsilon_{m n k}
    \left( c_i \, C \gamma_{\mu } c_j^T \right) 
    \left( \bar{s}_m \,\gamma_5\,\gamma^{\mu } C \bar{s}_n^T \right)$\\
\hline
{\it  Vector $1^-$}\\
 $T_{cc\bar u\bar d}$
 &  $ {\cal O}_{T_{ud}^{1^-}} = \frac{1}{\sqrt{2}}\epsilon_{i j k} \:\epsilon_{m n k} \left(
 c_i^T\, C \gamma^\mu \,c_j \right) \big{[} \left( \bar{u}_m\, 
  C \,\bar{d}_n^T\right) -  \left( \bar{d}_m\, 
  C \,\bar{u}_n^T\right)\big{]}$\\
      $T_{cc\bar u\bar s}$
 &  $ {\cal O}_{T^{1^-}_{us}}   = \epsilon_{i j k} \:\epsilon_{m n k}
    \left( c_i \, C \gamma^{\mu } c_j^T \right) 
    \left( \bar{u}_m \, C \bar{s}_n^T \right)$\\
   \hline\hline
  \vspace*{-0.5cm}
\end{tabular}}
 \caption{Interpolating operators describing the $T_{cc\bar q\bar q'}$-like states. We note that the current $ {\cal O}_{T^{0^-}_{ss} } $ gives a null contribution. }  

\label{tab:current}
\vspace*{-0.5cm}
\end{table}
} 
\end{center}
\vspace*{-0.5cm}
There are another possibilities for constructing such 
states by using the currents proposed in Ref.\,\cite{ZHUT}.
\vspace*{-1cm} 
\begin{center}
   {\scriptsize
\begin{table}[H]
\setlength{\tabcolsep}{2.pc}
    {\small
  \begin{tabular}{lll}
&\\
\hline
\hline
States&  Four-quark  Currents & Light quarks $q$\\
\hline
{\it  Pseudoscalar $0^-$}\\
 $T_{cc\bar q\bar q}$
 &  $ \eta_1 =  \left(
 c_a^T\, C \,c_b \right) \big{[} \left( \bar{q}_a\gamma_5 \, 
  C \,\bar{q}_b^T\right) +  \left( \bar{q}_b\gamma_5\, 
  C \,\bar{q}_a^T\right)\big{]}$&$u,d,s$\\

  $T_{cc\bar q\bar q}$
 &  $\eta_2 = \left(
 c_a^T\, C  \gamma_5 \,c_b \right) \big{[} \left( \bar{q}_a\,
  C \,\bar{q}_b^T\right) + \left( \bar{q}_b\,
  C \,\bar{q}_a^T\right)\big{]}$&$u,d,s$\\

 $T_{cc\bar q\bar u}$
 &  $\eta_4= \left(
 c_a^T\, C  \gamma_\mu \,c_b \right) \big{[} \left( \bar{u}_a\,\gamma^\mu\gamma_5\, 
  C \,\bar{q}_b^T\right) - \left( \bar{u}_b\,\gamma^\mu\gamma_5\, 
  C \,\bar{q}_a^T\right)\big{]}$&$d,s$\\
  $T_{cc\bar q\bar u}$
 &  $\eta_5 = \left(
 c_a^T\, C  \gamma_\mu \gamma_5\,c_b \right) \big{[} \left( \bar{u}_a\,\gamma^\mu\, 
  C \,\bar{q}_b^T\right) + \left( \bar{u}_b\,\gamma^\mu\, 
  C \,\bar{q}_a^T\right)\big{]}$&$d,s$\\
\hline
{\it Vector $1^-$}\\
 $T_{cc\bar q\bar q}$
 &  $ \eta_1 =  \left(
 c_a^T\, C \gamma^\mu\gamma_5 \,c_b \right) \big{[} \left( \bar{q}_a\gamma_5 \, 
  C \,\bar{q}_b^T\right) +  \left( \bar{q}_b\gamma_5\, 
  C \,\bar{q}_a^T\right)\big{]}$&$ u,d,s$ \\
   
$T_{cc\bar q\bar q}$
 &  $\eta_2 = \left(
 c_a^T\, C  \gamma_5 \,c_b \right) \big{[} \left( \bar{q}_a\,\gamma_\mu\gamma_5\, 
  C \,\bar{q}_b^T\right) + \left( \bar{q}_b\,\gamma_\mu\gamma_5\, 
  C \,\bar{q}_a^T\right)\big{]}$&$u,d,s$\\
  
   $T_{cc\bar u\bar q}$
 &  $ \eta_5 =  \left(
 c_a^T\, C \gamma^\mu \,c_b \right) \big{[} \left( \bar{u}_a \, 
  C \,\bar{q}_b^T\right) -  \left( \bar{u}_b\, 
  C \,\bar{q}_a^T\right)\big{]}$&$ d,s$\\
 
    $T_{cc\bar u\bar q}$
 &  $ \eta_6 =  \left(
 c_a^T\, C \,c_b \right) \big{[} \left( \bar{u}_a \, 
  \gamma^\mu C \,\bar{q}_b^T\right) +  \left( \bar{u}_b\, 
  \gamma^\mu C \,\bar{q}_a^T\right)\big{]}$&$d,s$\\
   \hline\hline
  \vspace*{-0.5cm}
\end{tabular}}
 \caption{Some other interpolating operators describing the $T_{cc\bar q\bar q'}$-like states used in Ref.\,\cite{ZHUT}.}

\label{tab:zhu}
\end{table}
}
\end{center}

We do not consider in this paper the molecule assignment of the $T_{QQ\bar q\bar q'}$ states as the method cannot distinguish the four-quark and molecule states within the errors. 

We complete the paper by re-analyzing the results in Ref.\,\cite{ZHUT} using the currents in Table\,\ref{tab:zhu} where $\eta_1, \eta_2$ are isovector $(I=1)$ currents in a $[6_f]_{\bar q\bar q}$ configuration, 
 while $\eta_4, \eta_5,\eta_6$ are isoscalar $(I=0)$ in a $[3_f]_{\bar q\bar q}$ configuration. 

\section{QCD input parameters}
\nin
The QCD parameters which shall be used here 
are the QCD coupling $\alpha_s$,  the charm and bottom quark running masses $\overline{m}_{c,b}$, the strange quark running mass $\overline{m}_{s}$ (we shall neglect the $u,d$ quark masses), the quark condensates, 
the gluon condensates $ \la\alpha_sG^2\ra$ and  $ \la g^3G^3\ra$.  Their values  are given in Table\,\ref{tab:param}. 
We shall use  $n_f$=4 (resp. 5) total number of flavours for the numerical value of $a_s\equiv\alpha_s/ \pi$ for the charm (resp. bottom) channel. 
\begin{table}[hbt]
\setlength{\tabcolsep}{1.8pc}
    {\small
  \begin{tabular}{llll}
&\\
\hline
\hline
Parameters&Values&Sources& Refs.    \\
\hline
$\alpha_s(M_Z)$& $0.1181(16)(3)$&$M_{\chi_{0c,b}-M_{\eta_{c,b}}}$&
\cite{SNparam,SNparam2,SNm20} \\
$\overline{m}_c(m_c)$ [MeV]&$1266(6)$ &$D, B_c \oplus {J/\psi}, \chi_{c1},\eta_{c}$&
\cite{SNm20,SNparam,SNbc20,SNmom18,SNFB13,SNH10,SNH11}\\
$\overline{m}_b(m_b)$ [MeV]&$4196(8)$ &$B_c\oplus{\Upsilon}$&
\cite{SNm20,SNH10,SNH11,SNH12,SNparam,SNbc20,SNmom18,SNFB13}\\
$\hat \mu_q$ [MeV]&$253(6)$ &Light Quarks&\,\cite{SNB1,SNp15} \\
$\hat m_s$ [MeV]&$114(6)$ &Light Quarks&\,\cite{SNB1,SNp15} \\
$\kappa\equiv\la \bar ss\ra/\la\bar dd\ra$& $0.74(6)$&Light \& Heavy Quarks&\cite{SNB1,SNp15,HBARYON1}\\
$M_0^2$ [GeV$^2$]&$0.8(2)$ &Light\& Heavy Quarks&\,\cite{SNB1,DOSCH,JAMI2a,JAMI2c,HEIDa,HEIDc,SNhl} \\
$\la\alpha_s G^2\ra$ [GeV$^4$]& $6.35(35) 10^{-2}$&Light \& Heavy Quarks &
 \cite{SNparam,SNm20}\\
${\la g^3  G^3\ra}/{\la\alpha_s G^2\ra}$& $8.2(1.0)$[GeV$^2$]&${J/\psi}$&\cite{SNH10,SNH11,SNH12}\\
$\rho \alpha_s\la \bar qq\ra^2$ [GeV$^6$]&$5.8(9) 10^{-4}$ &Light Quarks,$\tau$-decay&\cite{DOSCH,SNTAU,JAMI2a,JAMI2c,LNT,LAUNERb}\\
\hline\hline
\end{tabular}}
 \caption{QCD input parameters estimated from QSSR (Moments, LSR and ratios of sum rules) used here. 
 }  
\label{tab:param}
\end{table}
\section{QCD expressions of the spectral functions up to dimension-six}
\b The spectral functions are calculated explicitly to lowest order (LO) and up to dimension-six condensates using the Operator Product Expansion (OPE) \`a la SVZ. They are given in {\it some compact and integrated forms} in the Appendix. Unlike often used in the current literature, we do not include some classes of higher dimension condensates contributions. The reason is that the size of such condensates are not under
a good control: the estimate of the high-dimension gluon condensates based on a dilute gas instanton and the factorization assumption used to the estimates of these high dimension condensates can be violated (see e.g. the case of  the $\la G^3\ra$ and the four-quark condensate in Table\,\ref{tab:param}). In addition,  their structure is more complicated due to their mixing under renormalization (see e.g.\,\cite{SNTARRACH}). 

\b To guarantee the validity of our approximation , we always check that the dimension-six condensates remain small corrections of the previous low dimension condensate contributions in the OPE from which we may estimate the systematics induced by these unkown high-dimension condensates as :
\beq
\Delta OPE \simeq \frac{\tau}{3}m_c^2\times ({\rm d=6~contributions)}. 
\eeq 
In addition, we consider the contribution of a class of $d=8$ condensates retained in different papers for checking this systematic error, where, we shall see in the following that, these $d=8$ contributions are (almost) negligible. 

\b  We estimate the (factorized) Next-to-Leading Order (NLO) of the perturbative contribution using a convolution integral of the spectral functions from bilinear quark currents following the works in Refs.\,\cite{PICH,SNPIVO,HAGIWARA}. This procedure has been explained in details in our previous works. 
 The NLO QCD expressions of the bilinear currents have been e.g compiled in Ref.\,\cite{SNB2}. 
\section{Optimal results from the analysis}
Like in our previous works, we shall use stabilty criteria on the external variables:  $\tau$ (sum rule variable), $t_c$ (QCD continuum threshold) and $\mu$ (substraction constant) for extracting the optimal information from the sum rules. These stability points manifest either as plateau, minimum or inflexion points.  More detailed discussions about the stability criteria have been given in the books (e.g.\,\cite{SNB1,SNB2}) and in our previous works. 

We note that the value of the $\mu$-stability is (almost) universal in different heavy four-quark for the charm (resp. bottom) channels\,\cite{Tcc,MOLE16X,SU3,4Q,DK,Zc,Zb}:
\beq
\mu=(4.65\pm 0.05)~{\rm GeV} ~~~~({\rm resp.}~~  ~~(5.2\pm 0.05)~{\rm GeV}.
\eeq
 We shall use these numbers in the following analysis. 
 
To check or/and better restrict the conservative region of $(\tau,t_c)$, we shall also request that the Pole (Resonance) contribution to the spectral integral is larger than the QCD continuum one. This condition can be rigorously  formulated from e.g. the moment sum rule as:
\beq
R_{P/C} \equiv \frac{\int_{4M_c^2}^{t_c}dt\,e^{-t\tau}{\,\rm Im}\,\Pi(t)}{\int_{t_c}^{\infty}dt\,e^{-t\tau}{\,\rm Im}\,\Pi(t)}\geq 1.
\eeq
 Set of $(\tau,t_c)$ stability not satisfiying this requirement will be rejected. This feature can happen in some channels at low values of $(\tau,t_c)$ where some curves are flat or have some minimum at (too) low $(\tau,t_c)$-values.

\section{Importance of the Figures}
We emphasize the importance of the Figures for each channels though they look to be repetitive and almost similar.  They are necessary for justifying our choice of the external parameters $(\tau,t_c)$ for extracting the optimal results from the sum rules.
\section{The $T_{cc\bar u\bar d}$ pseudoscalar state}
\subsection*{\b  On the choice of Charm Quark Mass at LO from the current in Table\,\ref{tab:current} $\equiv \eta_4$ of Ref.\,\cite{ZHUT}}
\begin{figure}[hbt]
\begin{center}
\centerline {\hspace*{-7.5cm} \bf a)\hspace{8cm} b)}
\includegraphics[width=8cm]{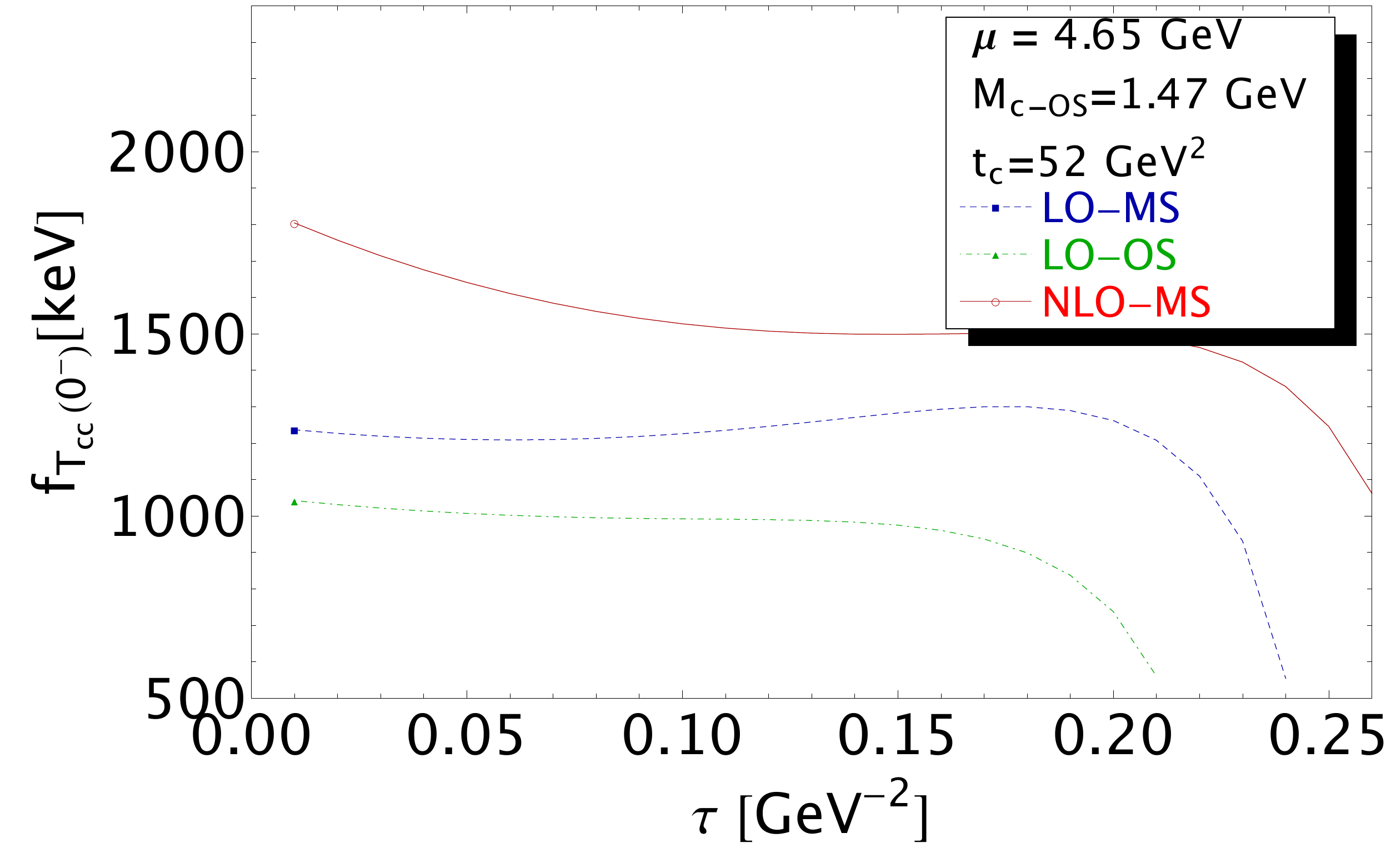}
\includegraphics[width=8cm]{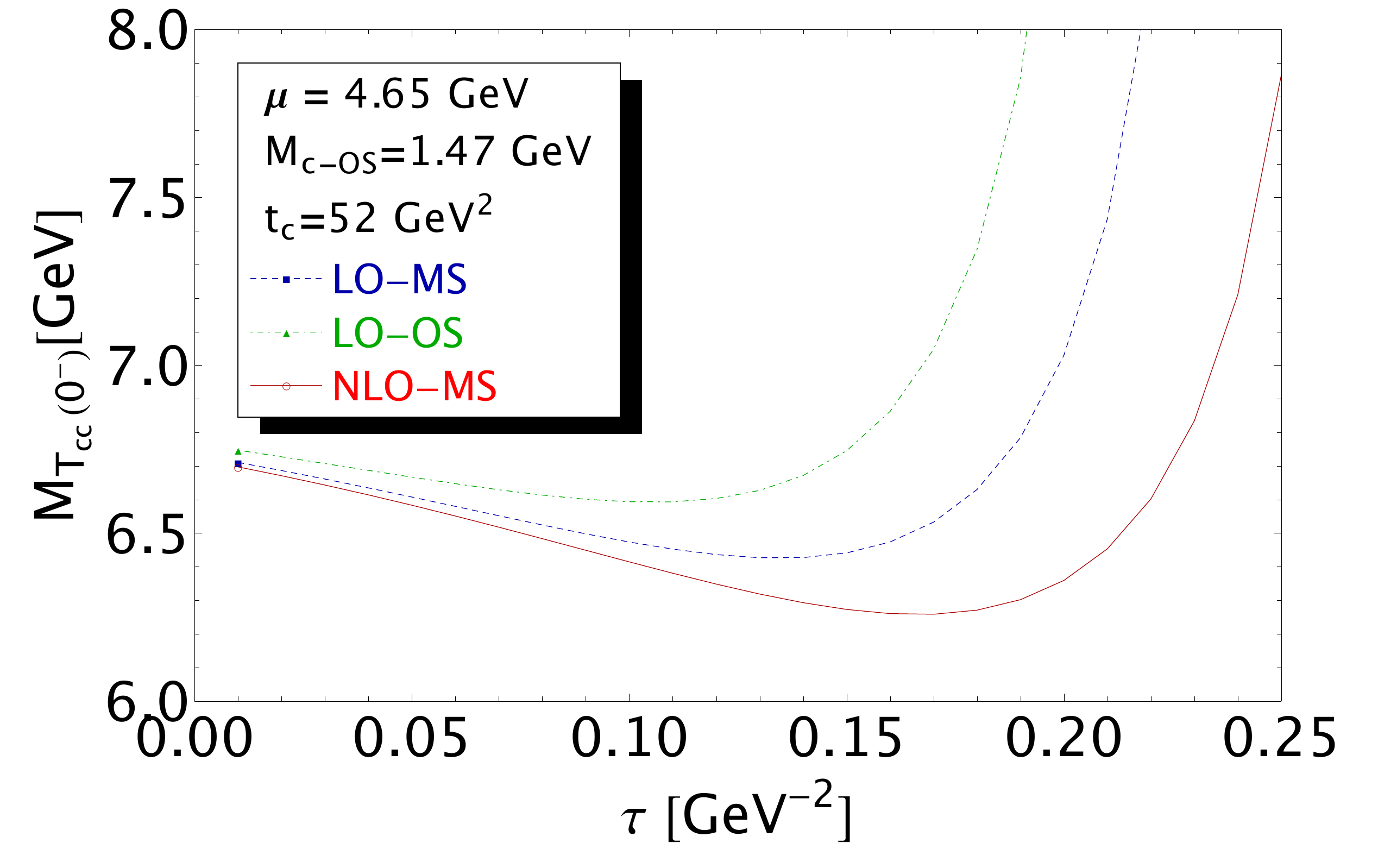}
\vspace*{-0.5cm}
\caption{\footnotesize  $f_{T_{cc\bar u\bar d}}$ and $M_{T_{cc\bar u\bar d}}$ as a function of $\tau$ at NLO and for  $t_c=52$ GeV$^2$. We use $\mu$=4.65 GeV and the QCD inputs in Table\,\ref{tab:param}. We use the on-shell mass: $M_c^{OS}=1.47$ GeV. } 
\label{fig:tcc-lo}
\end{center}
\vspace*{-0.5cm}
\end{figure} 

As stressed in the introduction and in our different previous works, the definition of the charm quark mass at LO is ill-defined and the (a priori) favoured choice of the running quark mass often used in the current sum rules literature is not justified. We study explicitly in Figs.\ref{fig:tcc-lo} the effect of the different choices (running, on-shell) quark mass at LO and compare the results with the NLO ones in the $\overline{MS}$-scheme for fixed value of $t_c=52$ GeV$^2$. One can notice that:

\d The uncertainty induced by the ambiguity for the choice of the quark mass at LO induces an error of about 100 keV for the coupling and of 84 MeV for the mass. 

\d The inclusion of the (factorized) NLO correction within the $\overline{MS}$-scheme increases the coupling by 300 keV and improves its $\tau$-stability while it decreases the mass by 168 MeV. 

\d The errors induced by the ill-definition of the charm quark mass at LO should be added as systematics in the quoted LO analysis in the current literature. 

\begin{figure}[hbt]
\begin{center}
\centerline {\hspace*{-7.5cm} \bf a)\hspace{8cm} b)}
\includegraphics[width=8cm]{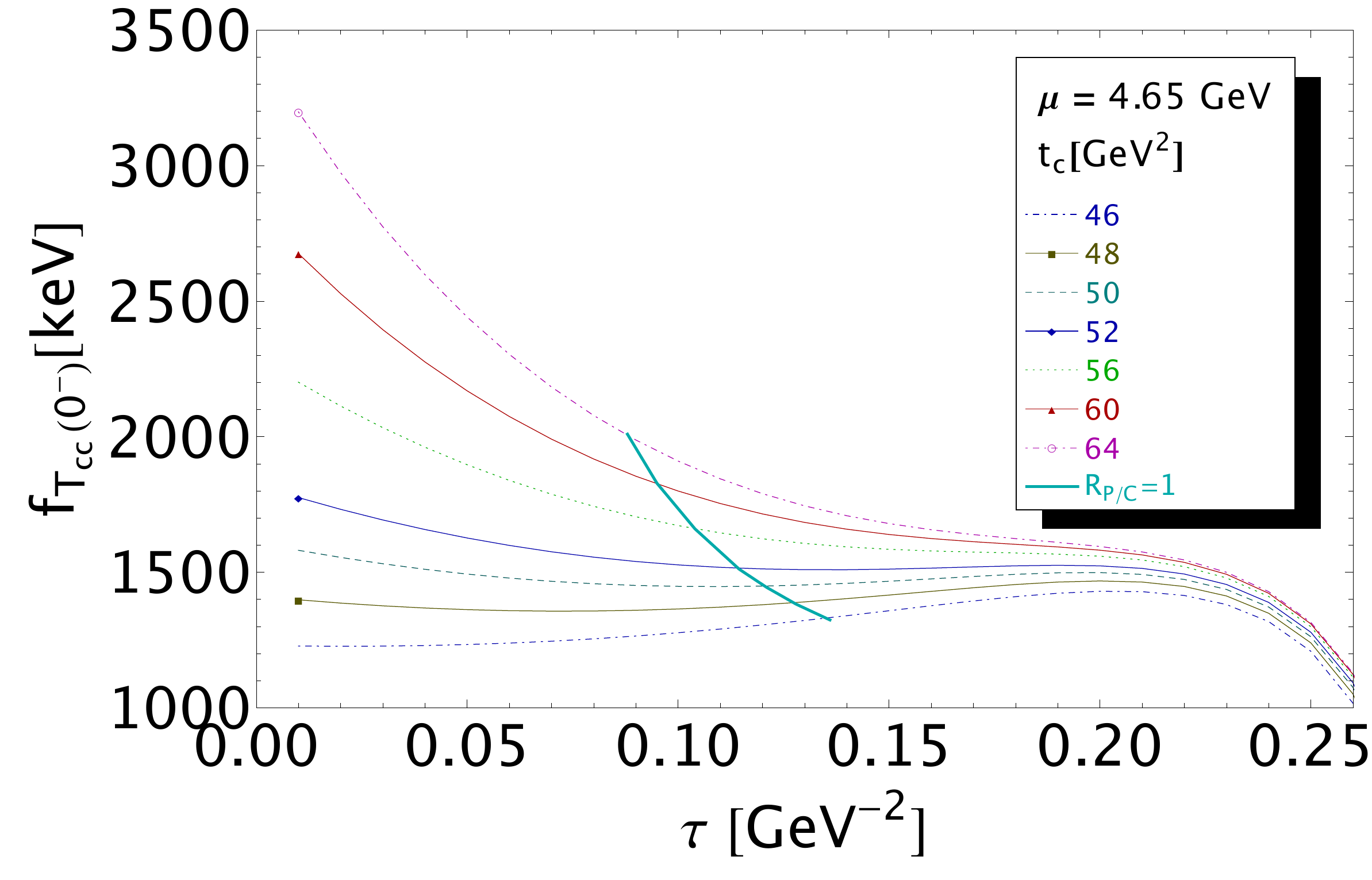}
\includegraphics[width=8cm]{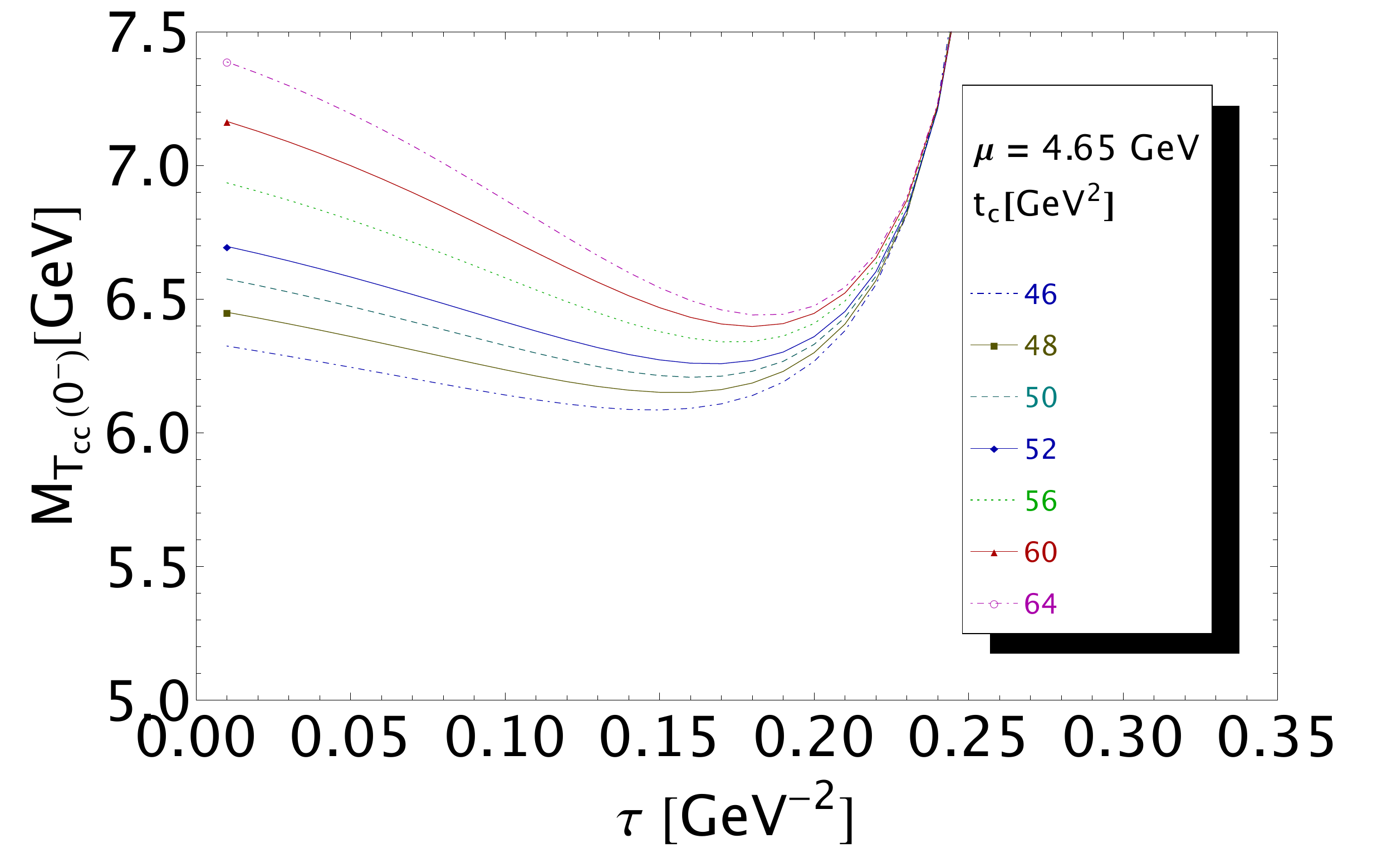}
\vspace*{-0.5cm}
\caption{\footnotesize  $f_{T_{cc\bar u\bar d}}$ and $M_{T_{cc\bar u\bar d}}$ as a function of $\tau$ at NLO and for different values of $t_c$. We use $\mu$=4.65 GeV and the QCD inputs in Table\,\ref{tab:param}. The curve $R_{P/C}=1$  is shown in a).  } 
\label{fig:tcc}
\end{center}
\vspace*{-0.5cm}
\end{figure} 
\subsection*{\b  Mass and decay constant from LSR at NLO  from the current in Table\,\ref{tab:current} $\equiv \eta_4$ of Ref.\,\cite{ZHUT}}
\subsection*{\d Analysis}
We study the behaviour of the mass and decay constant in Figs.\,
\ref{fig:tcc} for different values of $\tau$ and $t_c$ and for fixed value of $\mu=4.65$ GeV deduced  from our previous analysis of the four-quark states. We observe stabilites for the coupling for the sets  $(\tau , t_c)$ from (0.04, 46) to (0.18,60) (GeV$^{-2}$, GeV$^{2}$) and for the mass from  (0.15, 46) to (0.18,60)  (GeV$^{-2}$, GeV$^{2}$). Adding the condition that the ground state contribution to the moment ${\cal L}_0$ is bigger than the continuum one ($R_{P/C}\geq 1$), one can only retain the values of $t_c \geq $ 50 GeV$^2$, from which we deduce the optimal estimate for $(\tau , t_c)$ = (0.12, 50) to (0.18,60) (GeV$^{-2}$, GeV$^{2}$)(see Tables\,\ref{tab:resc}, \ref{tab:error-fc} and \ref{tab:error-mc}):
\beq
f_{T_{cc\bar u\bar d}(0^-)}= 1559(133)~{\rm keV}, ~~~~~~~~~~
M_{T_{cc\bar u\bar d}(0^-)}= 6303(123)~{\rm MeV}, 
\label{eq:tcc0}
\eeq
where we have used $\Delta\tau=\pm 0.02$ GeV$^{-2}$ around the minimum and taken the mean of the errors due to the change of $\tau$ for the two extremal values of $t_c$.
\subsection*{\d Convergence of the OPE}
We compare the contributions of each condensates at the stability points taking the mean from $t_c=50$ and 60 GeV$^2$. 

-- For the decay constant, the contribution of $d=6$ condensates is -6\%, of the PT$\oplus d=4$ ones, while the $d=8$ one is completely negligible\,\footnote{We note that all our contributions of the mixed $\la \bar qGq\ra$ and  the examples of a class of $d=8$  $\la \bar qq\ra\la\bar qGq\ra$ given in the Appendix disagree in size with Ref.\,\cite{ZHUT}. An unusual sign is also used in Ref.\,\cite{ZHUT} for parametrizing the mixed quark-gluon condensate.}.  

-- For the mass, the contribution of the $d=6$ condensates is +0.4\%, of the PT$\oplus d=4$ ones, while the $d=8$ one is completely negligible.

-- The good convergence of the OPE is expected at this relatively low value of $\tau$ where the sum rules stabilize. To be more conservative, we add a systematic  error due the trunction of the OPE as:
\beq
\Delta OPE=\ga\frac{m_c^2\tau}{3}\dr\times ({\rm d=6~ contributions}). 
\label{eq:trunc}
\eeq
The induced errors are given in Tables\,\ref{tab:error-fc} and \ref{tab:error-mc}. 
\subsection*{\d Mass-splittings between chiral multiplets}
 From the previous result, one can deduce the mass-splitting between the chiral multiplets:
\beq
M_{T_{cc\bar u\bar d}(0^-)}-M_{T_{cc}(0^+)}\simeq 2405~{\rm MeV},
\eeq
where we have used the $0^+$ mass $M_{T_{cc}(0^+)}\simeq 3898$ MeV from\,\cite{Tcc}.  This large mass-splitting is in line with our previous findings for the $X_c,Z_c$-like states obtained in\,\cite{MOLE16X}:
\beq
M_{Z_c(0^-)}-M_{Z_{c}(0^+)}\simeq 1852~{\rm MeV}.
\eeq
\subsection*{\b  Mass and decay constant at NLO from the $\eta_2$ current used in Ref.\cite{ZHUT} }
We recalculate the corresponding QCD spectral function and redo the analysis. The results at NLO are similar to Fig.\,\ref{fig:tcc} versus $\tau$, for different values of $t_c$ and fixing $\mu=4.65$ GeV. The coupling presents minimum for the set: $(\tau,t_c)$ = (0.12,46) and inflexion point  at (0.18,60)  (GeV$^{-2}$, GeV$^{2}$) while the mass presents minimum from (0.16,48) to  (0.18,60)  (GeV$^{-2}$, GeV$^{2}$). Then, we deduce at NLO:
\beq
f_{T_{cc\bar d\bar d}(0^-)(\eta_2)}= 1348(165) ~{\rm keV},  ~~~~~~~~~~
M_{T_{cc\bar d\bar d}(0^-)(\eta_2)}= 6267(148) ~{\rm MeV}, 
\label{eq:tcc0-eta2}
\eeq
with $\Delta\tau\simeq \pm 0.02$ GeV$^{-2}$. This result is in line with our previous results for the $X_c,Z_c$ states\,\cite{MOLE16X}. 
\subsection*{\b  Mass and decay constant at NLO from the $\eta_5$ current used in Ref.\,\cite{ZHUT} }
\subsection*{\d Analysis} 
\begin{figure}[hbt]
\begin{center}
\centerline {\hspace*{-7.5cm} \bf a)\hspace{8cm} b)}
\includegraphics[width=8cm]{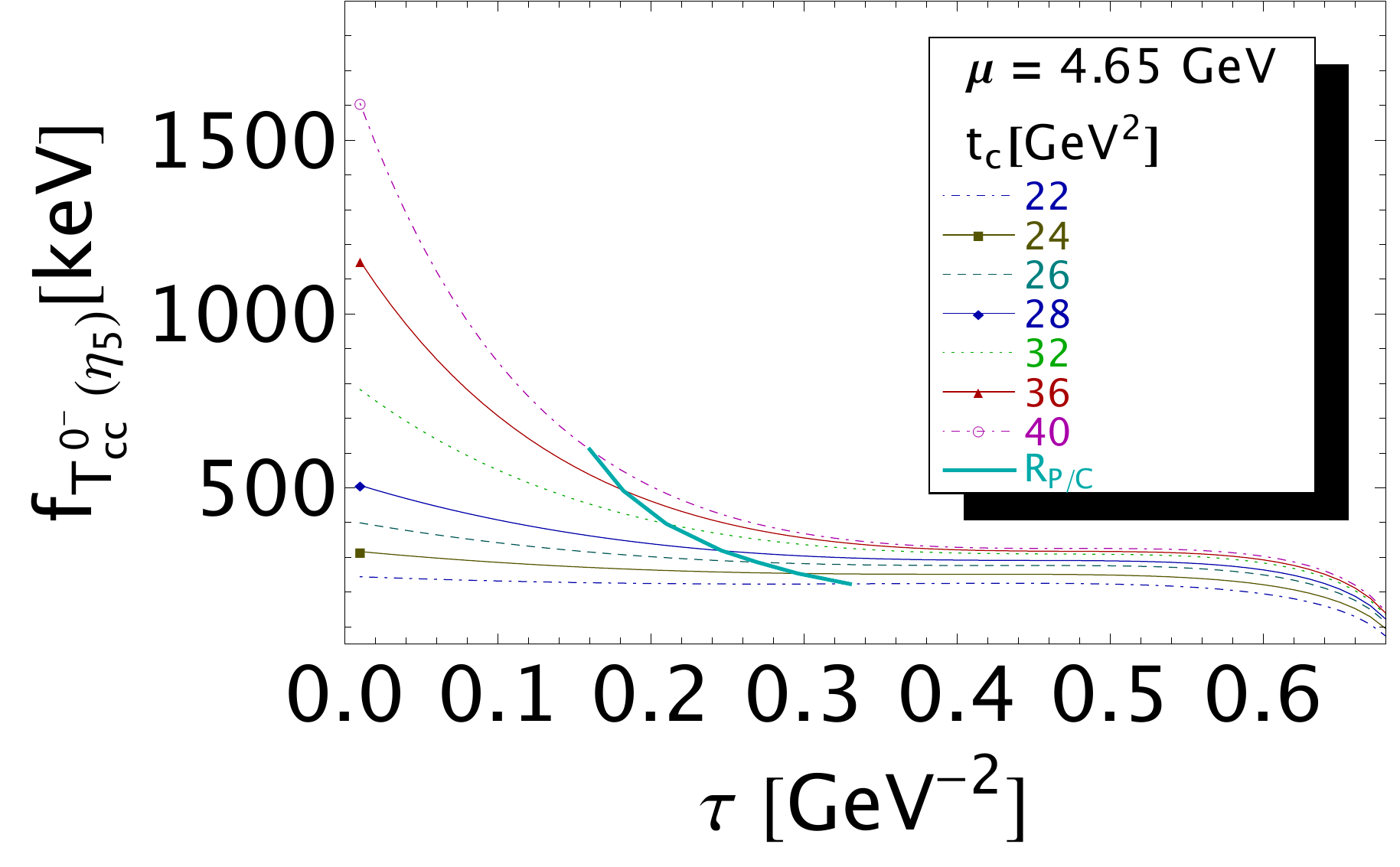}
\includegraphics[width=8cm]{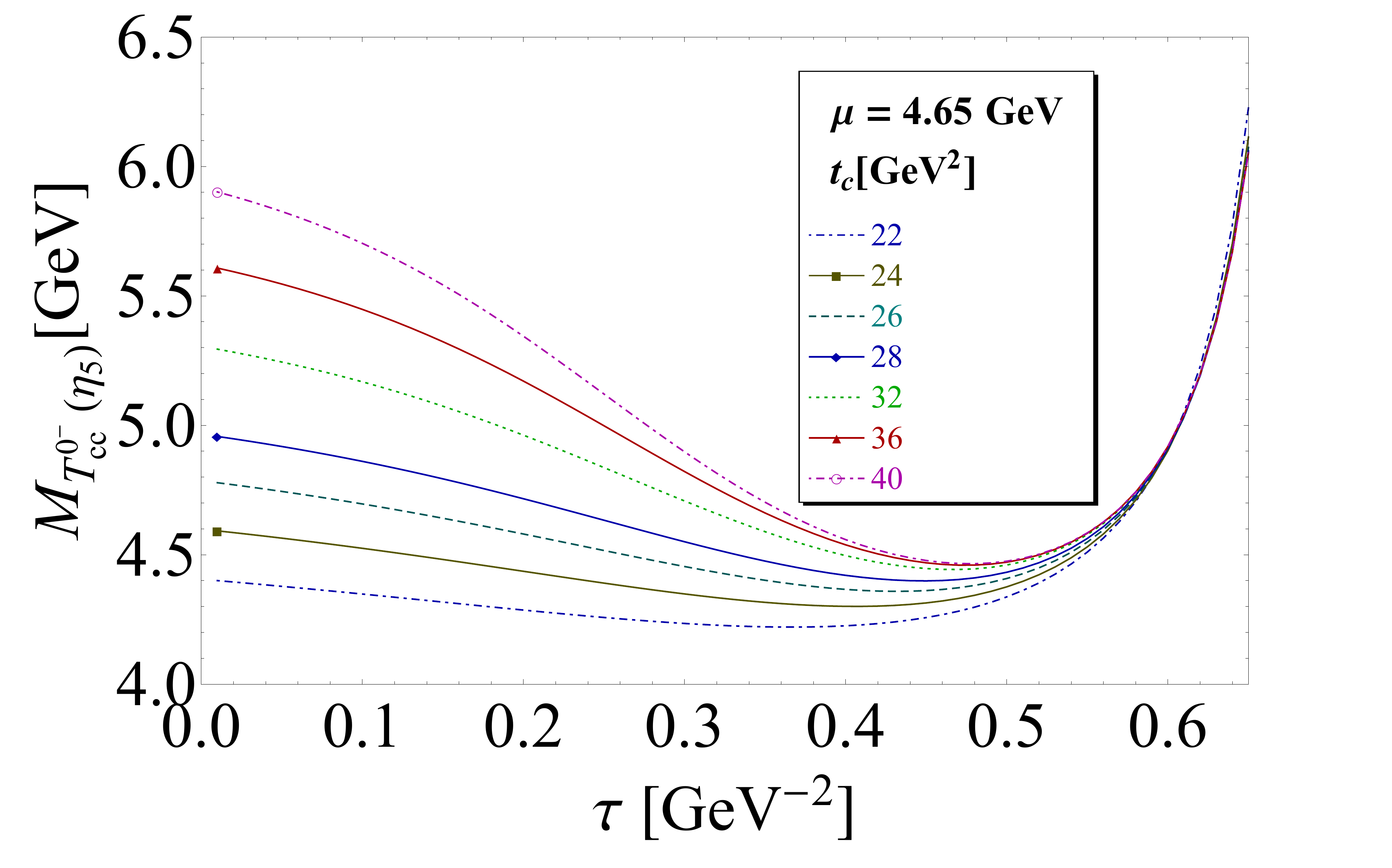}
\vspace*{-0.5cm}
\caption{\footnotesize  $f_{T_{cc\bar u\bar d}}$ and $M_{T_{cc\bar u\bar d}}$ for the $\eta_5$ current as a function of $\tau$ at NLO and for different values of $t_c$. We use $\mu$=4.65 GeV and the QCD inputs in Table\,\ref{tab:param}.  } 
\label{fig:eta5c}
\end{center}
\vspace*{-0.5cm}
\end{figure}

We also recalculate the corresponding QCD spectral function and redo the analysis. The results at NLO are shown in Fig.\,\ref{fig:eta5c} versus $\tau$, for different values of $t_c$ and fixing $\mu=4.65$ GeV. The coupling presents minimum for the set: $(\tau,t_c)$ from (0.32,24) and inflexion point for (0.45,36)  (GeV$^{-2}$, GeV$^{2}$) while the mass presents minimum from (0.37,22) to (0.47,36) (GeV$^{-2}$, GeV$^{2}$). The region for $t_c\geq 24$ GeV$^2$ satisfies the $R_{P/C}\geq 1$ condition. One deduces at NLO:
\beq
f_{T_{cc\bar u\bar d}(0^-)(\eta_5)}= 289(42) ~{\rm keV},  ~~~~~~~~~~
M_{T_{cc\bar u\bar d}(0^-)(\eta_5)}= 4380(128) ~{\rm MeV}.
\label{eq:eta5}
\eeq

\begin{figure}[hbt]
\begin{center}
\centerline {\hspace*{-7.5cm} \bf a)\hspace{8cm} b)}
\includegraphics[width=8cm]{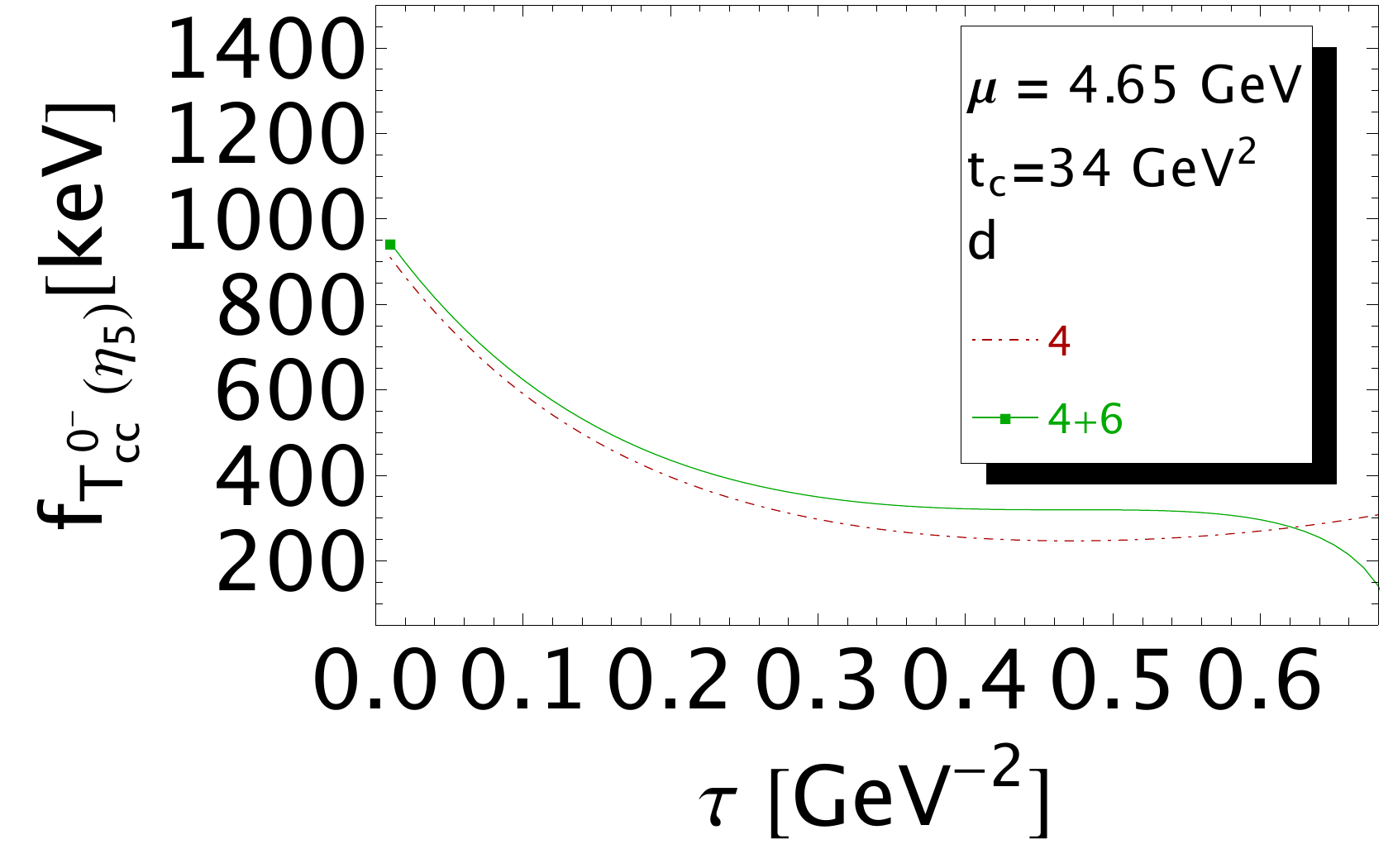}
\includegraphics[width=8cm]{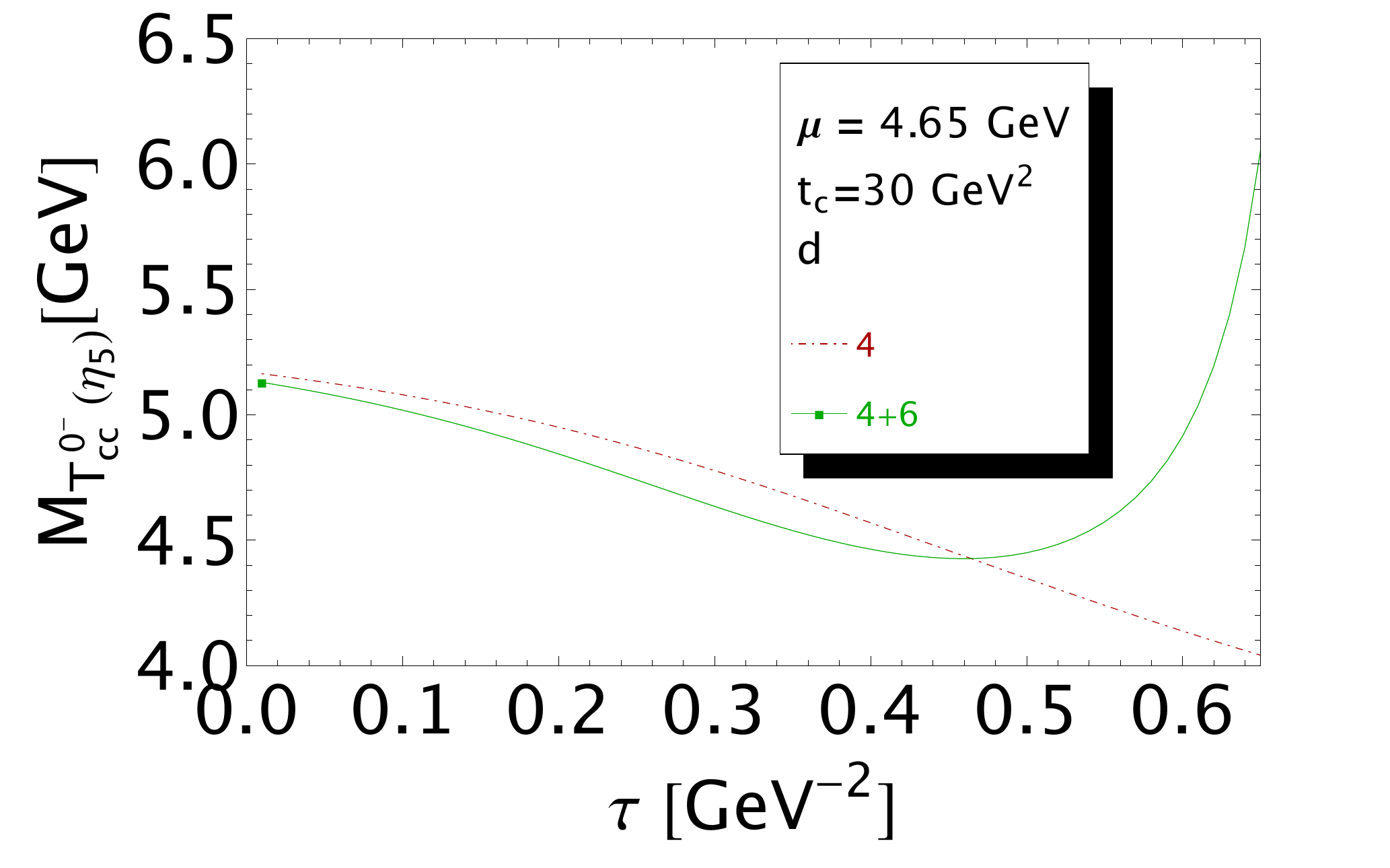}
\vspace*{-0.5cm}
\caption{\footnotesize  $f_{T_{cc\bar d\bar d}}$ and $M_{T_{cc\bar d\bar d}}$ for the $\eta_5$ current as a function of $\tau$ at NLO  for fixed values of $t_c$ and for different truncation of the OPE. We use $\mu$=4.65 GeV and the QCD inputs in Table\,\ref{tab:param}. } 
\label{fig:eta5-conv}
\end{center}
\vspace*{-0.5cm}
\end{figure} 
\subsection*{\d Convergence of the OPE}
-- As the stabilities are obtained at higher value of $\tau$ compared to the previous case, we check again the convergence of the OPE at this scale. In so doing, we take the mean $t_c=34$ GeV$^2$ for the coupling and 30 GeV$^2$ for the mass. The NLO result is shown in Figs.\,\ref{fig:eta5-conv}. One has 22\% contribution of the $d=6$ condensate for the coupling and 0.2\% for the mass indicating that the convergence of the OPE is quite satisfactory.  

-- We add a class of the $d=8$ condensate contribution given in the Appendix. We obtain: 
\beq
\Delta M^{d=8}_{T_{cc\bar u\bar d}(0^-)(\eta_5)} = \pm  24~{\rm MeV},  ~~~~~~~~~~ \Delta f^{d=8}_{T_{cc\bar u\bar d}(0^-)(\eta_5)}= \pm 4~{\rm keV},
\eeq
which is negligible

-- We estimate the error due the truncation of the OPE as in Eq.\,\ref{eq:trunc} which we quote in Tables\,\ref{tab:error-fc} and \ref{tab:error-mc}. 
\subsection*{\b  Mass and decay constant at NLO from the $\eta_1$ current used in Ref.\,\cite{ZHUT} }
\begin{figure}[hbt]
\begin{center}
\centerline {\hspace*{-7.5cm} \bf a)\hspace{8cm} b)}
\includegraphics[width=8cm]{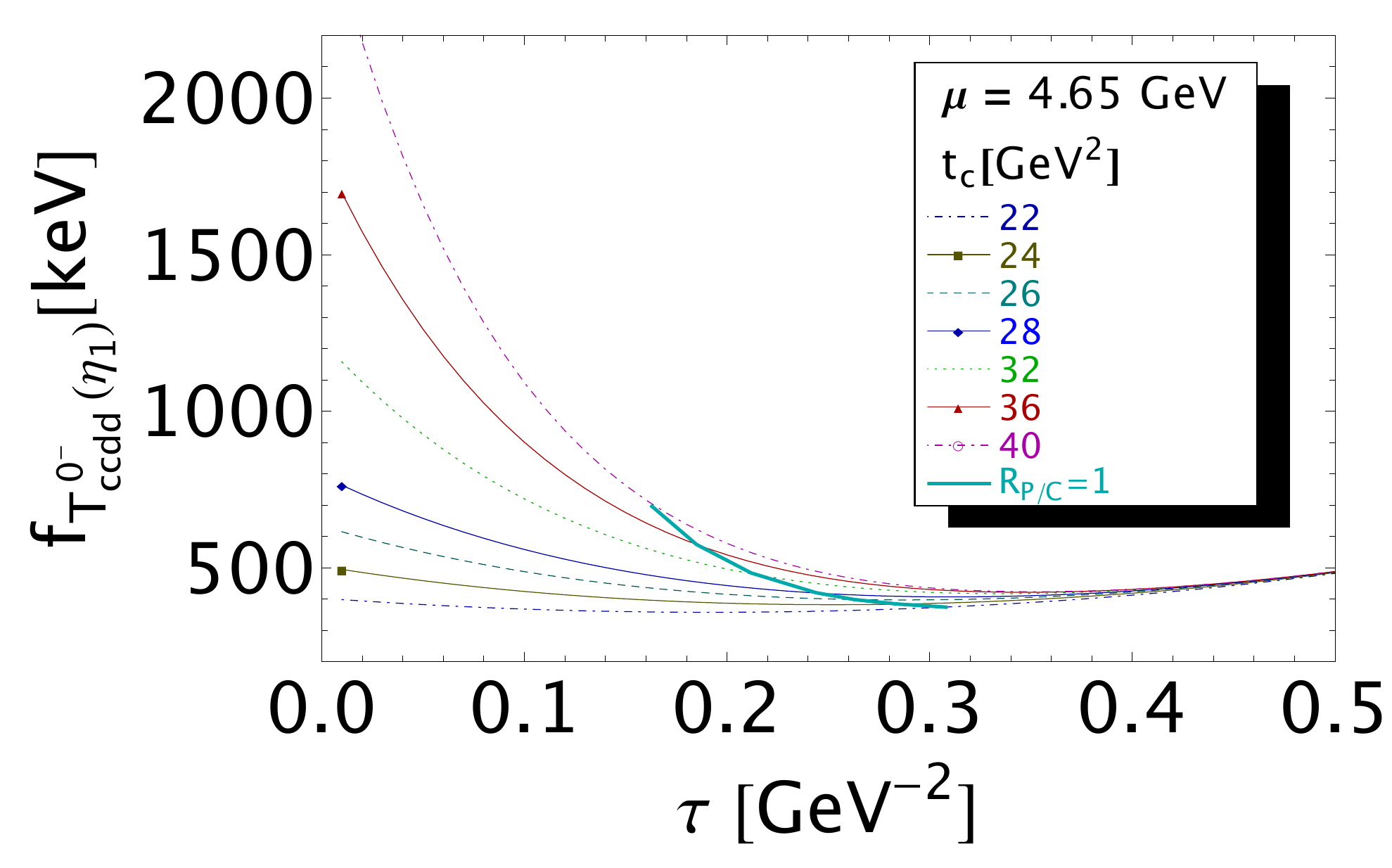}
\includegraphics[width=8cm]{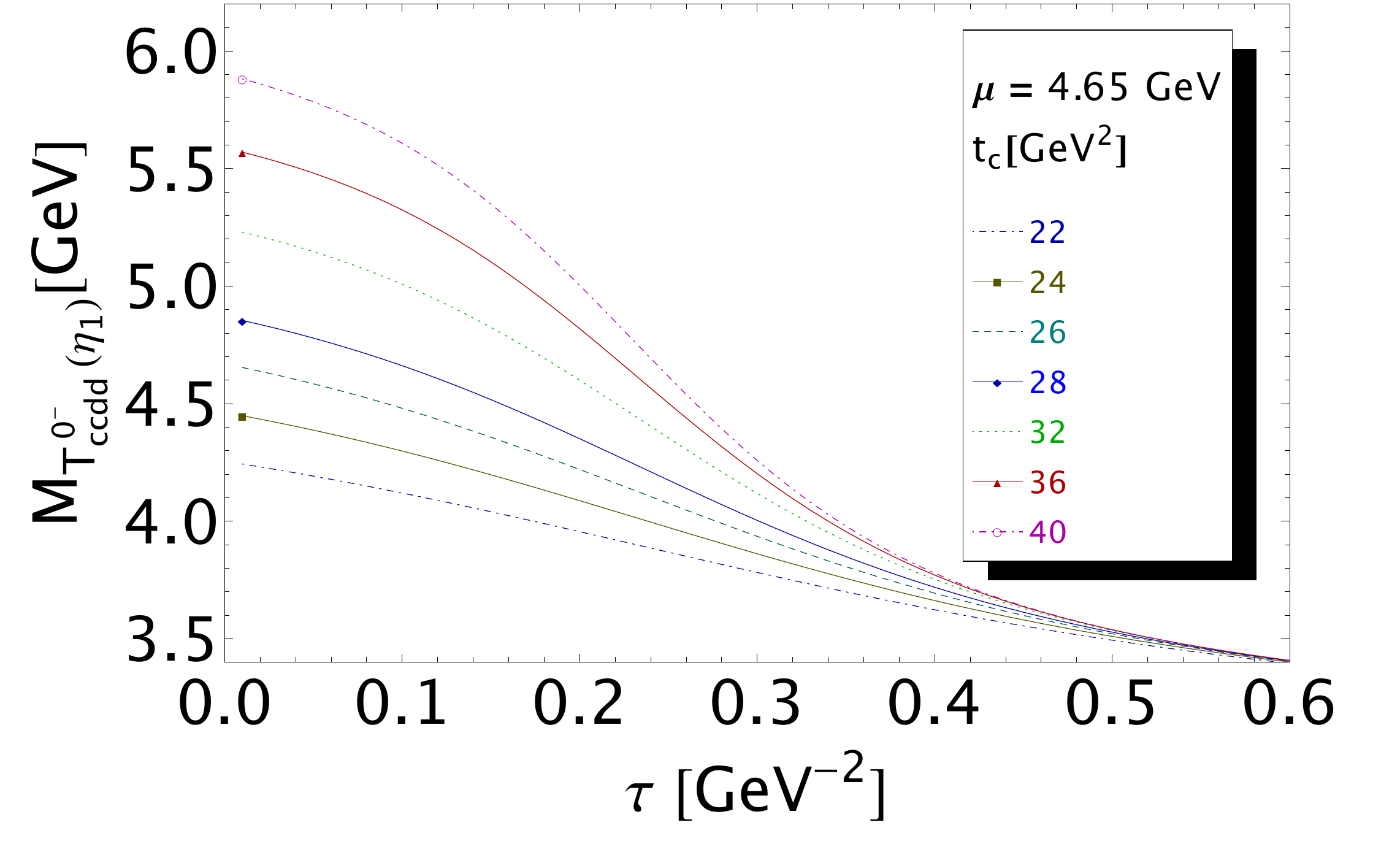}
\vspace*{-0.5cm}
\caption{\footnotesize  $f_{T_{cc\bar d\bar d}}$ and $M_{T_{cc\bar d\bar d}}$ for the $\eta_1$ current as a function of $\tau$ at NLO and for different values of $t_c$. We use $\mu$=4.65 GeV and the QCD inputs in Table\,\ref{tab:param}.  } 
\label{fig:eta1}
\end{center}
\vspace*{-0.5cm}
\end{figure}

We recalculate the corresponding QCD spectral function and redo the analysis. The results at NLO are shown in Fig.\,\ref{fig:eta1} versus $\tau$, for different values of $t_c$ and fixing $\mu=4.65$ GeV. The coupling presents minimum for the set: $(\tau,t_c)$ from (0.25,24) to (0.35,36)  (GeV$^{-2}$, GeV$^{2}$) while the mass presents inflexion points around such sets of values. One deduces at NLO:
\beq
f_{T_{cc\bar d\bar d}(0^-)(\eta_1)}= 402(45) ~{\rm keV},  ~~~~~~~~~~
M_{T_{cc\bar d\bar d}(0^-)(\eta_1)}= 3965(76)~{\rm MeV}.
\label{eq:tcc-eta1}
\eeq

\subsection*{\b Comments}
We have seen in the previous analysis that the mass of the $T_{cc}$ pseudoscalar state depends crucially on the choice of the interpolating currents. Our choice (Table\,\ref{tab:current}) and the $\eta_2,\eta_4$ currents of Ref\,\cite{ZHUT} lead to a high value of about 6 GeV of the mass while the other choices $\eta_1,\eta_5$ of Ref.\,\cite{ZHUT} to a mass around 4 GeV which can be easily reached by the experiments. However, it may well be that the physical $T_{cc}$ pseudoscalar state emerges from the mixing of these different interpolating currents,  where the study is beyond the scope of this paper.  Our results are compiled in Table\,\ref{tab:resc}.  For definiteness, we shall divide the currents into two classes : 
\subsection*{\d Class H0} 
This class corresponds to our current in Table\,\ref{tab:current} and $\eta_2,\eta_4$ of Ref.\,\cite{ZHUT} which lead to high mass values of $T_{cc}$ around 6 GeV. In the following, we shall consider our choice of current to be a representative of this Class H0 as these currents lead within the errors to the similar result.
\subsection*{\d Class L0} 
These are the currents $\eta_1$ and $\eta_5$ which lead to low mass values of $T_{cc}$ around 4 GeV.
\section{The $T_{cc\bar u\bar s}$ pseudoscalar state}
\subsection*{\b  Mass and decay constant from LSR at NLO from the {\it Class H0} currents}
We pursue our analysis by studying the effect of the $SU3$ breaking parameters ($m_s, \la\bar ss\ra$) on the coupling and mass of the $T_{cc\bar u\bar s}$ pseudoscalar state for the $\eta_4$ current. 
In so doing, we start with the usual moment ${\cal L}$ and ratio of moments ${\cal R}$. The analysis leads to a Figure similar to  Fig.\,\ref{fig:tcc}. The $(\tau,t_c)$ stability for the mass is obtained for (0.16,44) to (0.20,58)  (GeV$^{-2}$, GeV$^2$).  The coupling presents stability for (0.15,44) and inflexion point for (0.20,58)  (GeV$^{-2}$, GeV$^2$). 
Then, one deduces:
\beq
f_{T_{cc\bar u\bar s}(0^-)}= 983(147) ~{\rm keV},  ~~~~~~~~~~
M_{T_{cc\bar u\bar s}(0^-)}= 6102(172) ~{\rm MeV}. 
\eeq
Compared with the mass of the $T_{cc}$ state in Eq.\,\ref{eq:tcc0}, one deduces the mass-splitting:
\beq
M_{T_{cc\bar u\bar s}(0^-)}-M_{T_{cc\bar u\bar d}(0^-)}\simeq -201(172)~{\rm MeV},
\eeq
where the negative sign in the central value is mainly due to the ratio $\la\bar ss\ra/\la\bar uu\ra$ of condensates which is smaller than the one due to the $SU3$ breaking (see Table\,\ref{tab:param}). Unfortunately the error in the determination of the mass-splitting  is large which does not allow a decisive test of this phenomena\,\footnote{The double ratio of the sum rules (DRSR) is not helpful here as it does not present a clear $\tau$-stability.}. 

\subsection*{\b  Mass and decay constant from LSR at NLO from the {\it Class L0} currents}
This contribution comes from the $\eta_5$ current.
The curves are similar to the ones in Fig.\,\ref{fig:eta5}. The coupling presents stability for (0.50,24) up to  (0.56,36)  (GeV$^{-2}$, GeV$^2$) which satisfy the $R_{P/C}\geq 1$ condtion. The $(\tau,t_c)$ stability for the mass is obtained about the same set of values.   We deduce the results:
\bea
f_{T_{cc\bar u\bar s}(0^-)(\eta_5)}= 236(51) ~{\rm keV},  ~~~~~~~~~~
M_{T_{cc\bar u\bar s}(0^-)(\eta_5)}= 4251(127) ~{\rm MeV}. 
\eea

\d Compared with the mass of the $T_{cc\bar u\bar d}$ state in Eq.\,\ref{eq:eta5}, one deduces the mass-splitting:
\beq
M_{T_{cc\bar u\bar s}(0^-)(\eta_5)}-M_{T_{cc\bar u\bar d}(0^-)(\eta_5)}\simeq -129(127)~{\rm MeV}, 
\eeq
where like in the previous case, the negative sign in the central value is mainly due to the ratio $\la\bar ss\ra/\la\bar uu\ra$ of condensates. 
\section{The $T_{cc\bar s\bar s}$ pseudoscalar state}
\subsection*{\b  Mass and decay constant from the {\it Class H0} currents}
The contribution from our choice of current in Table\,\ref{tab:current} is identical to zero in this case. 
The analysis for the $\eta_2$ current gives curves similar to Fig.\,\ref{fig:eta5}. The $(\tau,t_c)$ stability for the mass is obtained for (0.18,42) to (0.22,58)  (GeV$^{-2}$, GeV$^2$).   The coupling presents stability for (0.16,42) and inflexion point for (0.22,58)  (GeV$^{-2}$, GeV$^2$). We deduce the results:
\beq
f_{T_{cc\bar s\bar s}(0^-)(\eta_2)}= 1164(168)~{\rm keV},  ~~~~~~~~~~
M_{T_{cc\bar s\bar s}(0^-)(\eta_2)}= 5946(180) ~{\rm MeV}. 
\eeq
\d Compared with the mass of the $T_{cc}$ state in Eq.\,\ref{eq:tcc0-eta2}, one deduces the mass-splitting:
\beq
M_{T_{cc\bar s\bar s}(0^-)(\eta_2)}-M_{T_{cc\bar d\bar d}(0^-)(\eta_2)}\simeq -321(157)~{\rm MeV}. 
\eeq
\subsection*{\b  Mass and decay constant from  the {\it Class L0} currents}
\subsection*{\d $\eta_1$ current}
Here, {\it Class L0} is represented by the current $\eta_1$. 
The analysis is similar to Fig.\,\ref{fig:eta1}. The $(\tau,t_c)$ inflexion points for the mass are obtained for (0.25,24) to (0.36,36)  (GeV$^{-2}$, GeV$^2$).The coupling presents minimum for (0.25,24) to (0.36,36)  (GeV$^{-2}$, GeV$^2$).  We deduce the results:
\beq
f_{T_{cc\bar s\bar s}(0^-)(\eta_1)}= 322(42) ~{\rm keV},  ~~~~~~~~~~
M_{T_{cc\bar s\bar s}(0^-)(\eta_1)}= 4096(96) ~{\rm MeV}. 
\eeq
\subsection*{\d SU3 mass-splitting}
 Compared with the mass of the $T_{cc}$ state in Eq.\,\ref{eq:tcc-eta1}, one deduces the mass-splitting:
\beq
M_{T_{cc\bar s\bar s}(0^-)(\eta_1)}-M_{T_{cc\bar d\bar d}(0^-)(\eta_1)}\simeq +131(71)~{\rm MeV}.  
\eeq
\section{The $T_{cc\bar q\bar q'}$ vector state}
\subsection*{\b The $T_{cc\bar u\bar d}$ vector state for the current in Table\,\ref{tab:current}}
 We study the behaviour of the mass and decay constant for different values of $\tau$ and $t_c$ and for fixed value of $\mu=4.65$ GeV deduced  from our previous analysis of the four-quark states. The curves are similar to the ones in Figs.\,\ref{fig:tcc}. We observe stabilities for the coupling for the sets  $(\tau , t_c)$ from (0.14, 46) to (0.18,60) (GeV$^{-2}$, GeV$^{2}$) and minimum for the mass from  (0.15, 46) to (0.18,60)  (GeV$^{-2}$, GeV$^{2}$) which (almost) coincide with the ones of the pseudoscalar state. Then, we deduce the conservative estimate :
\beq
f_{T_{cc\bar u\bar d}(1^-)}= 841(117) ~{\rm keV},  ~~~~~~~~~~
M_{T_{cc\bar u\bar d}(1^-)}= 6212(172) ~{\rm MeV}, 
\label{eq:tcc1}
\eeq
where we have used $\Delta\tau=\pm 0.02$ GeV$^{-2}$ around the minimum and taken the mean of the errors due to the change of $\tau$ for the two extremal values of $t_c$.

 From the previous result, one deduces the mass-splittings between the chiral multiplets:
\beq
M_{T_{cc\bar u\bar d}(1^-)}-M_{T_{cc\bar u\bar d}(1^+)}\simeq 2326~{\rm MeV},
\eeq
where we have used the $1^+$ mass $M_{T_{cc}(1^+)}\simeq 3886$ MeV from\,\cite{Tcc}.  This large mass-splitting is in line with our previous findings for the $X_c,Z_c$-like states obtained in\,\cite{MOLE16X}:
\beq
M_{Z_c(1^-)}-M_{Z_{c}(1^+)}\simeq 1449~{\rm MeV}.
\eeq
\subsection*{\b The $T_{cc\bar d\bar d}$ vector state from {\it Class H1} currents of Ref.\,\cite{ZHUT}}
In the $1^-$ channel, these {\it Class H1} currents are the $\eta_5$ (similar to ours) and $\eta_2$ currents. 
We study the behaviour of the mass and decay constant for the $\eta_2$ current for different values of $\tau$ and $t_c$ and for fixed value of $\mu=4.65$ GeV. The curves are similar to  Figs.\,
\ref{fig:tcc}. We observe stabilities for the coupling for the sets  $(\tau , t_c)$ from (0.13, 48) to (0.16,60) (GeV$^{-2}$, GeV$^{2}$) and for the mass from  (0.14, 48) to (0.16,60)  (GeV$^{-2}$, GeV$^{2}$) which (almost) coincide with the ones of the pseudoscalar state. Then, we deduce the conservative estimate :
\beq
f_{T_{cc\bar d\bar d}(1^-)(\eta_2)}= 1165(107)~{\rm keV},  ~~~~~~~~~~
M_{T_{cc\bar d\bar d}(1^-)(\eta_2)}= 6362(140) ~{\rm MeV}. 
\label{eq:tcc1-eta2}
\eeq
\subsection*{\b The $T_{cc\bar q\bar q'}$ vector state from {\it Class L1} currents}
These are the $\eta_1$ and $\eta_6$ currents of Ref.\,\cite{ZHUT}. 
\subsection*{\d  $\eta_1$ current}
The analysis is similar to Figs.\,\ref{fig:eta1} for different values of $\tau$ and $t_c$ and for fixed value of $\mu=4.65$ GeV. We observe stabilities for the coupling for the sets  $(\tau , t_c)$ from (0.23, 26) to (0.32,40) (GeV$^{-2}$, GeV$^{2}$) and for the mass presents inflexion points around these values which (almost) coincide with the ones of the pseudoscalar state. Then, we deduce the conservative estimate :
\beq
f_{T_{cc\bar d\bar d}(1^-)(\eta_1)}= 380(39) ~{\rm keV},  ~~~~~~~~~~
M_{T_{cc\bar d\bar d}(1^-)(\eta_1)}= 4170(110) ~{\rm MeV}. 
\label{eq:tcc1-eta1}
\eeq
\subsection*{\d  $\eta_6$  current}

{\it -- Analysis}

The analysis for different values of $\tau$ and $t_c$ and for fixed value of $\mu=4.65$ GeV is similar to  Figs.\,\ref{fig:eta1}. We observe stabilities for the coupling for the sets  $(\tau , t_c)$ from (0.22, 26) to (0.31,40) (GeV$^{-2}$, GeV$^{2}$) and inflexion points around these sets of values. Then, we deduce the conservative estimate :
\beq
f_{T_{cc\bar u\bar d}(1^-)(\eta_6)}= 293(51)~{\rm keV},  ~~~~~~~~~~
M_{T_{cc\bar u\bar d}(1^-)(\eta_6)}= 4113(108)~{\rm MeV}. 
\label{eq:tcc1-eta6}
\eeq

\section{The $T_{cc\bar s\bar u}$ vector state}
 \subsection*{\b Class H1: our  current in Table\,\ref{tab:current}}
 The analysis  for different values of $\tau$ and $t_c$ and for fixed value of $\mu=4.65$ GeV is similar to Figs.\,\ref{fig:tcc}. We observe stabilities for the coupling for the sets  $(\tau , t_c)$ from (0.16, 44) to (0.20,58) (GeV$^{-2}$, GeV$^{2}$) and inflexion points for the mass from  (0.17, 44) to (0.20,58)  (GeV$^{-2}$, GeV$^{2}$). Then, we deduce the conservative estimate :
\bea
f_{T_{cc\bar u\bar s}(1^-)}= 558(76) ~{\rm keV},  ~~~~~~~~~~
M_{T_{cc\bar u\bar s}(1^-)}= 6064(167)~{\rm MeV},
\label{eq:tccs1}
\eea
and the mass-splitting due to SU3 breakings:
\beq
M_{T_{cc\bar u\bar s}(1^-)}-M_{T_{cc}(1^-)}\simeq -147(167)~{\rm MeV},  
\eeq
\subsection*{\b  {\it Class L1}  $\eta_6$ current}
The analysis for different values of $\tau$ and $t_c$ and for fixed value of $\mu=4.65$ GeV is similar to Figs.\,\ref{fig:eta1}. We observe stabilities for the coupling for the sets  $(\tau , t_c)$ from (0.23, 26) to (0.32,40) (GeV$^{-2}$, GeV$^{2}$) and inflexion points for the mass around previous values. Then, we deduce the conservative estimate :
\bea
f_{T_{cc\bar u\bar s}(1^-)(\eta_6)}= 262(29) ~{\rm keV},  ~~~~~~~~~~
M_{T_{cc\bar u\bar s}(1^-)(\eta_6)}= 4157(100) ~{\rm MeV},
\label{eq:tccs1-eta6}
\eea
and the mass-splitting due to SU3 breakings:
\beq
M_{T_{cc\bar u\bar s}(1^-)(\eta_6)}-M_{T_{cc\bar u\bar d}(1^-)(\eta_6)}\simeq +44(100)~{\rm MeV}.  
\eeq

\section{The $T_{cc\bar s\bar s}$ vector state}
From our choice of the current, this contribution is trivialy zero. Using the currents in Ref.\,\cite{ZHUT}
 we have :
\subsection*{\b {\it Class H1} $\eta_2$  current}
The analysis for different values of $\tau$ and $t_c$ and for fixed value of $\mu=4.65$ GeV is similar to the one in Figs.\,\ref{fig:tcc}. We observe stabilities of the coupling for the sets  $(\tau , t_c)$ from (0.14, 44) to (0.20,58) (GeV$^{-2}$, GeV$^{2}$) and inflexion points for the mass from  (0.17, 46) to (0.20,58)  (GeV$^{-2}$, GeV$^{2}$). Then, we deduce the conservative estimate :
\beq
f_{T_{cc\bar s\bar s}(1^-)(\eta_2)}= 1085(109) ~{\rm keV},  ~~~~~~~~~~
M_{T_{cc\bar s\bar s}(1^-)(\eta_2)}= 6110(154) ~{\rm MeV},
\label{eq:tccss1-eta2}
\eeq
and the mass-splitting due to SU3 breakings:
\beq
M_{T_{cc\bar s\bar s}(1^-)(\eta_2)}-M_{T_{cc\bar d\bar d}(1^-)(\eta_2)}\simeq -252(154)~{\rm MeV}.  
\eeq
\subsection*{\b {\it Class L1} $\eta_1$  current}
The analysis for different values of $\tau$ and $t_c$ and for fixed value of $\mu=4.65$ GeV is similar to  Figs.\,\ref{fig:eta1}. We observe stabilities for the coupling for the sets  $(\tau , t_c)$ from (0.26, 26) to (0.34,40) (GeV$^{-2}$, GeV$^{2}$) and inflexion points for the mass around these points. Then, we deduce the conservative estimate:
\beq
f_{T_{cc\bar s\bar s}(1^-)(\eta_1)}= 289(37) ~{\rm keV}, ~~~~~~~~~~
M_{T_{cc\bar s\bar s}(1^-)(\eta_1)}= 4210(103) ~{\rm MeV},
\label{eq:tccss-eta1}
\eeq
and the mass-splitting due to SU3 breakings:
\beq
M_{T_{cc\bar s\bar s}(1^-)(\eta_1)}-M_{T_{cc\bar d\bar d}(1^-)(\eta_1)}\simeq+40(103)~{\rm MeV}.  
\eeq
\begin{figure}[hbt]
\begin{center}
\centerline {\hspace*{-7.5cm} \bf a)\hspace{8cm} b)}
\includegraphics[width=8cm]{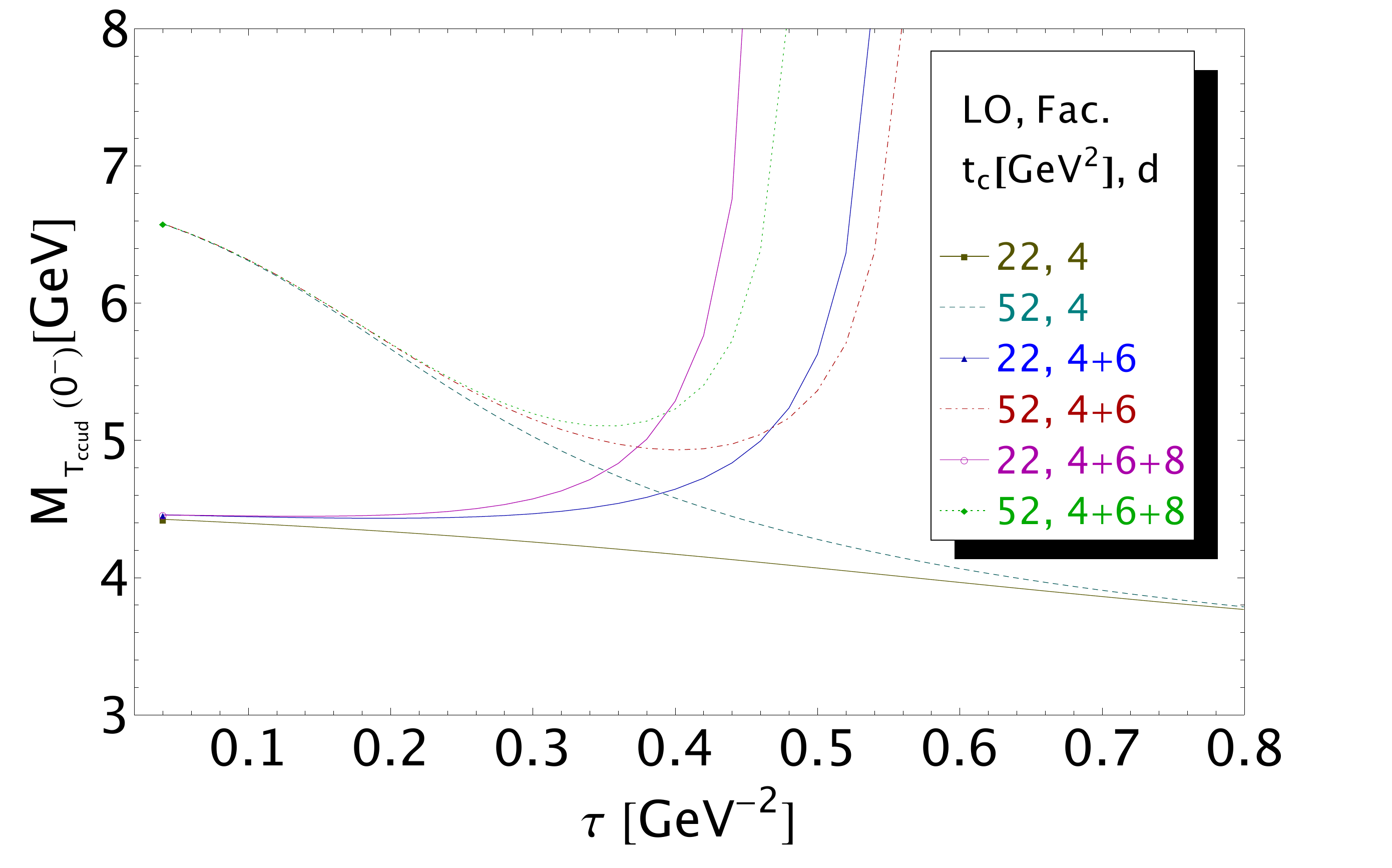}
\includegraphics[width=8cm]{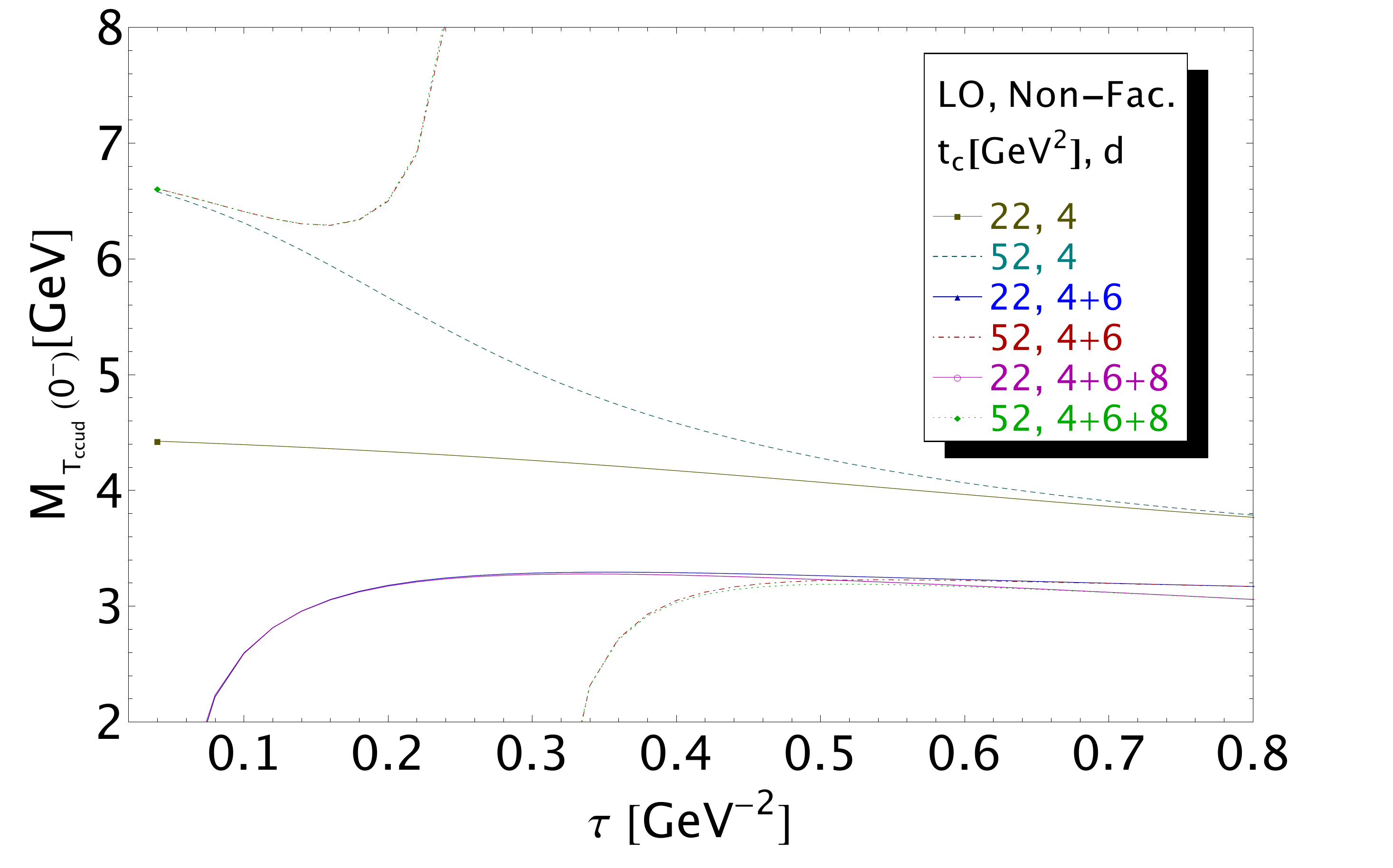}
\vspace*{-0.5cm}
\caption{\footnotesize  $M_{T_{cc\bar d\bar d}}$ for the $\eta_4$ current as a function of $\tau$ at LO  for two values of $t_c$ and for different truncations of the OPE ($d$ indicate the dimension of the condensate). We use the QCD inputs in Table\,\ref{tab:param}.  a) corresponds to the use of factorization of the four-quark condensate; b) the use of non-factorization with the value given inTable\,\ref{tab:param}. } 
\label{fig:zhu-lo}
\end{center}
\vspace*{-0.5cm}
\end{figure} 

\vspace*{-0.5cm} 
\begin{center}
   {\footnotesize
\begin{table}[H]
\setlength{\tabcolsep}{0.21pc}
    {
  \begin{tabular}{lllllll | lll}
&\\
\hline
\hline
\multicolumn{7}{c}{\bf Our Work}&\multicolumn{3}{c}{Ref.\,\cite{ZHUT}}\\
\hline
States&Current  &$t_c$ [GeV$^2$] &$\tau$  [GeV$^{-2}$]  & $f^{NLO} _{T_{ccqq'}}$&$\tau$  [GeV$^{-2}$] &$M^{NLO}_{T_{ccqq'}}$&$t_c$ & $\tau$  [GeV$^{-2}$] &$M^{LO}_{T_{ccqq'}}$\\ 
 \hline 
 \bf Class H &&&&&&&\\
{ \bma  $0^-$} &&&&&&&\\
$T_{cc\bar u\bar d}$&$ {\cal O}_T^{0^-},\eta_4$&$50 \to 60$&$0.12\to 0.18$&1559(133)&$0.16\to 0.18$&6303(123)&22&$0.29\to 0.37$& 4430(130) \\
$T_{cc\bar d\bar d}$ & $\eta_2$&$48\to 60$&$0.14\to 0.18$&1348(165)&$0.16\to 0.18$&6267(148)&&&-- \\
$T_{cc\bar u\bar s}$& $ {\cal O}_{T^{0^-}_{us}},\eta_4$&$44\to 58$&$0.16\to 0.20$&983(147)&$0.15\to 0.20$&6102(172)&&&-- \\
$T_{cc\bar s\bar s}$ & $\eta_2$&$42\to 58$&$0.16\to 0.22$&1164(168)&$0.18\to 0.22$&5946(180)&&&-- \\
{  $1^-$} &&&&&&&\\
 $T_{cc\bar u\bar d}$&  $ {\cal O}_T^{1^-},\eta_5 $&$46\to 60$&$0.14\to 0.18$&841(117)&$0.15\to 0.18$&6212(172)&&&--\\
  $T_{cc\bar d\bar d}$ &  $\eta_2 $&$48\to 60$&$0.13\to 0.16$&1165(107)&$0.14\to 0.16$&6362(140)&&&--\\
   $T_{cc\bar u\bar s}$& $ {\cal O}_{T^{1^-}_{us}},\eta_5$&$44\to 58$&$0.16\to 0.20$&558(76)&$0.17\to 0.20$&6064(167)&&&--\\
 $T_{cc\bar s\bar s}$&  $ \eta_2 $&$46\to 58$&$0.15\to 0.20$&1085(109)&$0.17\to 0.20$&6110(154)&&&-- \\
 
  \bf Class L&&&&&&&\\
  { \bma  $0^-$} &&&&&&&\\
$T_{cc\bar u\bar d}$&$ \eta_5$&$24\to 36$&$0.32\to 0.45$&289(42)&$0.41\to 0.47$&4380(128)&23&$0.27\to 0.40$&4410(140) \\
$T_{cc\bar d\bar d}$& $\eta_1$&$24\to 36$&$0.25\to 0.35$&402(45)&$0.25\to 0.35$&3965(76)&24&$0.26\to 0.33$&4430(120) \\
 $T_{cc\bar u\bar s}$& $ \eta_5$&$24\to 36$&$0.50\to 0.56$&236(51)&$0.50\to 0.56$& 4251(127)&24&$0.24\to 0.39$&4500(160) \\

  $T_{cc\bar s\bar s}$&$\eta_1$&$24\to 36$&$0.25\to 0.36$&322(42)&$0.25\to 0.36$&4096(96)&25&$0.25\to 0.36$&4460(130)\\
{  $1^-$} &&&&&&&\\
  $T_{cc\bar u\bar d}$&  $\eta_6 $&$26\to 40$&$0.22\to 0.31$&293(51)&$0.22\to 0.31$&4113(108)&23&$0.27\to 0.32$&4340(160)\\
 $T_{cc\bar d\bar d}$& $ \eta_1 $&$26\to 40$&$0.23\to 0.32$&380(39)&$0.23\to 0.32$&4170(110)&23&$0.28\to 0.33$&4350(140) \\
   $T_{cc\bar u\bar s}$& $ \eta_6$&$26\to 40$&$0.23\to 0.32$&262(29)&$0.23\to 0.32$&4157(100) &23&$0.27\to 0.35$&4350(160)\\

 $T_{cc\bar s\bar s}$&  $ \eta_1 $&$26\to 40$&$0.26\to 0.35$&289(37)&$0.26\to 0.35$&4210(103)&24&$0.27\to 0.36$&4470(130) \\

   \hline\hline
  \vspace*{-0.75cm}
\end{tabular}}
 \caption{Our predictions of the couplings [keV] and masses [MeV] are compared with the ones in Ref.\,\cite{ZHUT}. 
 For the {\it Class H} currents we found that the results quoted by Ref.\,\cite{ZHUT} using factorization of the four-quark condensates do not satisfy stability criteria while the ones from non-factorization have imaginary couplings. More details on the comparison with Ref.\,\cite{ZHUT} are given in the text.}  

\label{tab:resc}
\end{table}
} 
\end{center}
\vspace*{-0.75cm} 
\section{Class H pseudoscalar states in Table\,\ref{tab:resc}  compared with Ref.\cite{ZHUT}}
\subsection*{\b  Check of the mass at LO from our current  $\equiv\eta_4$ used in Ref.\cite{ZHUT} }
\d The $\eta_4$ current used in Ref.\,\cite{ZHUT} (see Table\,\ref{tab:zhu})  is similar to ours.
By comparing the QCD spectral function, we realize that the contributions of the mixed condensate differ in size in the two papers\,\footnote{Notice that Ref.\,\cite{ZHUT} uses an unusual negative sign for parametrizing the mixed condensate by the $\la \bar qq\ra$ one.}.

\d Comparing our previous result at LO  in the $\overline{MS}$ scheme with our set of parameters (see Fig.\,\ref{fig:tcc-lo}):
\beq
M^{LO}_{T_{cc\bar d\bar d}(0^-)}\simeq 6560~{\rm MeV}, 
\label{eq:tcc-lo}
\eeq
with the one in \,\cite{ZHUT} at the same order:
\beq
M^{LO}_{T_{cc\bar d\bar d}(0^-)(\eta_4)}= 4430(130)~{\rm MeV}, 
\eeq
 we note that the value of $t_c\simeq 22$ GeV$^2$ used there is much smaller than ours $(48\sim 60)$ GeV$^2$ while the optimal choice of $\tau\simeq 0.33$ GeV$^{-2}$ is much bigger than ours obtained at  the stability point: $\tau\simeq 0.13$ GeV$^{-2}$ at LO.  

\d We repeat the analysis by using the same set of parameters as Ref.\cite{ZHUT} at LO. We use $t_c=22$ GeV$^2$ (choice of Ref.\,\cite{ZHUT}) and $t_c= 52$ GeV$^2$ (a representative of our choice). The analysis is shown in Fig.\,\ref{fig:zhu-lo} versus the sum rule variable $\tau$ and for different truncations of the OPE ($d$ indicates the dimension of the condensate involved in the OPE):

\d We notice that, for the factorized case a), there is no $\tau$-stability in the region $\tau\simeq (0.29-0.37)$ GeV$^{-2}$) and $t_c=22$ GeV$^2$ favoured by Ref.\,\cite{ZHUT} corresponding to their quoted result 4450 MeV. On the contrary, a clear stability around $\tau=0.4$ GeV$^{-2}$ is obtained for our choice $t_c=52$ GeV$^2$ and truncating the OPE at $d=6$ to which corresponds a mass of about 4930 MeV. 
\begin{table}[H]
\setlength{\tabcolsep}{0.25pc}
{\scriptsize{
\begin{tabular}{ll ll  ll  ll ll ll ll ll c}
\hline
\hline
                States & Currents
                    &\multicolumn{1}{c}{$\Delta t_c$}
					&\multicolumn{1}{c}{$\Delta \tau$}
					&\multicolumn{1}{c}{$\Delta \mu$}
					&\multicolumn{1}{c}{$\Delta \alpha_s$}
					&\multicolumn{1}{c}{$\Delta PT$}
					&\multicolumn{1}{c}{$\Delta m_s$}
					&\multicolumn{1}{c}{$\Delta m_c$}
					&\multicolumn{1}{c}{$\Delta \overline{\psi}\psi$}
					&\multicolumn{1}{c}{$\Delta \kappa$}					
					&\multicolumn{1}{c}{$\Delta G^2$}
					&\multicolumn{1}{c}{$\Delta M^{2}_{0}$}
					&\multicolumn{1}{c}{$\Delta \overline{\psi}\psi^2$}
					&\multicolumn{1}{c}{$\Delta G^3$}
					&\multicolumn{1}{c}{$\Delta OPE$}
					&\multicolumn{1}{r}{Total[keV]}
\\
					
\hline
{\it Class H} &&&&&&&&&&\\
$0^-$ &&&&&&&&&&\\
$T_{cc\bar{u}\bar{d}}$ & ${\cal O}^{0^-}_{T}$, $\eta_4$ &125&6.00&2.21&9.86&38.3&$\cdots$&13.0&0.00&$\cdots$&0.00&0.00&12.7&0.09&6.84&133 \\
$T_{cc\bar{d}\bar{d}}$ & $\eta_2$ &133&2.00&1.26&5.36&97.1&$\cdots$&10.4&0.00&$\cdots$&0.25&0.00&9.18&0.03&4.25&165 \\
$T_{cc\bar{u}\bar{s}}$ & ${\cal O}^{0^-}_{T_{us}}$, $\eta_4$ &138&1.00&1.66&7.43&48.7&0.45&9.52&0.74&5.17&0.02&0.12&9.34&0.08&5.76&147 \\
$T_{cc\bar{s}\bar{s}}$ & $\eta_2$ &147&1.00&1.58&6.83&76.5&0.84&11.6&1.53&14.0&0.41&0.56&11.7&0.04&7.93&168 \\
$1^-$ &&&&&&&&&&\\
$T_{cc\bar{u}\bar{d}}$ & ${\cal O}^{1^-}_{T}$, $\eta_5$ &106&1.00&1.01&4.37&41.3&$\cdots$&7.40&0.00&$\cdots$&0.14&0.00&8.97&0.02&5.36&117\\
$T_{cc\bar{d}\bar{d}}$ & $\eta_2$ &98.0&2.00&1.10&4.67&39.0&$\cdots$&9.17&0.00&$\cdots$&0.08&0.00&14.9&0.03&8.08&107 \\
$T_{cc\bar{u}\bar{s}}$ & ${\cal O}^{1^-}_{T_{us}}$, $\eta_5$ &70.0&0.40&0.74&3.21&26.1&0.22&5.03&0.36&2.50&0.10&0.60&4.56&0.02&2.61&76 \\
$T_{cc\bar{s}\bar{s}}$ & $\eta_2$ &100&1.00&1.21&5.24&40.1&0.82&8.54&1.31&9.39&0.08&0.26&7.56&0.03&4.03&109\\
\\
{\it Class L} &&&&&&&&&&\\
$0^-$ &&&&&&&&&&\\
$T_{cc\bar{u}\bar{d}}$ & $\eta_5$ &36.0&4.00&2.25&10.5&4.11&$\cdots$&7.89&0.00&$\cdots$&0.09&0.00&8.46&0.05&10.7&42 \\
$T_{cc\bar{d}\bar{d}}$ & $\eta_1$ &20.0&0.80&1.88&8.97&6.19&$\cdots$&4.87&0.00&$\cdots$&0.23&0.00&26.0&0.09&29.6&45 \\
$T_{cc\bar{u}\bar{s}}$ & $\eta_5$ &43.0&0.00&3.13&14.5&8.71&0.34&11.1&0.83&3.94&0.03&0.00&8.79&0.07&15.6&51 \\
$T_{cc\bar{s}\bar{s}}$ & $\eta_1$ &23.0&1.00&1.47&6.97&5.09&0.32&4.04&0.27&18.9&0.30&0.22&17.9&0.12&21.0&42 \\
$1^-$ &&&&&&&&&&\\
$T_{cc\bar{u}\bar{d}}$ & $\eta_6$ &16.0&1.00&1.19&5.66&1.35&$\cdots$&2.88&0.00&$\cdots$&0.03&0.00&17.7&0.06&43.6&51 \\
$T_{cc\bar{d}\bar{d}}$ &$\eta_1$ &24.0&1.00&1.58&7.50&10.3&$\cdots$&4.40&0.00&$\cdots$&0.29&0.00&20.1&0.08&19.9&39\\
$T_{cc\bar{u}\bar{s}}$ & $\eta_6$ &16.0&1.00&1.06&5.08&1.57&0.11&2.62&0.11&7.80&0.02&0.20&15.1&0.06&17.2&29 \\
$T_{cc\bar{s}\bar{s}}$ & $\eta_1$ &20.0&1.00&1.31&6.18&15.9&0.18&3.86&0.26&15.0&0.34&0.18&14.1&0.10&15.8&37 \\
\\
\hline
\hline
\end{tabular}
}}
 \caption{Sources of errors of $T_{cc\bar u\bar d}$, $T_{cc\bar d\bar d}$, $T_{cc\bar u\bar s}$, $T_{cc\bar s\bar s}$ couplings.The error due to $M_{T_{cc\bar q\bar q'}}$ is intrinsically included in the estimate of the coupling as we use a mass corresponding to each value of  $t_c$. We take $\ve \Delta \mu\ve=0.05$ GeV and $\ve \Delta \tau\ve =0.02$ GeV$^{-2}$.}

\label{tab:error-fc}
\end{table}


\begin{table}[H]
\setlength{\tabcolsep}{0.22pc}
{\scriptsize{
\begin{tabular}{ll ll  ll  ll ll ll ll ll c}
\hline
\hline
                States & Currents
                    &\multicolumn{1}{c}{$\Delta t_c$}
					&\multicolumn{1}{c}{$\Delta \tau$}
					&\multicolumn{1}{c}{$\Delta \mu$}
					&\multicolumn{1}{c}{$\Delta \alpha_s$}
					&\multicolumn{1}{c}{$\Delta PT$}
					&\multicolumn{1}{c}{$\Delta m_s$}
					&\multicolumn{1}{c}{$\Delta m_c$}
					&\multicolumn{1}{c}{$\Delta \overline{\psi}\psi$}
					&\multicolumn{1}{c}{$\Delta \kappa$}					
					&\multicolumn{1}{c}{$\Delta G^2$}
					&\multicolumn{1}{c}{$\Delta M^{2}_{0}$}
					&\multicolumn{1}{c}{$\Delta \overline{\psi}\psi^2$}
					&\multicolumn{1}{c}{$\Delta G^3$}
					&\multicolumn{1}{c}{$\Delta OPE$}
					&\multicolumn{1}{r}{Total [MeV]}
\\
					
\hline
{\it Class H} &&&&&&&&&&\\
$0^-$ &&&&&&&&&&\\
$T_{cc\bar{u}\bar{d}}$ & ${\cal O}^{0^-}_{T}$, $\eta_4$ &95.0&32.0&1.16&4.69&9.42&$\cdots$&11.5&0.00&$\cdots$&0.23&0.00&61.5&0.42&39.7&123 \\
$T_{cc\bar{d}\bar{d}}$ & $\eta_2$ &121&31.0&0.68&2.20&27.2&$\cdots$&10.9&0.00&$\cdots$&0.88&0.00&61.1&0.16&38.6&148 \\
$T_{cc\bar{u}\bar{s}}$ & ${\cal O}^{0^-}_{T_{us}}$, $\eta_4$ &150&27.0&1.37&5.65&5.89&0.98&12.5&2.77&31.1&0.32&0.87&58.1&0.53&40.9&172 \\
$T_{cc\bar{s}\bar{s}}$ & $\eta_2$ &143&23.0&1.09&4.22&29.3&1.05&13.2&4.67&67.3&1.34&2.92&59.6&0.22&46.9&180 \\
$1^-$ &&&&&&&&&&\\
$T_{cc\bar{u}\bar{d}}$ & ${\cal O}^{1^-}_{T}$, $\eta_5$ &149&26.0&1.01&3.99&19.3&$\cdots$&11.7&0.00&$\cdots$&0.50&0.00&65.1&0.17&43.2&172\\
$T_{cc\bar{d}\bar{d}}$ & $\eta_2$ &113&28.0&0.50&1.61&20.0&$\cdots$&9.06&0.00&$\cdots$&0.23&0.00&63.1&0.14&37.4&140 \\
$T_{cc\bar{u}\bar{s}}$ & ${\cal O}^{1^-}_{T_{us}}$, $\eta_5$ &139&26.0&1.22&4.97&22.5&0.83&12.8&2.53&33.4&0.62&0.80&62.7&0.20&45.2&167 \\
$T_{cc\bar{s}\bar{s}}$ & $\eta_2$ &109&25.0&0.87&3.30&19.2&1.81&11.5&5.21&70.3&0.36&2.05&61.9&0.21&44.9&154 \\
\\
{\it Class L} &&&&&&&&&&\\
$0^-$ &&&&&&&&&&\\
$T_{cc\bar{u}\bar{d}}$ & $\eta_5$ &80.0&8.00&16.5&76.7&8.23&$\cdots$&60.0&0.00&$\cdots$&0.34&0.00&1.33&0.24&2.24&128 \\
$T_{cc\bar{d}\bar{d}}$ & $\eta_1$ &8.00&66.0&1.62&7.85&2.08&$\cdots$&2.87&0.00&$\cdots$&0.46&0.00&24.5&0.06&27.1&76 \\
$T_{cc\bar{u}\bar{s}}$ & $\eta_5$ &40.0&1.40&19.8&91.8&1.70&7.60&72.8&5.23&1.31&0.68&0.65&8.79&0.35&17.0&127 \\
$T_{cc\bar{s}\bar{s}}$ & $\eta_1$ &11.0&78.0&1.24&6.18&0.50&3.06&0.51&0.29&29.9&0.47&0.36&29.2&0.17&35.1&96 \\
$1^-$ &&&&&&&&&&\\
$T_{cc\bar{u}\bar{d}}$ & $\eta_6$ &14.0&94.0&1.27&6.06&3.25&$\cdots$&3.08&0.00&$\cdots$&0.36&0.00&4.29&0.00&51.0&108 \\
$T_{cc\bar{d}\bar{d}}$ &$\eta_1$ &18.0&101&1.58&7.77&9.90&$\cdots$&1.44&0.00&$\cdots$&0.49&0.00&11.9&0.18&35.5&110\\
$T_{cc\bar{u}\bar{s}}$ & $\eta_6$ &15.0&94.0&1.16&5.57&5.92&1.51&2.48&0.37&7.25&0.41&0.17&7.63&0.01&28.3&100 \\
$T_{cc\bar{s}\bar{s}}$ & $\eta_1$ &21.0&90.0&1.33&6.65&15.2&2.70&0.40&0.34&5.39&0.42&0.46&6.64&0.34&41.4&103 \\
\\
\hline
\hline
\end{tabular}
}}
 \caption{Sources of errors of $T_{cc\bar u\bar d}$, $T_{cc\bar d\bar d}$, $T_{cc\bar u\bar s}$, $T_{cc\bar s\bar s}$ masses. We take $\ve \Delta \mu\ve=0.05$ GeV and $\ve \Delta \tau\ve =0.02$ GeV$^{-2}$.}

\label{tab:error-mc}
\end{table}

\d For the non-factorized four-quark condensate in case b), we find that the result for the choice $t_c=22$ GeV$^2$ of Ref.\,\cite{ZHUT} stabilizes for $\tau\approx 0.3$ GeV$^{-2}$ corresponding to a mass of about 3290 MeV. One notices that, for this choice of $t_c$ and $\tau$, the contribution of the $d\geq 6$ condensates is about  26\% of the lowest dimension one which may question the convergence of the OPE at this scale. 

\d For the non-factorized four-quark condensate in case b) and for our choice $t_c=52$ GeV$^2$, one has two stabilities : 

-- $\tau\simeq 0.16$ GeV$^{-2}$ where the contribution of high-dimension condensates is only 6\% of the lowest dimension ones which guarantees the convergence of the OPE at this scale. One obtains a mass:
\beq
M^{LO}_{T_{cc\bar u\bar d}(0^-)(\eta_4)}\simeq 6294~{\rm MeV}, 
\eeq
for the set of QCD parameters used in Ref.\,\cite{ZHUT} apart the four-quark condensate. This mass differs slightly from the one in Eq.\,\ref{eq:tcc-lo} due to the different QCD inputs used here and in Table\,\ref{tab:param}.  At this scale the convergence of the OPE is quite satisfactory leading to a systematics:
\beq
\Delta M^{d=8}_{T_{cc\bar u\bar d}(0^-)(\eta_4)} = \pm 3~{\rm MeV}. 
\eeq

-- $\tau\simeq 0.5$ GeV$^{-2}$ where the mass 3170 MeV coincides with the one obtained from the choice $t_c=22$ GeV$^2$ in Ref.\,\cite{ZHUT} but again the contribution of the $d\geq 6$ condensates is about  26\% of the lowest dimension one which may question the convergence of the OPE at this scale.

\subsection*{\b  Check of the coupling at LO from our current $\equiv\eta_4$ used in Ref.\,\cite{ZHUT} }
\subsection*{\d Factorized four-quark condensate}
For this case and fixing $M_{T_{cc\bar u\bar d}(0^-)}\simeq 4.45$ GeV from Ref.\,\cite{ZHUT}, we
do not find $\tau$-stability for the coupling for the range $\tau\simeq (0.3\sim 0.4)$ GeV$^{-2}$ favoured by Ref.\,\cite{ZHUT} where the coupling is about $(200\sim 590)$ keV. 
\subsection*{\d Non-factorized four-quark condensate}
We find that the inclusion of the $d=6,8$ condensate leads to an imaginary value of the coupling  for the large value of $\tau\simeq 0.5$ GeV$^{-2}$ and low mass $M_{T_{cc}(0^-)}\simeq $ 4 GeV. Therefore, we shall not consider this 2nd choice which is not physical. 

\subsection*{\b Comments on previous analysis}
\d From the previous analysis, we have considered as optimal results the ones coming from the non-factorized four-quark condensate obtained at low values of $\tau\simeq 0.25$ GeV$^{-2}$.  A similar choice has been done  for the  current $\eta_2$. One finds that the masses of the states associated to the {\it Class H currents} are in the range of 6 GeV which are mainly due to the violation of the four-quark condensates. These results for the mass differ significantly from the one in\,\cite{ZHUT} which for the $\eta_2$ current is\,:
\beq
M^{LO}_{T_{cc\bar u\bar s}(0^-)(\eta_2)}= 4680(120)~{\rm MeV},
\eeq
 \d The low mass of 4.4 GeV  obtained in Ref.\,\cite{ZHUT} from e.g the $\eta_4$ current using factorization of the four-quark condensate does not statisfy the stability criteria while, for the non-factorized case, the choice of $t_c=22$ GeV$^2$ and  $\tau$ around 0.3 GeV$^{-2}$ of Ref.\,\cite{ZHUT} leads to an imaginary decay constant. 

\d The (non)-findings of these high-mass states can be an alternative test of the (non)-violation of factorization.

\section{Class L pseudoscalar states in Table\,\ref{tab:resc} compared with Refs.\cite{ZHUT,TURC}}
\subsection*{\b $\eta_5$ current}
Our result in Eq.\,\ref{eq:eta5}  is comparable with the LO mass obtained in Ref.,\cite{ZHUT} within factorization:
\beq
M^{LO}_{T_{cc\bar u\bar d}(0^-)(\eta_5)}= 4410(140)~{\rm MeV}. 
\eeq

\subsection*{\b  $\eta_1$ current}

\d Our NLO result in Eq.\,\ref{eq:tcc-eta1} is lower by about 448 MeV compared to the LO result within factorization in Ref.\,\cite{ZHUT} :
\beq
M^{LO}_{T_{cc\bar d\bar d}(0^-)(\eta_1)}= 4430(120)~{\rm MeV},
\eeq
obtained at the lowest value of $t_c=24$ GeV$^2$.  

\d  Using factorization of the four-quark condensate as in Ref.\,\cite{ZHUT}, we obtain at LO for the sets $(\tau,t_c)$ = (0.27,24) to (0.40,44) (GeV$^{-2}$, GeV$^2$) where the coupling exhibits minimum and the mass inflexion points: 
\beq
M^{LO}_{T_{cc\bar d\bar d}(0^-)(\eta_1)}\vert_{Fac}= 4382(40)_{t_c}~{\rm MeV}, 
\eeq
which is comparable with  the one of Ref.\,\cite{ZHUT} indicating that the effect of the $d=8$ condensate given there is negligible.  
This LO result is also in the range of the one in Ref.\,\cite{TURC} taking into account the fact that the SU3 breakings are tiny. 
However, using again factorization, we add the NLO corrections. Then, the result  becomes:
\beq
M^{NLO}_{T_{cc\bar d\bar d}(0^-)(\eta_1)}\vert_{Fac}= 3784(34)_{t_c}~{\rm MeV}, 
\eeq
which is approximately the same as our previous NLO result in Table\,\ref{tab:resc} without using factorization. This feature indicates that in this channel the NLO corrections are important.
Thus, we conclude  that the origin of this difference is mainly due to the NLO corrections.

 \d The low mass of 4.43 GeV  obtained in Ref.\,\cite{ZHUT} from e.g the $\eta_4$ current using factorization of the four-quark condensate does not statisfy the stability criteria while, for the non-factorized case, the choice of $t_c=22$ GeV$^2$ and  $\tau$ around 0.3 GeV$^{-2}$ of Ref.\,\cite{ZHUT} leads to an imaginary decay constant ! 
 
\section {Class L vector states in Table\,\ref{tab:resc}  and comparison with Ref\,\cite{ZHUT}}
 
\subsection* {\b $\eta_1$ current}
\d Our NLO result for  $T_{cc\bar u\bar d}$ is lower by about 109 MeV than the LO one of Ref.\,\cite{ZHUT} using factorization but agrees within the error. In this case the  SU3 breakings increase slightly the mass by 48 MeV while it is about 120 MeV for \cite{ZHUT}.

\d We find that the size of SU3 breakings is about + 48 MeV  while it is 120 MeV from Ref.\,\cite{ZHUT}. This result is similar to the case of pseudoscalar states.
 
\subsection* {\b $\eta_6$ current}
 \d To compare our result  in Eq.\,\ref{eq:tcc1-eta6} with Ref.\,\cite{ZHUT}, 
we use the factorization of the four-quark condensate and the OPE up to dimension-6, we obtain to LO:
\beq
M^{LO}_{T_{cc\bar u\bar d}(1^-)(\eta_6)}=3899(8)_{t_c}
\eeq
while including NLO, one obtains :
\beq
M^{NLO}_{T_{cc\bar u\bar d}(1^-)(\eta_6)}=3904(0)_{t_c}
\eeq
which indicates that the NLO correction is negligible. 
By comparing this result with the one in Eq.\,\ref{eq:tcc1-eta6}, one finds that the violation of factorization decreases the mass by about 71 MeV. However, such results are much lower than the one :
\beq
M^{LO}_{T_{cc\bar u\bar d}(1^-)(\eta_6)}=4340(160)~{\rm MeV}
\eeq
of Ref.\,\cite{ZHUT}. 
 
\d We have also checked that the effect of the $d=8$ condensate  on the result for the  $T_{cc\bar u\bar d}$ mass  is almost negligible which leads to a value of 4.17 GeV for the choice $t_c=23$ GeV$^2$ of Ref.\,\cite{ZHUT} and taking $\tau=0.22$ GeV$^{-2}$ at the minimum of the coupling.
 Therefore, we fail to understand the origin of the value 4.34 GeV quoted by Ref.\,\cite{ZHUT} for this current. 
 
  
\d Our results suggest that the vector state associated to the $\eta_6$ current is below the $D^*D^*_0, DD_1$ thresholds and will not be observed through its pure hadronic decays.

\section{$SU(3)$ breakings}

\subsection* {\b Class H currents}
 The mass-splitting between the $T_{cc\bar s\bar s}$ and $T_{cc\bar d\bar d}$ pseudoscalar (resp. vector)  states is large --258 (resp. --283) MeV. It is +140 (resp. +148) MeV 
for the  pseudoscalar (resp. vector) $T_{cc\bar s\bar u}$ and $T_{cc\bar d\bar u}$ masses.

\subsection* {\b Class L currents}
The mass-splittings between the  $T_{cc\bar s\bar s}$ and $T_{cc\bar s\bar u}$ states with their analogue in the chiral limit are  +13 (resp.+ 48 MeV) for the vector states. It is --125 (resp. --130) MeV for the pseudoscalar states.     
    
  \section{The $T_{bb\bar u\bar d}$ pseudoscalar states}
  We extend the previous analysis to the $b$-quark sector. 
  \subsection*{\b  {\it Class H0 }  currents}
  These are our current ($\equiv \eta_4$) in Table\,\ref{tab:current}  and $\eta_2$.
   \subsection*{\d Our current $\equiv \eta_4$ in Table\,\ref{tab:current}}
\begin{figure}[hbt]
\begin{center}
\centerline {\hspace*{-7.5cm} \bf a)\hspace{8cm} b)}
\includegraphics[width=8cm]{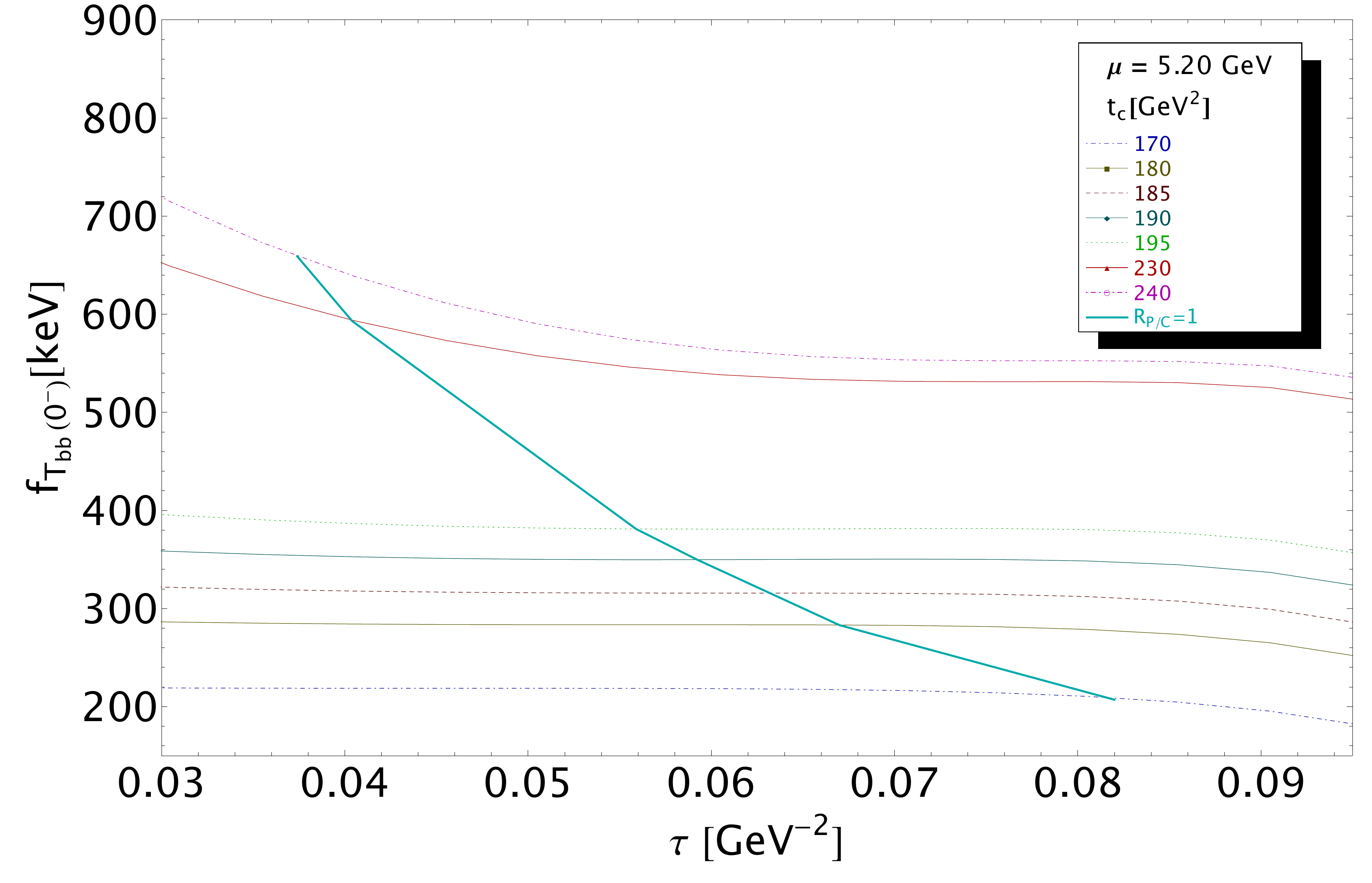}
\includegraphics[width=8cm]{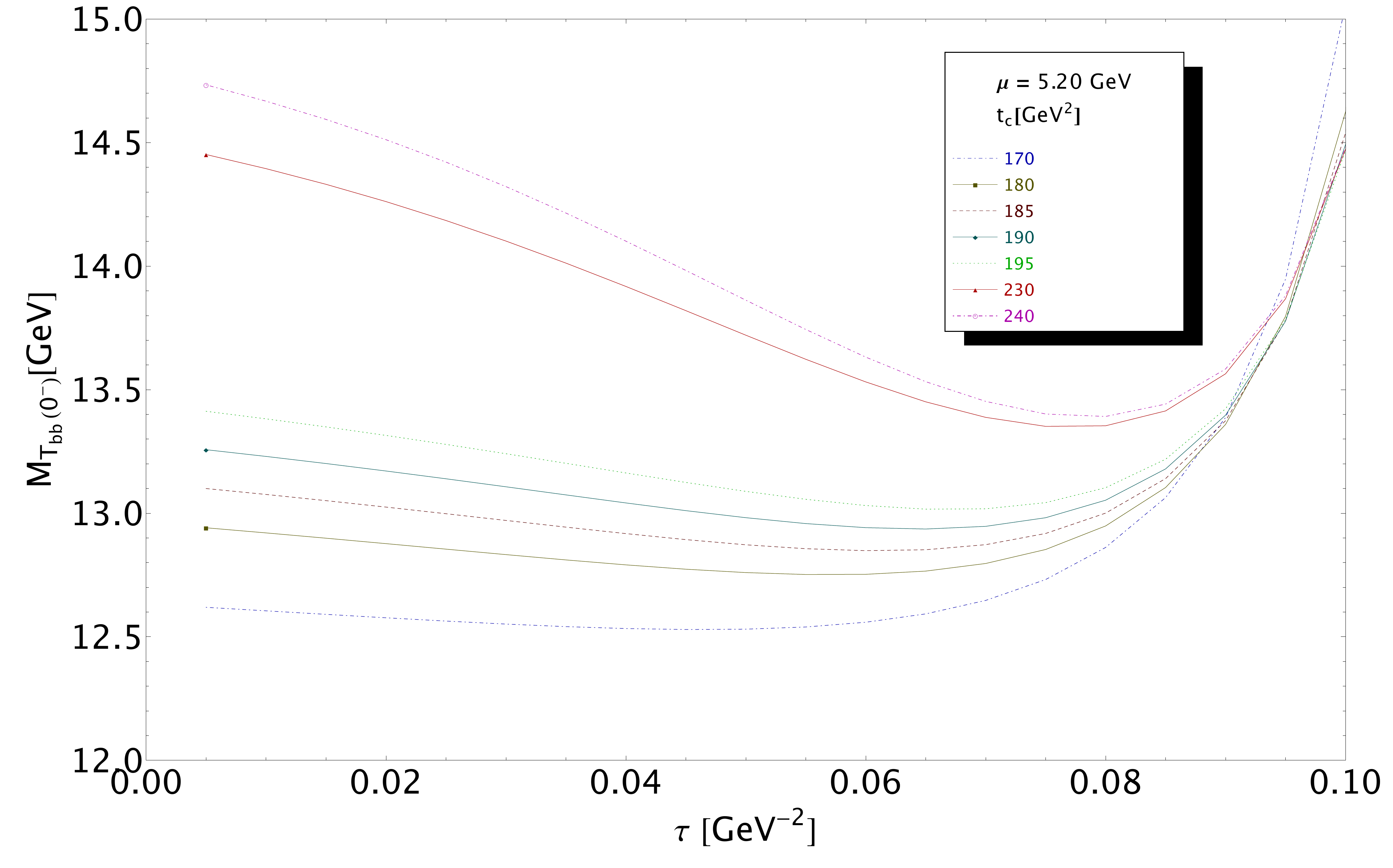}
\vspace*{-0.5cm}
\caption{\footnotesize  $f_{T_{bb\bar u\bar d}}$ and $M_{T_{bb\bar u\bar d}}$ as a function of $\tau$ at NLO and for different values of $t_c$. We use $\mu$=5.2 GeV and the QCD inputs in Table\,\ref{tab:param}. The curve $R_{P/C}=1$ is shown in a). } 
\label{fig:tbb0}
\end{center}
\vspace*{-0.5cm}
\end{figure}

  We illustrate the analysis by our current which is shown in Fig.\,\ref{fig:tbb0}.\label{eq:tbb0}
 The mass presents minimum from (0.045,170) to (0.075,230) (GeV$^{-2}$, GeV$^2$). The coupling presents stability for ($\tau,t_c$) = (0.045,170) and an inflexion point up to (0.075,230) (GeV$^{-2}$, GeV$^2$). To constrain more the values of $t_c$, we select the region where $R_{P/C}\geq 1$ in the lowest moment ${\cal L}_0$. The limited region is shown in Fig.\,\ref{fig:tbb0}a) which excludes the values of $t_c\leq 185$ GeV$^2$ where the minimum are outside the allowed region  (RHS of the curve  $R_{P/C}$).  Then, one deduces the optimal result inside the region $(t_c,\tau)=(185,0.060)\to (230,0.075)$ (GeV$^2$, GeV$^{-2}$)  (see Table\,\ref{tab:resb}):
\beq
f_{T_{bb\bar u\bar d}(0^-)}=424(108) ~{\rm keV}, ~~~~~~~~~~
M_{T_{bb\bar u\bar d}(0^-)}= 13100(278) ~{\rm MeV}, 
\eeq
where we have used $\Delta\tau=\pm 0.01$ GeV$^{-2}$ around the minimum and taken the mean of the errors due to the change of $\tau$ for the two extremal values of $t_c$. The different sources of the errors are given in Tables\,\ref{tab:error-fb} and\,\ref{tab:error-mb}.  

   \subsection*{\d $\eta_5$ current\,\footnote{Note that in the $b$-quark channel, the $\eta_5$ becomes a Class H0 current.}}
\begin{figure}[hbt]
\begin{center}
\centerline {\hspace*{-7.5cm} \bf a)\hspace{8cm} b)}
\includegraphics[width=8cm]{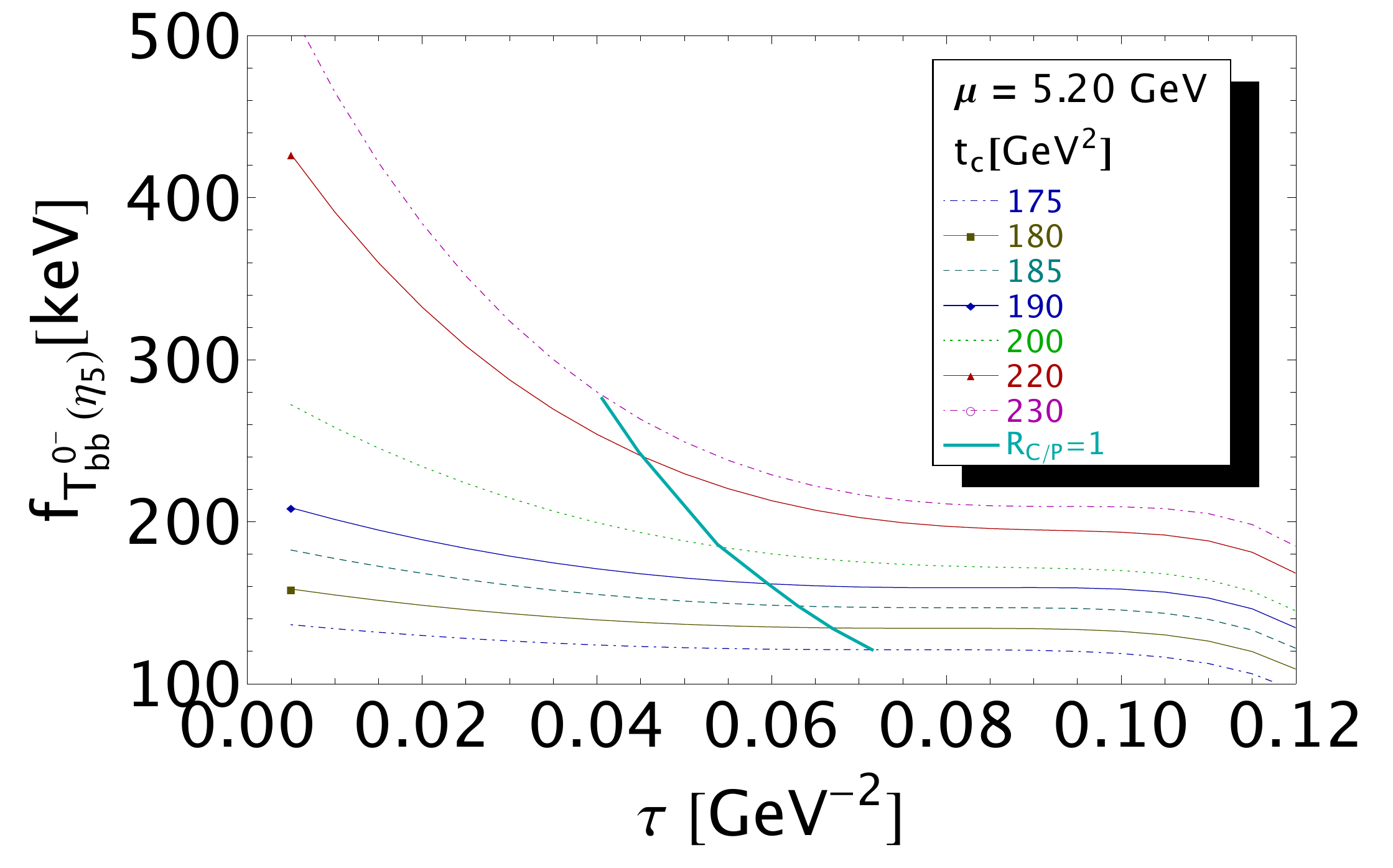}
\includegraphics[width=8cm]{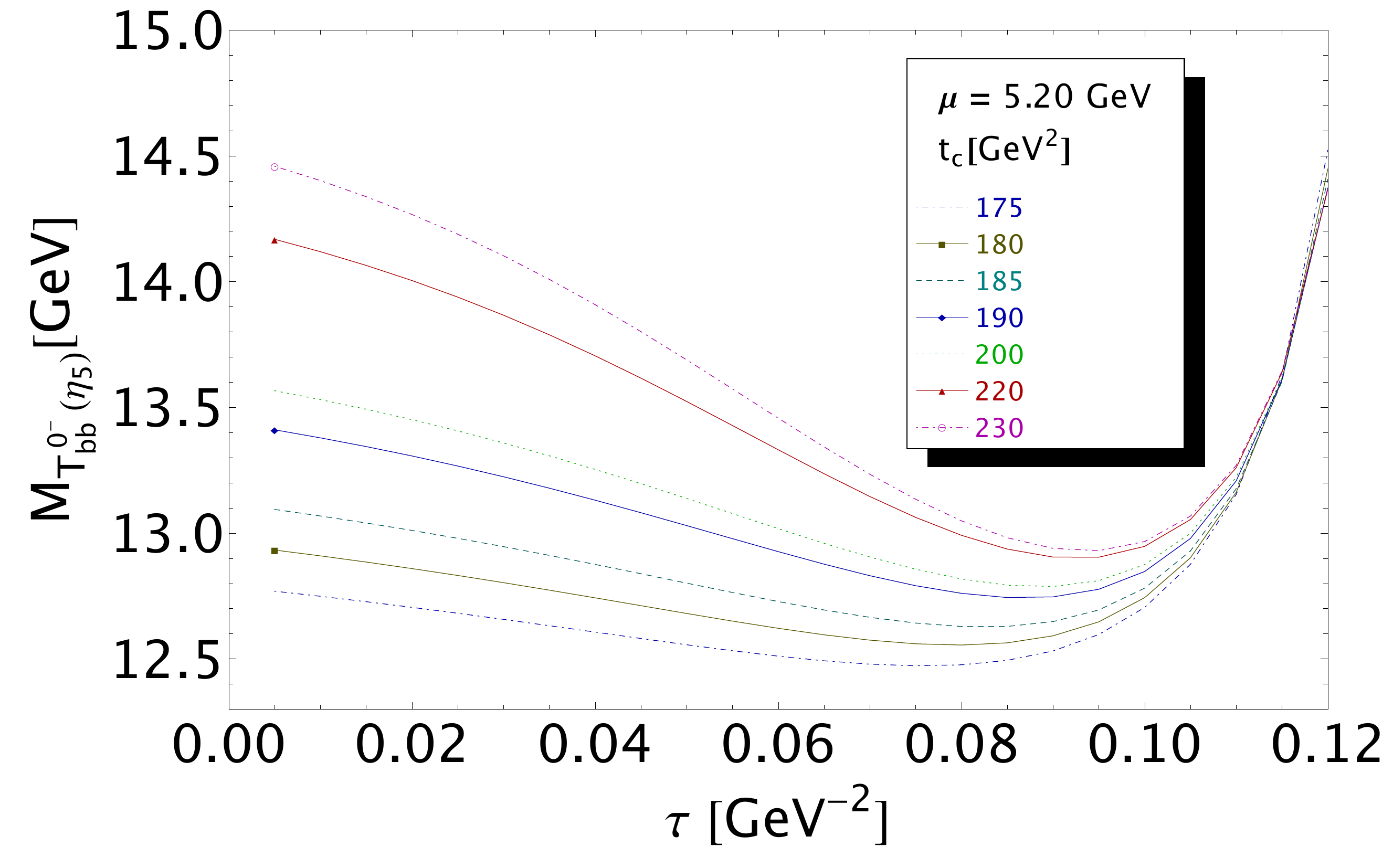}
\vspace*{-0.5cm}
\caption{\footnotesize  $f_{T_{bb\bar u\bar d}}$ and $M_{T_{bb\bar u\bar d}}$ for the $\eta_5$ current as a function of $\tau$ at NLO and for different values of $t_c$. The OPE is truncated at $d=6$ condensates.  } 
\label{fig:eta5}
\end{center}
\vspace*{-0.5cm}
\end{figure}

  The analysis is similar to the one in Fig.\,\ref{fig:tbb0}. We show the result in Fig.\,\ref{fig:eta5}.

  The coupling presents stability for ($\tau,t_c$) = (0.080,180) and an inflexion point up to (0.090,220) (GeV$^{-2}$, GeV$^2$)  while the mass shows minimum from (0.075,175) to (0.095,220) (GeV$^{-2}$, GeV$^2$).  We show in Fig.\,\ref{fig:eta5}a the limitng curve $R_{P/C}=1$ which excludes value of $t_c$ below 180 GeV$^2$. One deduces the optimal result  (see Table\,\ref{tab:resb}):
\beq
f_{T_{bb\bar u\bar d}(0^-)(\eta_5)}=165(31)~{\rm keV}, ~~~~~~~~~~
M_{T_{bb\bar u\bar d}(0^-)(\eta_5)}= 12730(207)~{\rm MeV}, 
\label{eq:tbb0-eta5}
\eeq
where we have used $\Delta\tau\simeq 0.01$ GeV$^{-2}$.

   \subsection*{\d  $\eta_2$ current}
  The analysis is similar to the one in Fig.\,\ref{fig:tbb0}. 
  The minimum of the coupling stability allowed by $R_{P/C}\geq 1$ starts from  ($\tau,t_c$) = (0.06,185) up to an inflexion point at (0.08,230) (GeV$^{-2}$, GeV$^2$). 
  The mass shows minimum from (0.065,185) to (0.08,230) (GeV$^{-2}$, GeV$^2$). One deduces the optimal result  compiled in Table\,\ref{tab:resb}:
\beq
f_{T_{bb\bar d\bar d}(0^-)(\eta_2)}= 373(90)~{\rm keV}, ~~~~~~~~~~
M_{T_{bb\bar d\bar d}(0^-)(\eta_2)}=13039(258) ~{\rm MeV}, 
\label{eq:tbb0-eta2}
\eeq
where we have used $\Delta\tau\simeq 0.01$ GeV$^{-2}$. 
  \subsection*{\b  {\it Class L0 }  currents}

   \subsection*{\d $\eta_1$ current}
\begin{figure}[hbt]
\begin{center}
\centerline {\hspace*{-7.5cm} \bf a)\hspace{8cm} b)}
\includegraphics[width=8cm]{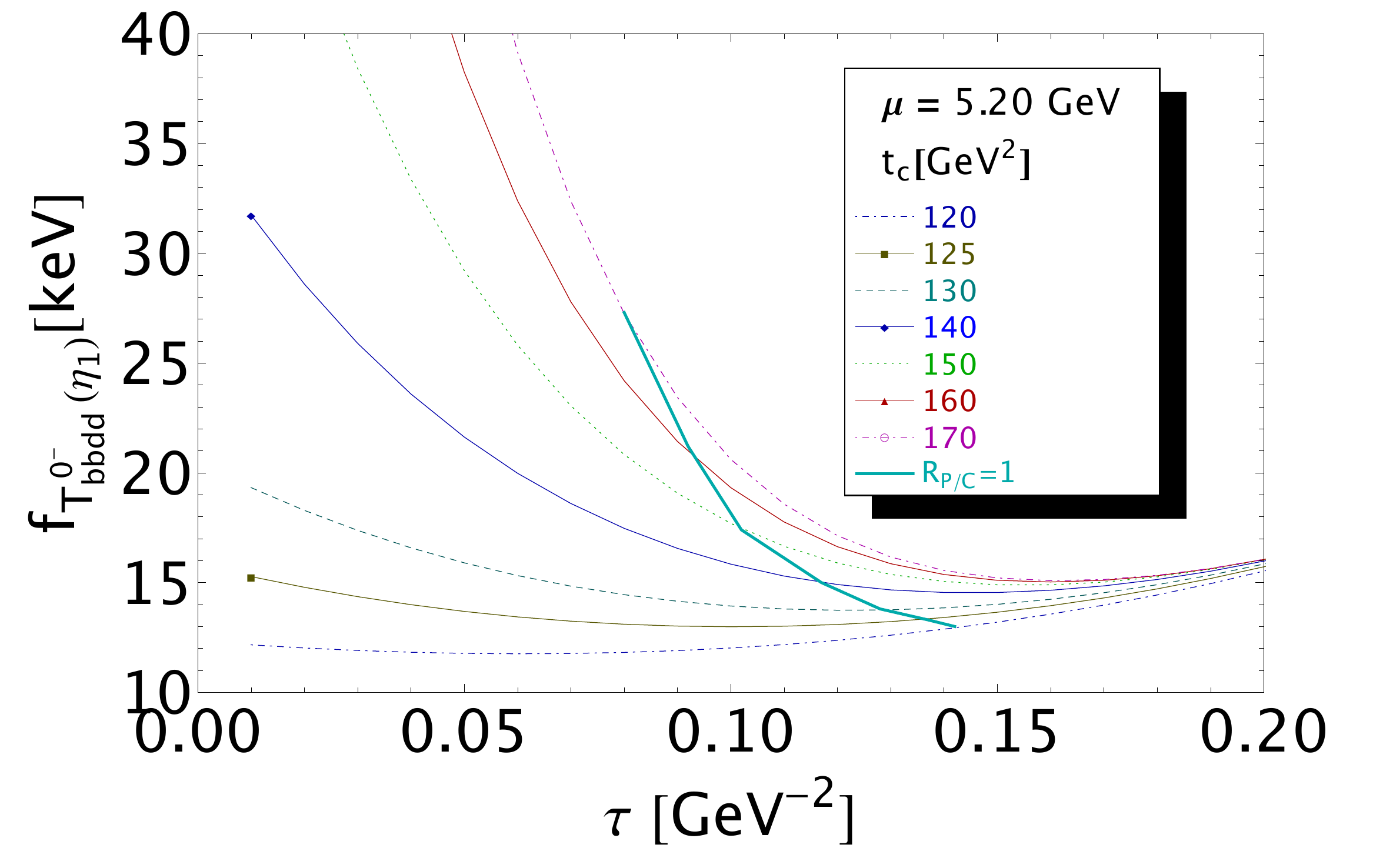}
\includegraphics[width=8cm]{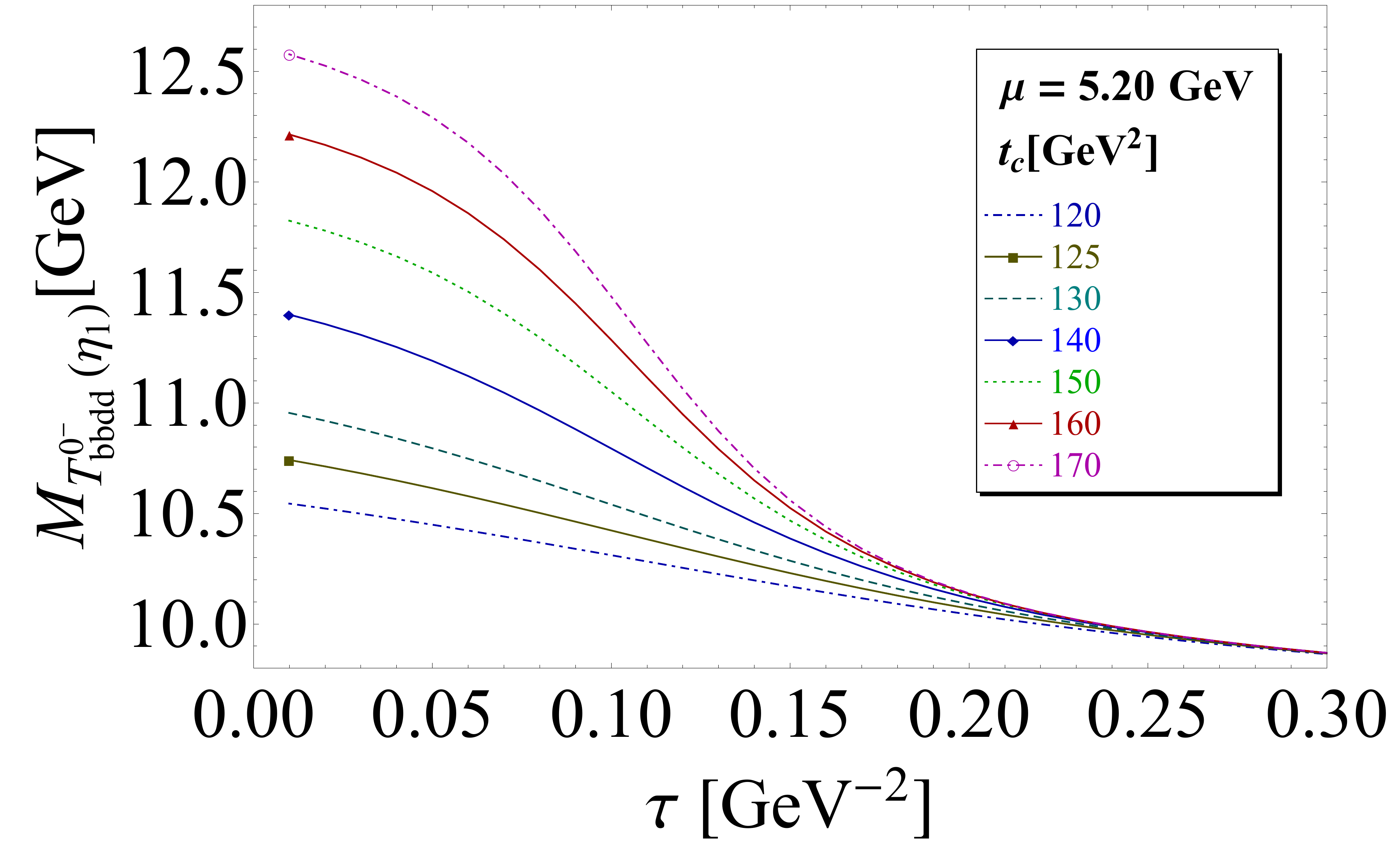}
\vspace*{-0.5cm}
\caption{\footnotesize  $f_{T_{bb\bar u\bar d}}$ and $M_{T_{bb\bar u\bar d}}$ for the $\eta_1$ current as a function of $\tau$ at NLO and for different values of $t_c$. We use $\mu$=5.2 GeV and the QCD inputs in Table\,\ref{tab:param}. $R_{P/C}=1$ delimits the region where the ground state contribution is larger than the one of the contiuum (RHS).} 
\label{fig:tbb0-eta1}
\end{center}
\vspace*{-0.5cm}
\end{figure}

  The analysis is shown in Fig.\,\ref{fig:tbb0-eta1}. The coupling presents minimum for ($\tau,t_c$) = (0.10,125) up to (0.16,160) (GeV$^{-2}$, GeV$^2$)  while the mass shows inflexion points around these values. The condition $R_{P/C}\geq 1$ eliminates values of $t_c$ below 130 GeV$^2$ (see Fig.\,\ref{fig:tbb0-eta1}). We deduce the optimal result  (see Table\,\ref{tab:resb}):
\beq
f_{T_{bb\bar d\bar d}(0^-)(\eta_1)}=14(2)~{\rm keV}, ~~~~~~~~~~
M_{T_{bb\bar d\bar d}(0^-)(\eta_1)}= 10407(124)~{\rm MeV}, 
\label{eq:tbb0-eta1}
\eeq
inside the region $(t_c,\tau)$=(130,0.12)$ \to$(160,0.16)  (GeV$^2$,GeV$^{-2}$), 
where we have used $\Delta\tau\simeq 0.01$ GeV$^{-2}$. 
\section{The $T_{bb\bar u\bar s}$ pseudoscalar state}
   \subsection*{\b {\it Class H0:} our current $\equiv \eta_4$ in Table\,\ref{tab:current}}
The analysis is similar to the one in Fig.\,\ref{fig:tbb0}. The coupling presents stability for ($\tau,t_c$) = (0.06,185) and an inflexion point up to (0.085,230) (GeV$^{-2}$, GeV$^2$) while the mass shows minimum from (0.07,185) to (0.08,230) (GeV$^{-2}$, GeV$^2$). One obtains the optimal results:
\beq
f_{T_{bb\bar u\bar s}(0^-)}= 283(65) ~{\rm keV}, ~~~~~~~~~~
M_{T_{bb\bar u\bar s}(0^-)}= 13005(247) ~{\rm MeV}, 
\label{eq:tbbs0}
\eeq
where we have used $\Delta\tau\simeq 0.01$ GeV$^{-2}$. 
   \subsection*{\b {\it Class H0} $\eta_5$ current}
The analysis is similar to the one in Fig.\,\ref{fig:tbb0}. The coupling presents stability for ($\tau,t_c$) = (0.035,175) and an inflexion point up to (0.09,210) (GeV$^{-2}$, GeV$^2$) while the mass shows minimum from (0.085,175) to (0.10,210) (GeV$^{-2}$, GeV$^2$). One obtains the optimal results:
\beq
f_{T_{bb\bar u\bar s}(0^-)(\eta_5)}= 144(40) ~{\rm keV}, ~~~~~~~~~~
M_{T_{bb\bar u\bar s}(0^-)(\eta_5)}= 12553(252) ~{\rm MeV}, 
\label{eq:tbbs0-eta5}
\eeq
where we have used $\Delta\tau\simeq 0.01$ GeV$^{-2}$. 
\section{The $T_{bb\bar s\bar s}$ pseudoscalar state}
   \subsection*{\b Class H0 $ \eta_2$  current}
The analysis is similar to the one in Fig.\,\ref{fig:tbb0}. The coupling presents stability for ($\tau,t_c$) = (0.06,185) and an inflexion point up to (0.085,230) (GeV$^{-2}$, GeV$^2$) while the mass shows minimum from (0.08,185) to (0.085,230) (GeV$^{-2}$, GeV$^2$). One obtains the optimal results:
\beq
f_{T_{bb\bar s\bar s}(0^-)(\eta_2)}= 302(82)~{\rm keV},~~~~~~~~~~
M_{T_{bb\bar s\bar s}(0^-)(\eta_2)}= 12745(274)~{\rm MeV}, 
\label{eq:tbbss0-eta2}
\eeq
   \subsection*{\b {\it Class L0} $ \eta_1$    current}
  The analysis is similar to Fig.\,\ref{fig:tbb0-eta1}. The coupling presents minimum for ($\tau,t_c$) = (0.10,125) up to (0.17,160) (GeV$^{-2}$, GeV$^2$)
  while the mass shows inflexion points around these values. One deduces the optimal result :
\beq
f_{T_{bb\bar s\bar s}(0^-)(\eta_1)}=12(2)~{\rm keV}, ~~~~~~~~~~
M_{T_{bb\bar s\bar s}(0^-)(\eta_1)}= 10437(109) ~{\rm MeV}, 
\label{eq:tbbss0-eta1}
\eeq
where we have used $\Delta\tau\simeq 0.01$ GeV$^{-2}$. 

\section{The $T_{bb\bar u\bar d}$ vector state}
  \subsection*{\b  {\it Class H1 }  currents}
  These are our current  in Table\,\ref{tab:current} ($\equiv \eta_5$) and $\eta_2$.
   \subsection*{\d Our current in Table\,\ref{tab:current}$\equiv \eta_5$}
  The analysis is similar to the one in Fig.\,\ref{fig:tbb0}. The region of the stability of the coupling allowed by the condition $R_{P/C}\geq 1$ is from  ($\tau,t_c$) = (0.065,190) while there is an inflexion point up to (0.075,230) (GeV$^{-2}$, GeV$^2$). 
  The mass shows minimum from (0.065,190) to (0.080,230) (GeV$^{-2}$, GeV$^2$). One deduces the optimal result :
\beq
f_{T_{bb\bar u\bar d}(1^-)}= 246(59) ~{\rm keV}, ~~~~~~~~~~
M_{T_{bb\bar u\bar d}(1^-)}= 13053(262) ~{\rm MeV}, 
\label{eq:tbb1}
\eeq
where we have used $\Delta\tau=\pm 0.01$ GeV$^{-2}$ around the minimum and taken the mean of the errors due to the change of $\tau$ for the two extremal values of $t_c$.
   \subsection*{\d  $\eta_2$ current}
  The analysis is similar to the one in Fig.\,\ref{fig:tbb0}. The stability of the coupling consistent with $R_{P/C}\geq 1$ is above ($\tau,t_c$) = (0.065,190) and an inflexion point up to (0.075,230) (GeV$^{-2}$, GeV$^2$)
  while the mass shows minimum from (0.065,190) to (0.075,230) (GeV$^{-2}$, GeV$^2$). One deduces the optimal result :
\beq
f_{T_{bb\bar d\bar d}(1^-)(\eta_2)}=361(75)~{\rm keV},~~~~~~~~~~
M_{T_{bb\bar d\bar d}(1^-)(\eta_2)}= 13131(240) ~{\rm MeV}, 
\label{eq:tbb1-eta2}
\eeq
where we have used $\Delta\tau\simeq 0.01$ GeV$^{-2}$. 
  \subsection*{\b  {\it Class L1 }  currents}
\subsection*{\b The $T_{bb}$ vector state from {\it Class L1} currents}
These are the $\eta_1$ and $\eta_6$ currents of Ref.\,\cite{ZHUT}. 
\subsection*{\d  $\eta_1$ current}
The analysis is similar to Figs.\,\ref{fig:tbb0-eta1} for different values of $\tau$ and $t_c$ and for fixed value of $\mu=5.2$ GeV. We observe stabilities for the coupling for the sets  $(\tau , t_c)$ from (0.06,120) to (0.16,160) (GeV$^{-2}$, GeV$^{2}$) while the mass presents inflexion points around these values. Then, we deduce the conservative estimate :
\beq
f_{T_{bb\bar d\bar d}(1^-)(\eta_1)}= 12(2) ~{\rm keV}, ~~~~~~~~~~
M_{T_{bb\bar d\bar d}(1^-)(\eta_1)}= 10412(99) ~{\rm MeV}. 
\label{eq:tbb1-eta1}
\eeq
We have used $\Delta\tau=\pm 0.01$ GeV$^{-2}$. 
\subsection*{\d  $\eta_6$  current}
The analysis is similar to Figs.\,\ref{fig:tbb0-eta1} for different values of $\tau$ and $t_c$ and for fixed value of $\mu=5.2$ GeV. We observe stabilities for the coupling for the sets  $(\tau , t_c)$ from (0.04,120) to (0.15,160) (GeV$^{-2}$, GeV$^{2}$) while the mass presents inflexion points around these values. Then, we deduce the conservative estimate :
\beq
f_{T_{bb\bar u\bar d}(1^-)(\eta_6)}= 11(2) ~{\rm keV}, ~~~~~~~~~~
M_{T_{bb\bar u\bar d}(1^-)(\eta_6)}= 10469(110) ~{\rm MeV}. 
\label{eq:tbb1-eta6}
\eeq
We have used $\Delta\tau=\pm 0.01$ GeV$^{-2}$. 
\section{The $T_{bb\bar u\bar s}$ vector states}
   \subsection*{\b {\it Class H1:} Our current in Table\,\ref{tab:current}$\equiv \eta_5$ }
The analysis is similar to the one in Fig.\,\ref{fig:tbb0}. The coupling allowed stability starts from  ($\tau,t_c$) = (0.07,180) and inflexion point up to (0.085,230) (GeV$^{-2}$, GeV$^2$) while the mass shows minimum from (0.07,180) to (0.085,230) (GeV$^{-2}$, GeV$^2$). One obtains the optimal results:
\beq
f_{T_{bb\bar u\bar s}(1^-)}= 158(41) ~{\rm keV}, ~~~~~~~~~~
M_{T_{bb\bar u\bar s}(1^-)}= 12910(268) ~{\rm MeV}, 
\label{eq:tbbs1}
\eeq
where we have used $\Delta\tau\simeq 0.01$ GeV$^{-2}$. 
   \subsection*{\b {\it Class L1} $\eta_6$ current}
The analysis is similar to the one in Fig.\,\ref{fig:tbb0-eta1}. The coupling presents stability for ($\tau,t_c$) = (0.10,125) and a minimum up to (0.16,160) (GeV$^{-2}$, GeV$^2$) while the mass shows inflexion points around these values. One obtains the optimal results:
\beq
f_{T_{bb\bar u\bar s}(1^-)(\eta_6)}= 10(2) ~{\rm keV}, ~~~~~~~~~~
M_{T_{bb\bar u\bar s}(1^-)(\eta_6)}= 10438(97) ~{\rm MeV}, 
\label{eq:tbbs1-eta5}
\eeq
where we have used $\Delta\tau\simeq 0.01$ GeV$^{-2}$. 
\section{The $T_{bb\bar s\bar s}$ vector state}
   \subsection*{\b Class H1 $ \eta_2$  current}
The analysis is similar to the one in Fig.\,\ref{fig:tbb0}. The coupling allowed stability is from ($\tau,t_c$) = (0.075,180) and an inflexion point up to (0.09,230) (GeV$^{-2}$, GeV$^2$) while the mass shows minimum from (0.075,185) to (0.09,230) (GeV$^{-2}$, GeV$^2$). One obtains the optimal results:
\beq
f_{T_{bb\bar s\bar s}(1^-)(\eta_2)}= 294(69) ~{\rm keV}, 
M_{T_{bb\bar s\bar s}(1^-)(\eta_2)}= 12834(246) ~{\rm MeV}, 
\label{eq:tbbss1-eta2}
\eeq
   \subsection*{\b {\it Class L1} $ \eta_1$    current}
  The analysis is similar to Fig.\,\ref{fig:tbb0-eta1}. The coupling presents minimum for ($\tau,t_c$) = (0.06,120) up to (0.17,160) (GeV$^{-2}$, GeV$^2$)
  while the mass shows inflexion points around these values. One deduces the optimal result :
\beq
f_{T_{bb\bar s\bar s}(1^-)(\eta_1)}=10(2) ~{\rm keV}, ~~~~~~~~~~
M_{T_{bb\bar s\bar s}(1^-)(\eta_1)}= 10445(116) ~{\rm MeV}, 
\label{eq:tbbss1-eta1}
\eeq
where we have used $\Delta\tau\simeq 0.01$ GeV$^{-2}$. 
\section{Comments on Table\,\ref{tab:resb} and on the results of Ref.\,\cite{ZHUT}}
 One can deduce from Table\,\ref{tab:resb} that :
\subsection* {\b Class H currents}
\d The masses of the associated particles are in the range of 13 GeV. 
 The SU3 breakings tend to decrease the mass of the states by about (140 -174) MeV which is of the order of the uncertainties of the approach while the coupling decreases by (40-160 keV). This class of currents is very sensitive to the use of factorization for the four-quark condensates both for the pseudoscalar and vector states. 

\vspace*{-0.5cm} 
\begin{center}
   {\footnotesize
\begin{table}[H]
\setlength{\tabcolsep}{0.1pc}
    {
  \begin{tabular}{lllllll | lll}
&\\
\hline
\hline
\multicolumn{7}{c}{\bf Our Work}&\multicolumn{3}{c}{Ref.\,\cite{ZHUT}}\\
\hline
States&Current  &$t_c$ [GeV$^2$] &$\tau$  [GeV$^{-2}$]  & $f^{NLO} _{T_{bbqq'}}$&$\tau$  [GeV$^{-2}$] &$M^{NLO}_{T_{bbqq'}}$&$t_c$ & $\tau$  [GeV$^{-2}$] &$M^{LO}_{T_{ccqq'}}$\\ 
 \hline 
 \bf Class H &&&&&&&\\
{ \bma  $0^-$} &&&&&&&\\
$T_{bb\bar u\bar d}$&$ {\cal O}_{T_{ud}^{0^-}},\eta_4$&$185\to 230$&$0.060\to 0.075$&424(108)&$0.060\to 0.075$&13100(278)&--&&  ?\\
$T_{bb\bar u\bar d}$&$ \eta_5$&$180\to 220$&$0.08\to 0.09$&165(31)&$0.08\to 0.095$&12730(207)&115&$0.13 \to 0.14$& 10300(300) \\
$T_{bb\bar d\bar d}$ & $\eta_2$&$185\to 230$&$0.060\to 0.080$&373(90)&$0.065\to 0.080$&13039(258)&&&--- \\
 $T_{bb\bar u\bar s}$& $ {\cal O}_{T^{0^-}_{us}},\eta_4$&$185\to 230$&$0.070\to 0.080$&283(65)&$0.070\to 0.080$&13005(247)&&&-- \\
  $T_{bb\bar u\bar s}$& $ \eta_5$&$175\to 210$&$0.085\to 0.090$&144(40)&$0.080\to 0.100$& 12553(252)&115&$0.13 \to 0.14$&10400(200) \\
  $T_{bb\bar s\bar s}$ & $\eta_2$&$185\to 230$&$0.06\to 0.085$&302(82)&$0.08\to 0.085$&12745(274)&&&--- \\
  { \bma  $1^-$} &&&&&&&\\
$T_{bb\bar u\bar d}$&$ {\cal O}_{T_{ud}^{1^-}},\eta_5$&$190\to 230$&$0.065\to 0.080$&246(59)&$0.065\to 0.080$&13053(262)&&&?\\
$T_{bb\bar d\bar d}$ & $\eta_2$&$190\to 230$&$0.065\to 0.075$&361(75)&$0.065\to 0.075$&13131(240)&&&-- \\
 $T_{bb\bar u\bar s}$& $ {\cal O}_{T^{1^-}_{us}},\eta_5$&$180\to 230$&$0.070\to 0.085$&158(41)&$0.07\to 0.085$&12910(268)&&&-- \\

  $T_{bb\bar s\bar s}$ & $\eta_2$&$180\to 230$&$0.075\to 0.090$&294(69)&$0.075\to 0.090$&12834(246)&&&-- \\

  \bf Class L&&&&&&&\\
  {  $0^-$} &&&&&&&\\
$T_{bb\bar d\bar d}$& $\eta_1$&$130\to 160$&$0.120\to 0.160$&14(2)&$0.120\to 0.160$&10407(124)&125&$0.10 \to 0.14$& 10600(300) \\
$T_{bb\bar s\bar s}$&$\eta_1$&$130\to 160$&$0.130\to 0.170$&12(2)&$0.130\to 0.170$&10437(109)&125&$0.10 \to 0.15$& 10600(300)\\
  {  $1^-$} &&&&&&&\\
  $T_{bb\bar u\bar d}$&  $\eta_6 $&$130\to 160$&$0.110\to 0.150$&11(2)&$0.110\to 0.150$&10469(110)&120&$0.11 \to 0.14$& 10400(200)\\
 $T_{bb\bar d\bar d}$& $ \eta_1 $&$130\to 160$&$0.120\to 0.160$&12(2)&$0.120\to 0.160$&10412(99)&125&$0.11 \to 0.14$& 10600(300) \\
   $T_{bb\bar u\bar s}$& $ \eta_6$&$130\to 160$&$0.120\to 0.160$&10(2)&$0.120\to 0.160$&10438(97)&120&$0.11 \to 0.14$&10400(200)\\

 $T_{bb\bar s\bar s}$&  $ \eta_1 $&$130\to 160$&$0.130\to 0.170$&10(2)&$0.130\to 0.170$&10445(116)& 125 &$0.10 \to 0.15$& 10600(300) \\

   \hline\hline
  \vspace*{-0.5cm}
\end{tabular}}
 \caption{Our predictions of the couplings [keV] and masses [MeV] compared with the ones in Ref.\,\cite{ZHUT} (more details are given in the text). 
 The different sources of the errors are given in Tables\,\ref{tab:error-fb} and \,\ref{tab:error-mb}.  Note that in this $b$-quark channel, the $\eta_5$ current becomes a Class $H0$. }  

\label{tab:resb}
\end{table}
} 
\end{center}
\vspace*{-1cm}
\subsection* {\b Class L Pseudoscalar states}
\subsection* {\d  $\eta_5$ current}
\begin{figure}[hbt]
\begin{center}
\centerline {\hspace*{-7.5cm} \bf a)\hspace{8cm} b)}
\includegraphics[width=8cm]{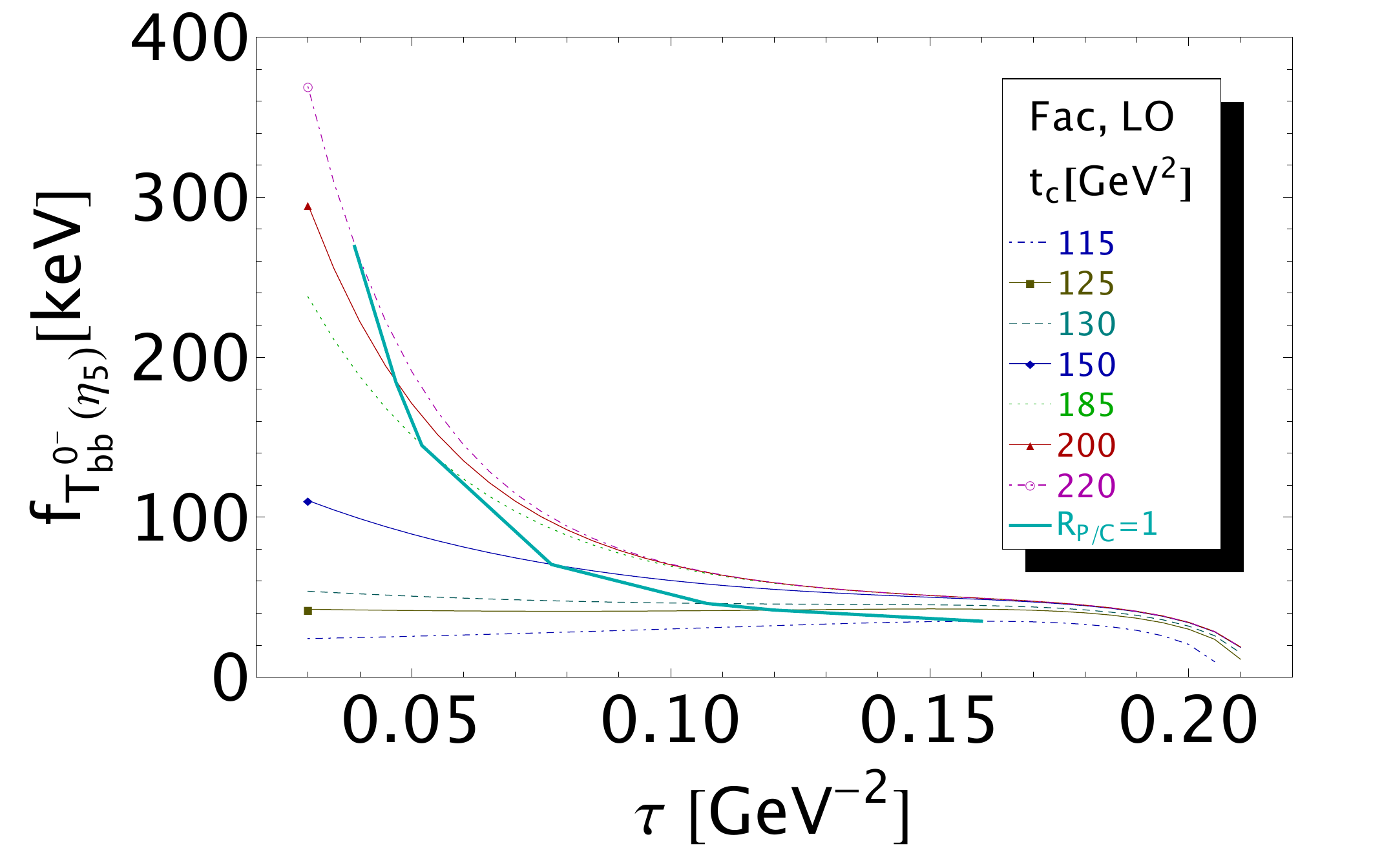}
\includegraphics[width=8cm]{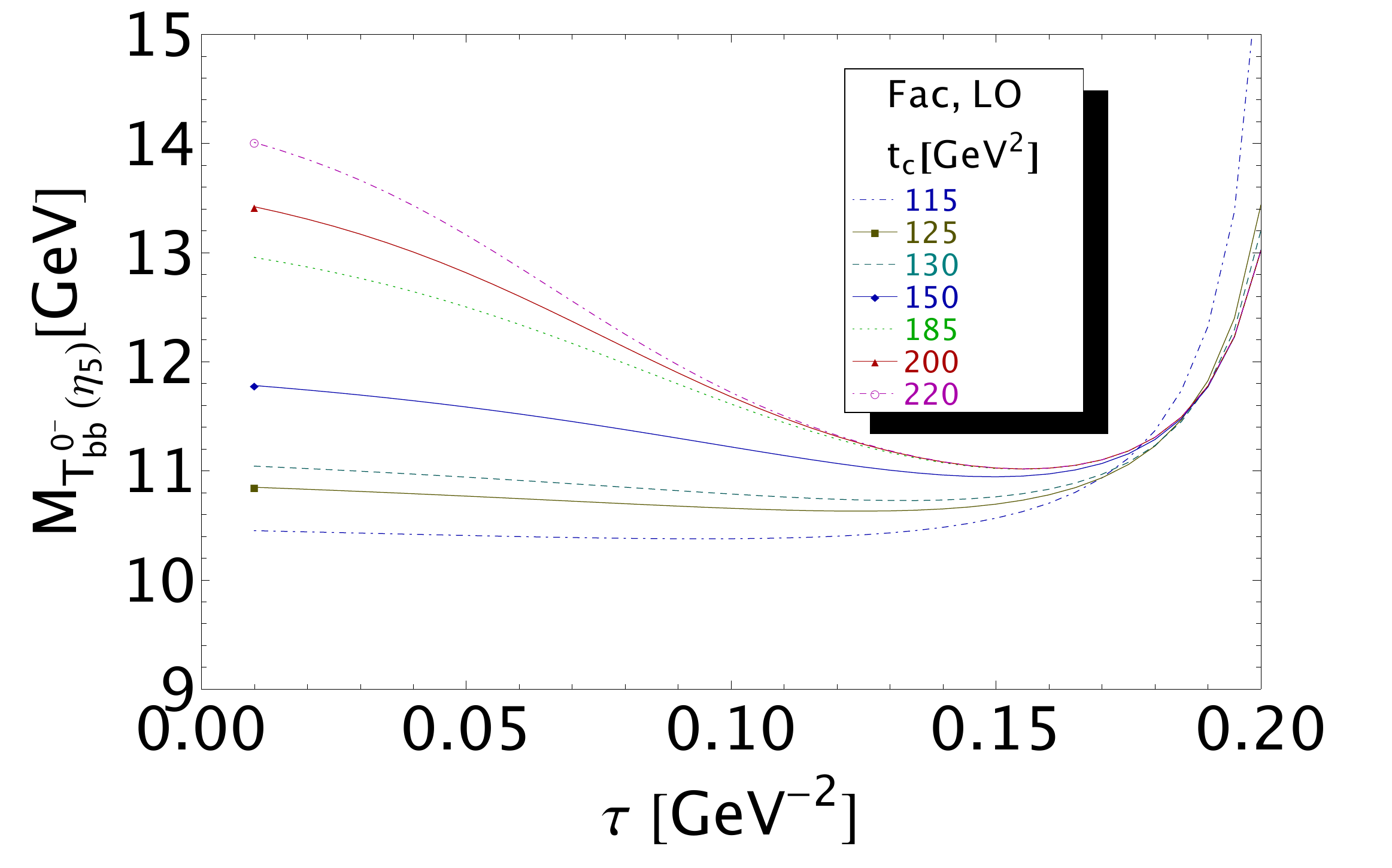}
\vspace*{-0.5cm}
\caption{\footnotesize  $f_{T_{bb\bar u\bar d}}$ and $M_{T_{bb\bar u\bar d}}$ for the $\eta_5$ current as a function of $\tau$ at LO and for different values of $t_c$. We use 
 the 
factorization of the four-quark condensates. The OPE is truncated at $d=6$ condensates. The limiting curve $R_{P/C}=1$ is shown in a).  } 
\label{fig:eta5-fac}
\end{center}
\vspace*{-0.5cm}
\end{figure}

 -- Our NLO results disagree with the one of Ref.\,\cite{ZHUT} obtained at LO and using factorization.  In order to understand this discrepancy, we work at LO and use the factorization of the four-quark condensate.  
 We show the results of the analysis in Fig.\,\ref{fig:eta5-fac}. One can notice, for the choice of $t_c=115$ GeV$^2$ used in Ref.\,\cite{ZHUT}, that the value of the mass is stable versus $\tau$ with a value  about 10377 MeV but the coupling does not. However, one can also see that the mass and the coupling increase with the value of $t_c$.
\begin{table}[H]
\setlength{\tabcolsep}{0.3pc}
{\scriptsize{
\begin{tabular}{ll ll  ll  ll ll ll ll ll c}
\hline
\hline
                States & Currents
                    &\multicolumn{1}{c}{$\Delta t_c$}
					&\multicolumn{1}{c}{$\Delta \tau$}
					&\multicolumn{1}{c}{$\Delta \mu$}
					&\multicolumn{1}{c}{$\Delta \alpha_s$}
					&\multicolumn{1}{c}{$\Delta PT$}
					&\multicolumn{1}{c}{$\Delta m_s$}
					&\multicolumn{1}{c}{$\Delta m_c$}
					&\multicolumn{1}{c}{$\Delta \overline{\psi}\psi$}
					&\multicolumn{1}{c}{$\Delta \kappa$}					
					&\multicolumn{1}{c}{$\Delta G^2$}
					&\multicolumn{1}{c}{$\Delta M^{2}_{0}$}
					&\multicolumn{1}{c}{$\Delta \overline{\psi}\psi^2$}
					&\multicolumn{1}{c}{$\Delta G^3$}
					&\multicolumn{1}{c}{$\Delta OPE$}
					&\multicolumn{1}{r}{Total [keV]}
\\
					
\hline
{\bf Class H} &&&&&&&&&&\\
$0^-$ &&&&&&&&&\\
$T_{bb\bar{u}\bar{d}}$ & ${\cal O}^{0^-}_{T}$, $\eta_4$ &107&0.70&0.69&5.40&1.36&$\cdots$&5.11&0.00&$\cdots$&0.02&0.00&3.20&0.01&10.9&108 \\
$T_{bb\bar{u}\bar{d}}$ & $\eta_5$ &30.0&1.10&0.31&2.40&0.18&$\cdots$&2.23&0.00&$\cdots$&0.02&0.00&1.31&0.00&5.61&31 \\
$T_{bb\bar{d}\bar{d}}$ & $\eta_2$ &88.0&1.00&0.60&4.84&8.61&$\cdots$&4.88&0.00&$\cdots$&0.05&0.00&3.70&0.00&11.4&90 \\
$T_{bb\bar{u}\bar{s}}$ & ${\cal O}^{0^-}_{T_{us}}$, $\eta_4$ &64.0&0.70&0.51&4.03&0.25&0.09&3.84&0.17&1.76&0.02&0.05&3.27&0.01&8.02&65 \\
$T_{bb\bar{u}\bar{s}}$ & $\eta_5$ &40.0&0.30&0.30&2.33&1.37&0.02&2.11&0.05&0.37&0.01&0.01&0.64&0.00&3.68&40 \\
$T_{bb\bar{s}\bar{s}}$ & $\eta_2$ &81.0&0.70&0.58&4.60&3.09&0.13&4.45&0.25&2.57&0.05&0.09&2.19&0.00&5.32&82 \\
$1^-$ &&&&&&&&&&\\
$T_{bb\bar{u}\bar{d}}$ & ${\cal O}^{1^-}_{T}$, $\eta_5$ &59.0&0.70&0.38&3.06&3.14&$\cdots$&3.00&0.00&$\cdots$&0.03&0.00&1.75&0.00&3.98&59\\
$T_{bb\bar{d}\bar{d}}$ & $\eta_2$ &74.0&0.80&0.52&4.18&4.53&$\cdots$&4.15&0.00&$\cdots$&0.00&0.00&2.73&0.00&6.35&75 \\
$T_{bb\bar{u}\bar{s}}$ & ${\cal O}^{1^-}_{T_{us}}$, $\eta_5$ &41.0&0.40&0.26&2.10&1.41&0.04&2.06&0.08&0.78&0.03&0.02&1.45&0.00&2.66&41 \\
$T_{bb\bar{s}\bar{s}}$ & $\eta_2$ &68.0&0.70&0.48&3.77&0.13&0.11&3.59&0.21&1.85&0.00&0.05&1.54&0.00&5.41&69 \\
\\
{\bf Class L} &&&&&&&&&&\\
$0^-$ &&&&&&&&&&\\
$T_{bb\bar{d}\bar{d}}$ & $\eta_1$ &0.50&0.10&0.08&0.50&0.33&$\cdots$&0.30&0.00&$\cdots$&0.00&0.00&0.95&0.00&2.02&2 \\
$T_{bb\bar{s}\bar{s}}$ & $\eta_1$ &0.60&0.10&0.06&0.43&0.44&0.01&0.26&0.01&0.74&0.00&0.01&0.70&0.00&1.59&2 \\
$1^-$ &&&&&&&&&&\\
$T_{bb\bar{u}\bar{d}}$ & $\eta_6$ &0.70&0.10&0.05&0.33&0.11&$\cdots$&0.20&0.00&$\cdots$&0.01&0.00&0.65&0.00&1.04&2 \\
$T_{bb\bar{d}\bar{d}}$ &$\eta_1$ &0.50&0.10&0.06&0.38&0.10&$\cdots$&0.23&0.00&$\cdots$&0.00&0.00&0.75&0.00&1.26&2\\
$T_{bb\bar{u}\bar{s}}$ & $\eta_6$ &0.40&0.00&0.05&0.30&0.13&0.00&0.18&0.00&0.28&0.01&0.01&0.54&0.00&0.95&2 \\
$T_{bb\bar{s}\bar{s}}$ & $\eta_1$ &0.50&0.00&0.05&0.32&0.13&0.00&0.20&0.01&0.58&0.00&0.01&0.55&0.00&1.01&2 \\
\\
\hline
\hline
\end{tabular}
}}
 \caption{Sources of errors of $T_{bb\bar u\bar d}$, $T_{bb\bar d\bar d}$, $T_{bb\bar u\bar s}$, $T_{bb\bar s\bar s}$ couplings. We take $\ve \Delta \mu\ve=0.05$ GeV and $\ve \Delta \tau\ve =0.01$ GeV$^{-2}$. $\Delta OPE $ has been estimated assuming that the high-dimension condensates is given as $m^{2}_{b}(\tau/3)\times d=6$ contributions. The error due to $M_{T_{cc\bar q\bar q'}}$ is intrinsically included in the estimate of the coupling as we use a mass corresponding to each value of  $t_c$.}

\label{tab:error-fb}
\end{table}

 -- Plateau and inflexion points for the coupling are reached for $t_c\geq 130$ GeV$^2$ while the mass presents minimum in this region of $t_c$. Moreover, the $R_{P/C}\geq 1$ condition is only satisfied for $t_c\geq 130$ GeV$^2$. Therefore, we consider as optimal results at LO within factorization the one obtained from $t_c=130-220$ GeV$^2$:
 \beq
f^{LO}_{T_{bb\bar u\bar d}(0^-)(\eta_5)}\vert_{fac}=46(5)_{t_c} ~{\rm keV}, 
~~~~~~~~~~
M^{LO}_{T_{bb\bar u\bar d}(0^-)(\eta_5)}\vert_{fac}= 10875(145)_{t_c}~{\rm MeV},
\label{eq:tbb0-eta5-fac}
\eeq 
which shows that the LO mass value is underestimated by about 575 MeV in Ref.\,\cite{ZHUT}. 

-- Including the NLO corrections, the stability becomes $t_c = 175\to 220$ GeV$^2$ and $\tau\simeq 0.08$ GeV$^{-2}$. One obtains :
 \beq
f^{NLO}_{T_{bb\bar u\bar d}(0^-)(\eta_5)}\vert_{fac}=126(19)_{t_c}(0.3) ~{\rm keV}, ~~~~~~~~~~
M^{NLO}_{T_{bb\bar u\bar d}(0^-)(\eta_5)}\vert_{fac}= 12310(100)_{t_c}(47)~{\rm MeV},
\label{eq:tbb0-eta5-fac1}
\eeq 
which shows that the NLO corrections are large for the mass and coupling. It does not favour the LO result 10.3 GeV of Ref.\,\cite{ZHUT}. Comparing with the NLO result in Eq.\,\ref{eq:tbb0-eta5} without factorization, one can see that the violation of factorization increases the mass by about 395 MeV and the coupling by 41 MeV in this channel.


\begin{table}[H]
\setlength{\tabcolsep}{0.3pc}
{\scriptsize{
\begin{tabular}{ll ll  ll  ll ll ll ll ll c}
\hline
\hline
                States & Currents
                    &\multicolumn{1}{c}{$\Delta t_c$}
					&\multicolumn{1}{c}{$\Delta \tau$}
					&\multicolumn{1}{c}{$\Delta \mu$}
					&\multicolumn{1}{c}{$\Delta \alpha_s$}
					&\multicolumn{1}{c}{$\Delta PT$}
					&\multicolumn{1}{c}{$\Delta m_s$}
					&\multicolumn{1}{c}{$\Delta m_c$}
					&\multicolumn{1}{c}{$\Delta \overline{\psi}\psi$}
					&\multicolumn{1}{c}{$\Delta \kappa$}					
					&\multicolumn{1}{c}{$\Delta G^2$}
					&\multicolumn{1}{c}{$\Delta M^{2}_{0}$}
					&\multicolumn{1}{c}{$\Delta \overline{\psi}\psi^2$}
					&\multicolumn{1}{c}{$\Delta G^3$}
					&\multicolumn{1}{c}{$\Delta OPE$}
					&\multicolumn{1}{r}{Total [MeV]}
\\
					
\hline
{\bf Class H} &&&&&&&&&&\\
$0^-$ &&&&&&&&&&\\
$T_{bb\bar{u}\bar{d}}$ & ${\cal O}^{0^-}_{T}$, $\eta_4$ &252&64.0&0.60&6.55&17.1&$\cdots$&9.89&0.00&$\cdots$&0.21&0.00&59.2&0.08&76.0&278 \\
$T_{bb\bar{u}\bar{d}}$ & $\eta_5$ &174&60.0&0.71&7.76&19.9&$\cdots$&10.9&0.00&$\cdots$&0.48&0.00&50.5&0.00&75.3&207 \\
$T_{bb\bar{d}\bar{d}}$ & $\eta_2$ &226&58.0&0.66&6.11&25.1&$\cdots$&10.6&0.00&$\cdots$&0.44&0.00&61.6&0.00&88.2&258 \\
$T_{bb\bar{u}\bar{s}}$ & ${\cal O}^{0^-}_{T_{us}}$, $\eta_4$ &214&69.0&0.68&6.02&18.3&0.70&10.9&2.05&30.9&0.31&1.09&56.7&0.12&76.4&247 \\
$T_{bb\bar{u}\bar{s}}$ & $\eta_5$ &221&47.0&0.76&8.05&18.3&0.05&11.6&1.79&28.3&0.60&0.66&50.2&0.00&94.3&252 \\
$T_{bb\bar{s}\bar{s}}$ & $\eta_2$ &230&55.0&0.72&6.21&29.0&0.28&11.5&3.51&65.4&0.69&2.66&59.5&0.00&101&274 \\
$1^-$ &&&&&&&&&&\\
$T_{bb\bar{u}\bar{d}}$ & ${\cal O}^{1^-}_{T}$, $\eta_5$ &233&66.0&0.64&6.00&20.5&$\cdots$&10.4&0.00&$\cdots$&0.54&0.00&61.2&0.00&76.1&262\\
$T_{bb\bar{d}\bar{d}}$ & $\eta_2$ &203&54.0&0.61&6.08&21.0&$\cdots$&9.90&0.00&$\cdots$&0.00&0.00&61.0&0.00&95.0&240 \\
$T_{bb\bar{u}\bar{s}}$ & ${\cal O}^{1^-}_{T_{us}}$, $\eta_5$ &238&58.0&0.65&6.44&21.9&0.63&10.6&1.85&31.6&0.65&0.69&61.2&0.00&82.0&268 \\
$T_{bb\bar{s}\bar{s}}$ & $\eta_2$ &205&55.0&0.68&6.17&27.9&0.73&10.9&3.77&63.9&0.01&1.93&57.5&0.01&81.5&246 \\
\\
{\bf Class L} &&&&&&&&&&\\
$0^-$ &&&&&&&&&&\\
$T_{bb\bar{d}\bar{d}}$ & $\eta_1$ &28.0&75.0&2.36&15.6&0.49&$\cdots$&9.11&0.00&$\cdots$&0.16&0.00&11.5&0.00&93.0&124 \\
$T_{bb\bar{s}\bar{s}}$ & $\eta_1$ &21.0&70.0&2.40&18.5&1.04&1.68&5.14&0.38&19.7&0.21&0.26&7.46&0.01&74.9&109 \\
$1^-$ &&&&&&&&&&\\
$T_{bb\bar{u}\bar{d}}$ & $\eta_6$ &25.0&89.0&1.98&15.7&1.61&$\cdots$&4.52&0.00&$\cdots$&0.01&0.00&17.6&0.00&55.1&110\\
$T_{bb\bar{d}\bar{d}}$ &$\eta_1$ &26.0&68.0&2.11&10.5&0.32&$\cdots$&8.12&0.00&$\cdots$&0.17&0.00&10.6&0.00&65.8&99\\
$T_{bb\bar{u}\bar{s}}$ & $\eta_6$ &24.0&68.0&3.56&15.1&0.10&0.43&4.75&0.27&5.90&0.01&0.19&13.8&0.00&61.0&97 \\
$T_{bb\bar{s}\bar{s}}$ & $\eta_1$ &21.0&97.0&2.13&9.21&0.17&1.48&4.14&0.34&8.00&0.21&0.23&7.53&0.01&58.4&116 \\
\\
\hline
\hline
\end{tabular}
}}
 \caption{Sources of errors of $T_{bb\bar u\bar d}$, $T_{bb\bar d\bar d}$, $T_{bb\bar u\bar s}$, $T_{bb\bar s\bar s}$ masses. We take $\ve \Delta \mu\ve=0.05$ GeV and $\ve \Delta \tau\ve =0.01$ GeV$^{-2}$. $\Delta OPE $ has been estimated assuming the high-dimension condensates is given as $m^{2}_{b}(\tau/3)\times d=6$ contributions.}

\label{tab:error-mb}
\end{table}

-- Now, we abandon factorization but still work to LO.  The analysis shown in Fig.\,\ref{fig:eta5-non-fac}
 is similar to Fig.\,\ref{fig:eta5-fac}. Though the mass has minimum from $t_c=145$ GeV$^2$, the coupling starts  to stabilize from $t_c=170 $ GeV$^2$ where also the requirement that $R_{P/C}\geq 1$ is fullfilled. One should also note that, for low values of $t_c$ not shown in the figure, the coupling becomes imaginary.  

One can remark that the mass is shifted to higher values where the minimum is reached for $t_c\simeq (145\to 220)$ GeV$^2$ while the coupling stabilizes in the range $t_c\simeq (160\to 220)$ GeV$^2$. The requirement $R_{P/C}\geq 1$ reduces the $(t_c,\tau)$ region 
from $(170, 0.075) \to (220,0.085)$  (GeV$^2$, GeV$^{-2}$) , where we extract the optimal estimate:

\begin{figure}[H]
\begin{center}
\centerline {\hspace*{-7.5cm} \bf a)\hspace{8cm} b)}
\includegraphics[width=8cm]{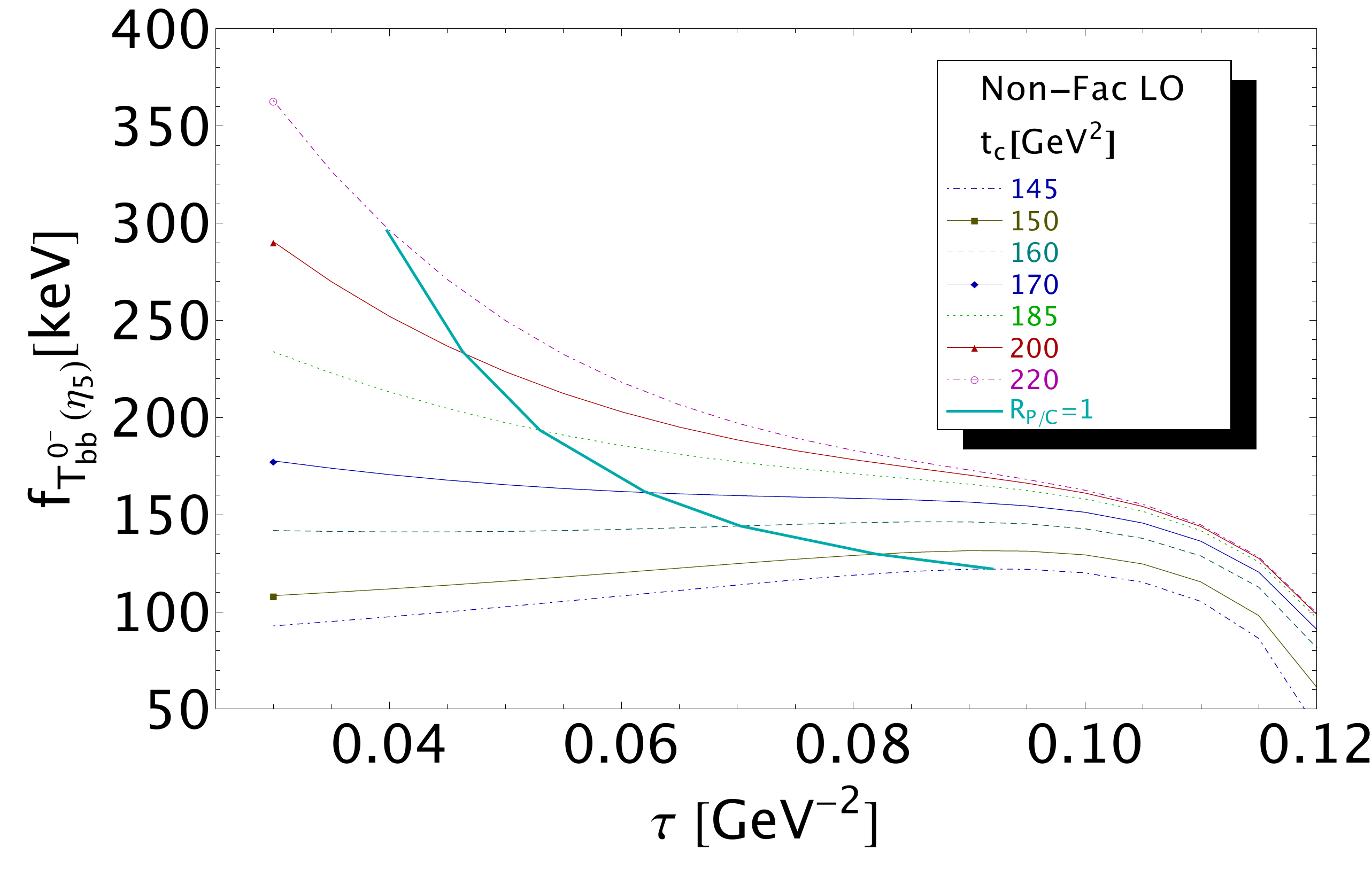}
\includegraphics[width=8cm]{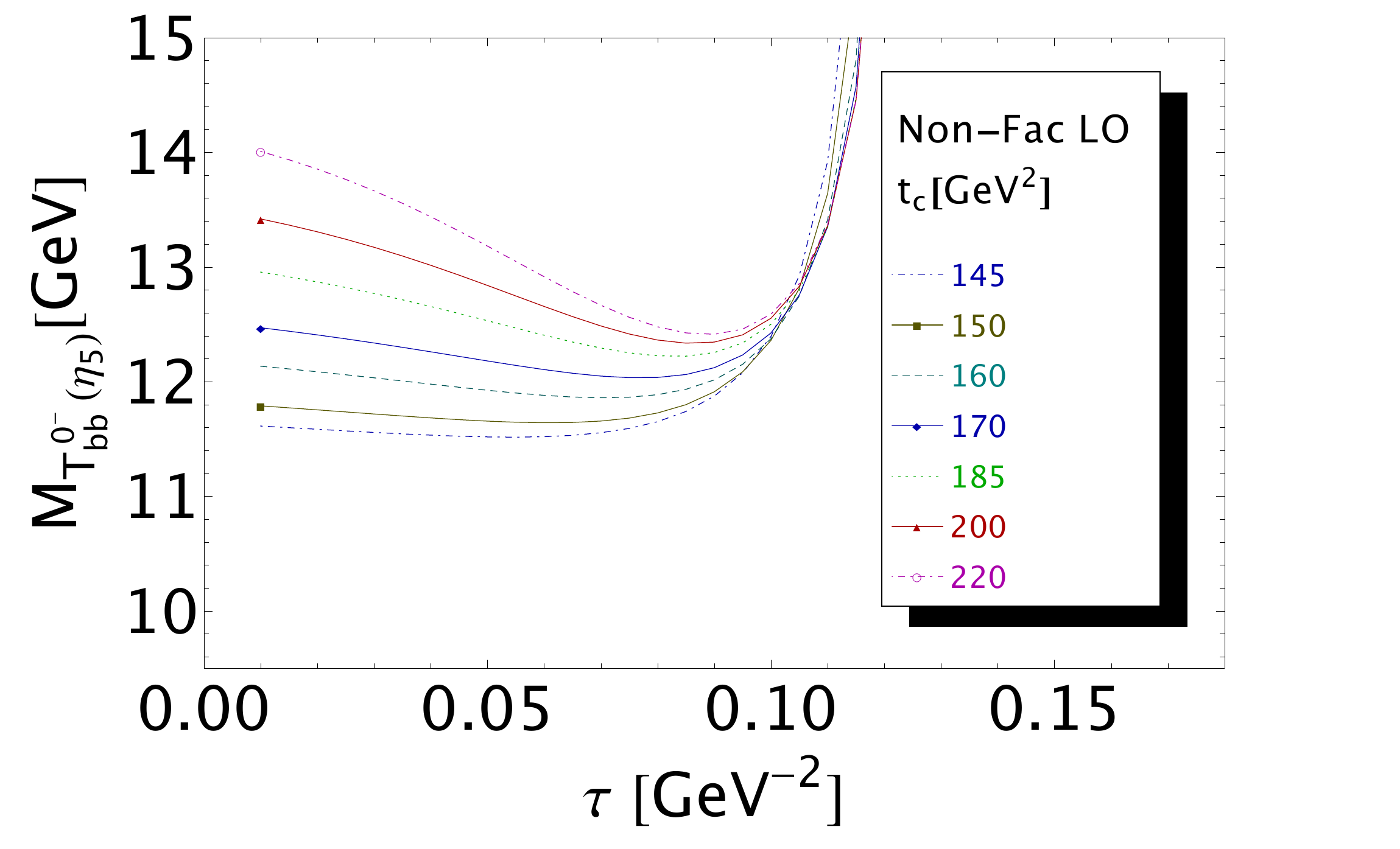}
\vspace*{-0.5cm}
\caption{\footnotesize  $f_{T_{bb\bar u\bar d}}$ and $M_{T_{bb\bar u\bar d}}$ for the $\eta_5$ current as a function of $\tau$ at LO and for different values of $t_c$. We use  the QCD inputs of Ref\,\cite{ZHUT} but abandon the factorization of the four-quark condensates. The OPE is truncated at $d=6$ condensates.  } 
\label{fig:eta5-non-fac}
\end{center}
\vspace*{-0.5cm}
\end{figure} 

 \beq
f^{LO}_{T_{bb\bar u\bar d}(0^-)(\eta_5)}\vert_{non-fac}=167(31)~{\rm keV}, ~~~~~~~~~~
M^{LO}_{T_{bb\bar u\bar d}(0^-)(\eta_5)}\vert_{non-fac}= 12188(216)_{t_c}~{\rm MeV}.
\label{eq:tbb0-eta5-non-fac}
\eeq 

Adding to this LO result the NLO corrections, we obtain the NLO results quoted in Eq.\,\ref{eq:tbb0-eta5} and in Table\,\ref{tab:resb} where one notice that the NLO corrections increases the LO mass value by 517 MeV while the coupling is (almost) unaffected. 

One can also notice that low values of the  mass are not favoured by the non-factorized four-quark condensate where the corresponding coupling becomes imaginary for $\tau\geq 0.15$ GeV$^{-2}$ (not shown in the figure).

Comparing the LO without and within factorization, one can also notice that the violation of factorization increases the mass by 1.3 GeV and the coupling by a factor 3.8.
\subsection* {\d  $\eta_1$ current}
  
 We check the LO result of Ref.\,\cite{ZHUT} using factorization and using the same QCD parameters in the case of $T_{bb\bar d\bar d}$ for the choice $t_c=125$ GeV$^2$ and $\tau=0.10-0.14$ GeV$^{-2}$  . We find 9.97 GeV which is  630 MeV below the one of Ref.\,\cite{ZHUT}. This difference might be due to the different size of the mixed condensate in the two papers.  We add NLO corrections and find :
  \beq
  M^{NLO}_{T_{bb\bar d\bar d}(0^-)(\eta_1)}\vert_{fac}= 10505(100)_{t_c}~{\rm MeV},
  \eeq
  which is comparable with the NLO result of 10407 MeV in Table\,\ref{tab:resb} obtained without using four-quark factorization. This result indicates that corrections due to factorization are negligible here while the NLO ones are large. 
  
\subsection* {\d  SU3 breakings}
  
  For the pseudoscalar states, SU3 breakings decrease the central values of the masses by about (139-191) MeV for the Class H current while for Class L, the effect is tiny (an increase of about 4-45 MeV). 

\subsection* {\b Class L Vector states}
 
\subsection* {\d  $\eta_1$ current}
  
 To LO and using factorization, we find that the coupling exhibits minimum from $t_c=120\to 160$ GeV$^2$ at $\tau=0.05\to 0.16$ GeV$^{-2}$. We obtain :
  \beq
f^{LO}_{T_{bb\bar d\bar d}(1^-)(\eta_1)}\vert_{fac}=6(1)_{t_c} (0)_\tau~{\rm keV}, ~~~~~~~~~~
M^{LO}_{T_{bb\bar d\bar d}(1^-)(\eta_1)}\vert_{fac}= 10430(12)_{t_c}(113)_{\tau}~{\rm MeV},
\label{eq:tbb1-eta1-fac}
\eeq 
compared to the LO result without factorization obtained from $\tau=0.05\to 0.14$ GeV$^{-2}$:
  \beq
f^{LO}_{T_{bb\bar d\bar d}(1^-)(\eta_1)}=10(1)_{t_c}(0)_{\tau} ~{\rm keV}, ~~~~~
M^{LO}_{T_{bb\bar d\bar d}(1^-)(\eta_1)}= 10438(23)_{t_c}(97)_{\tau}~{\rm MeV},
\label{eq:tbb1-eta1-fac1}
\eeq 
where the effect of the factorization is negligible.  Adding NLO corrections, to the LO result within factorization, we obtain:
\beq
f^{NLO}_{T_{bb\bar d\bar d}(1^-)(\eta_1)}\vert_{fac}=7(1)_{t_c}(0)_\tau ~{\rm keV}, ~~~~~
M^{NLO}_{T_{bb\bar d\bar d}(1^-)(\eta_1)}\vert_{fac}= 10434(14)_{t_c}(69)_{\tau}~{\rm MeV},
\label{eq:tbb1-eta1-fac-nlo}
\eeq
where the NLO corrections are negligible.  We find that Ref.\,\cite{ZHUT} overestimates  the central value of the mass by about 200 MeV while the $\tau$-stabillity starts at earlier value $t_c=120$ GeV$^2$ in our case.

\subsection* {\d  $\eta_6$ current}
 
 To LO and using factorization, we find that the coupling exhibits minimum from $t_c=120\to 160$ GeV$^2$ at $\tau=0.04\to 0.15$ GeV$^{-2}$. Within the previous range, we obtain :
  \beq
f^{LO}_{T_{bb\bar u\bar d}(1^-)(\eta_6)}\vert_{fac}=6(1)_{t_c}(0.1)_{\tau} ~{\rm keV}, ~~~~~~~~~~
M^{LO}_{T_{bb\bar u\bar d}(1^-)(\eta_6)}\vert_{fac}= 10494(13)_{t_c}(68)_{\tau}~{\rm MeV},
\label{eq:tbb1-eta6-fac}
\eeq 
compared to the LO result without factorization obtained from $\tau=0.03\to 0.14$ GeV$^{-2}$:
  \beq
f^{LO}_{T_{bb\bar u\bar d}(1^-)(\eta_6)}=10(2)_{t_c}(0)_{\tau} ~{\rm keV}, ~~~~~
M^{LO}_{T_{bb\bar u\bar d}(1^-)(\eta_6)}= 10463(18)_{t_c}(58)_{\tau}~{\rm MeV},
\label{eq:tbb1-eta1-fac2}
\eeq 
where the effect of the factorization of 31 MeV is negligible. We notice that the $\tau$ stability for the choice $t_c=120$ GeV$^2$ is reached earlier than the one 0.11 GeV$^{-2}$ quoted in Ref.\,\cite{ZHUT}.  Adding NLO corrections, to the LO result within factorization, $\tau$ moves in the region $(0.05\to 0.17)$ GeV$^{-2}$. We obtain:
\beq
f^{NLO}_{T_{bb\bar u\bar d}(1^-)(\eta_6)}\vert_{fac}=6.7(1)_{t_c}(0.1)_\tau ~{\rm keV}, ~~~~~~~~~~
M^{NLO}_{T_{bb\bar u\bar d}(1^-)(\eta_6)}\vert_{fac}= 10393(58)_{t_c}(51)_{\tau}~{\rm MeV},
\label{eq:tbb1-eta6-fac-nlo}
\eeq
where the NLO corrections have decreased the LO result by 101 MeV. 

\subsection* {\d  SU3 breakings}

For vector currents, SU3 breakings are tiny and decrease slightly the mass of the states associated to the Class H and Class L currents.

\section{Radial excitations of the $T_{QQ\bar u\bar d}$ and $T_{QQ\bar d\bar d}$ states}
We  complete the analysis by estimating the masses and couplings of the first radial excitations of the $T_{QQ\bar u\bar d}$ and $T_{QQ\bar d\bar d}$ states. In so doing, we parametrize the spectral function by two resonances $\oplus$ QCD continuum contributions. We use the parameters of the lowest resonances obtained previously. The analysis is very similar to the previous ones.
\subsection*{\b  $T_{cc\bar u\bar d}$ and $T_{cc\bar d\bar d}$ states}
\subsubsection*{\d  Class H states}
\begin{figure}[hbt]
\begin{center}
\centerline {\hspace*{-7.5cm} \bf a)\hspace{8cm} b)}
\includegraphics[width=8cm]{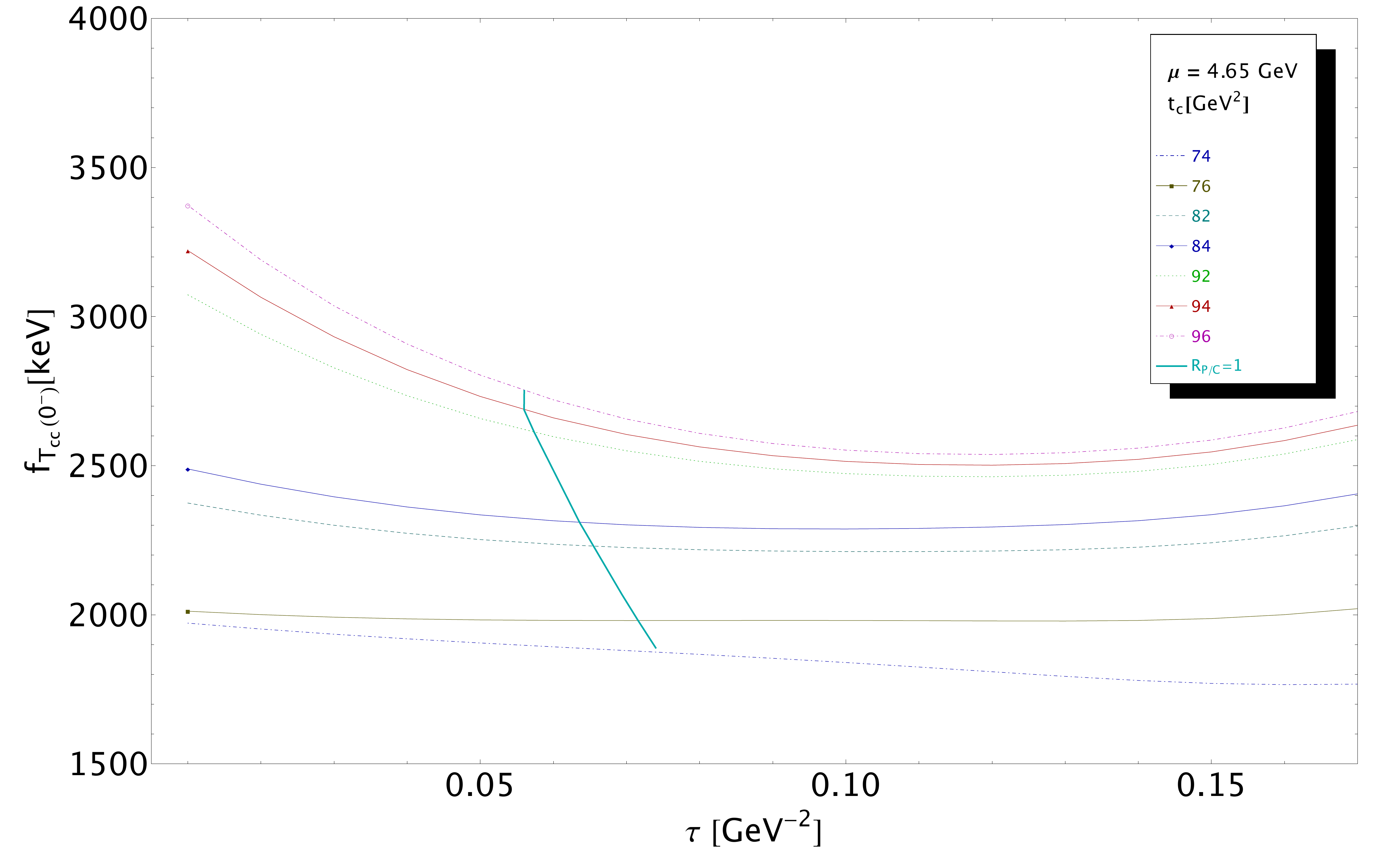}
\includegraphics[width=8cm]{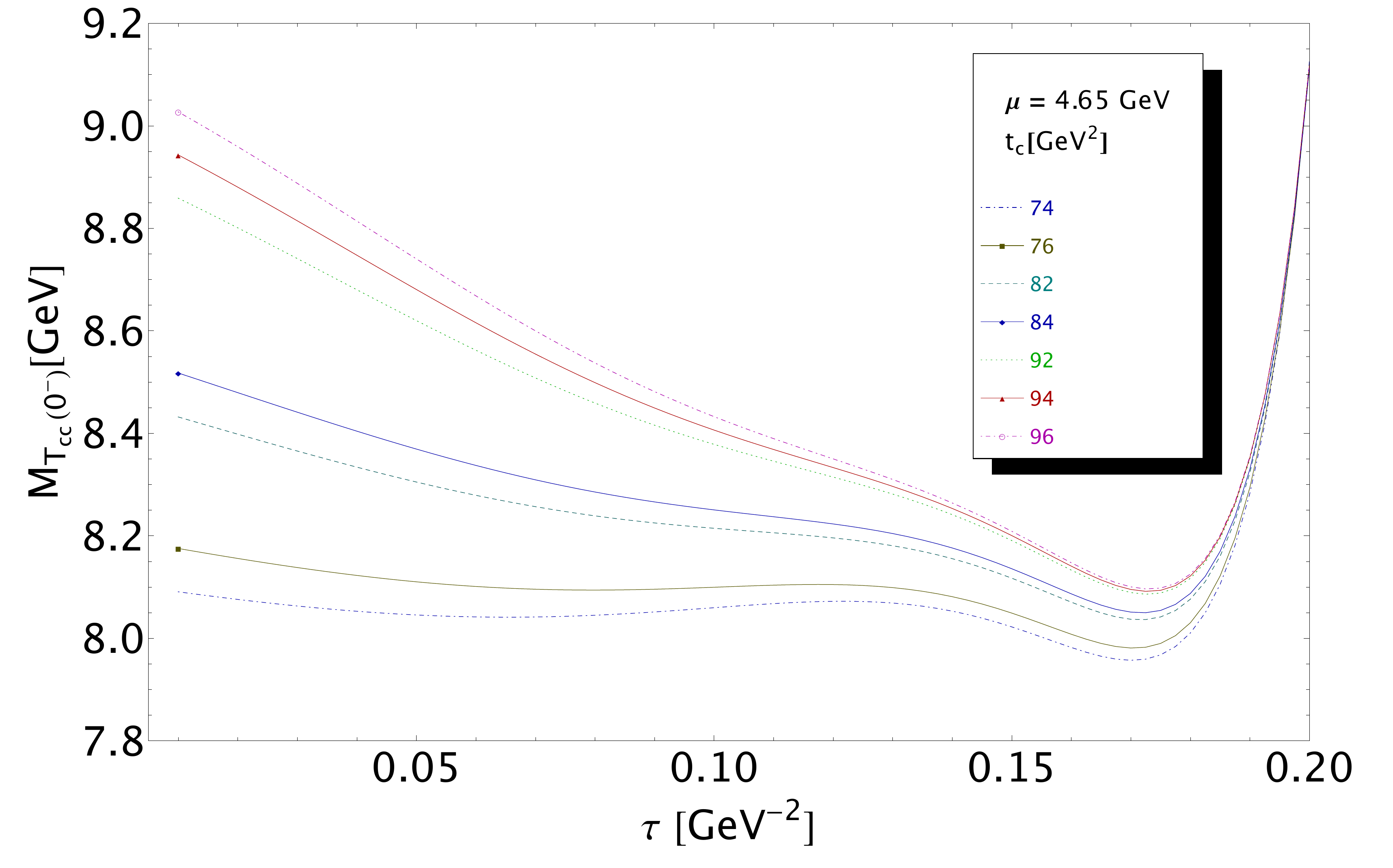}
\vspace*{-0.5cm}
\caption{\footnotesize  $f_{T^{0^-}_{cc\bar u\bar d}}$ and $M_{T^{0^-}_{cc\bar u\bar d}}$ 1st radial excitation coupling and mass for the $\eta_4$ pseudoscalar current as a function of $\tau$ at NLO and for different values of $t_c$.   } 
\label{fig:rad0H}
\end{center}
\vspace*{-0.5cm}
\end{figure}

-- We show explicitly the analysis of the $\eta_4$ pseudoscalar state  in Fig.\,\ref{fig:rad0H} as a representative of the Class H heavy states.  The mass presents plateau with a slight minimum at $\tau\simeq 0.07$ GeV$^{-2}$ for low values of $t_c$ and inflexion point at $\tau\simeq 0.12$ GeV$^{-2}$ for large $t_c$ values. These values of $\tau$ also coincide with the minimum of the coupling. A (misleading) minimum, which we shall not consider, appears at larger value of $\tau\simeq 0.17$ GeV$^{-2}$ where the OPE does not converge.  The behaviour of the curves for the other Class H currents is similar to the one of the $\eta_4$.

-- Our results are summarized in Table\,\ref{tab:res-rad-c} where the different sources of the errors can be found in Table\,\ref{tab:error-rad-c}.  We conclude that the 1st radial excitations of the Class H states are in the region of 8 GeV which are quite high (about 2 GeV above the corresponding ground states) rendering them difficult for detection. 

-- One can also notice their relative large couplings  to the currents which are about the same strength as the lowest ground state ones. A feature which is unusual compared to the case of the ordinary mesons ($\rho,\rho',\cdots$). However, we should mention that the LSR approach cannot detect an unusual weakly coupled radial excitation state or an almost degenerate one with the ground state.

-- The common sum rule scale for these states is:
$
\tau\simeq (0.07\sim .13)~{\rm GeV}^{-2}
$, 
which is smaller than that of the corresponding ground states: $\tau\simeq (0.12\sim .18)~{\rm GeV}^{-2}$. 
\vspace*{-0.5cm} 
\begin{center}
   {\scriptsize
\begin{table}[hbt]
\setlength{\tabcolsep}{0.9pc}
    {\small
  \begin{tabular}{lllllll }
&\\
\hline
\hline
States&Current  &$t_c$ [GeV$^2$] &$\tau$  [GeV$^{-2}$]  & $f^{NLO} _{T_{ccqq'}}$ [keV]& $\tau$  [GeV$^{-2}$] &
$M^{NLO}_{T_{ccqq'}}$ [MeV]\\ 
 \hline 
\bf Class H&&&&&&\\
  { \bma  $0^-$} &&&&&&\\
  $T_{cc\bar u\bar d}$&$ {\cal O}_T^{0^-},\eta_4$&$76\to 94$&$0.07\to 0.12$&2101(215)&$0.08\to 0.12$&8214(214)\\
  
$T_{cc\bar d\bar d}$ & $\eta_2$&$74\to130$&$0.07\to 0.12$&2026(514)&$0.08\to 0.12$&8212(353) \\
 { \bma  $1^-$} &&&&&&\\
 $T_{cc\bar u\bar d}$& $ {\cal O}_T^{1^-},\eta_5 $&$74\to 94$&$0.10\to 0.10$&1262(224)&$0.10\to 0.10$&8110(325)\\
  $T_{cc\bar d\bar d}$& $\eta_2 $&$74\to 94$&$0.12\to 0.12$&1571(423)&$0.12\to 0.12$&8111(316)\\

  \bf Class L&&&&&&\\
  { \bma  $0^-$} &&&&&&\\
$T_{cc\bar u\bar d}$&$ \eta_5$&$38\to 50$&$0.20\to 0.31$&596(135)&$0.20\to 0.31$&5787(213) \\
$T_{cc\bar d\bar d}$& $\eta_1$&$40\to 60$&$0.17\to 0.17$&580(169)&$0.16\to 0.23$&5957(251) \\
  { \bma  $1^-$} &&&&&&\\
  $T_{cc\bar u\bar d}$&  $\eta_6 $&$44\to 62$&$0.26\to 0.21$&451(100)&$0.14\to 0.21$&6217(349)\\
 $T_{cc\bar d\bar d}$& $ \eta_1 $&$46\to 62$&$0.14\to 0.16$&731(135)&$0.13\to 0.16$&6415(264) \\
   \hline\hline
  \vspace*{-0.5cm}
\end{tabular}}
 \caption{Masses and couplings of the 1st radial excitations of the $T_{cc\bar u\bar d}$ and $T_{cc\bar d\bar d}$ states at NLO. The different sources of the errors are in Table\,\ref{tab:error-rad-c}. }  

\label{tab:res-rad-c}
\end{table}
} 
\end{center}

\subsubsection*{\d  Class L states}
\begin{figure}[hbt]
\begin{center}
\centerline {\hspace*{-7.5cm} \bf a)\hspace{8cm} b)}
\includegraphics[width=8cm]{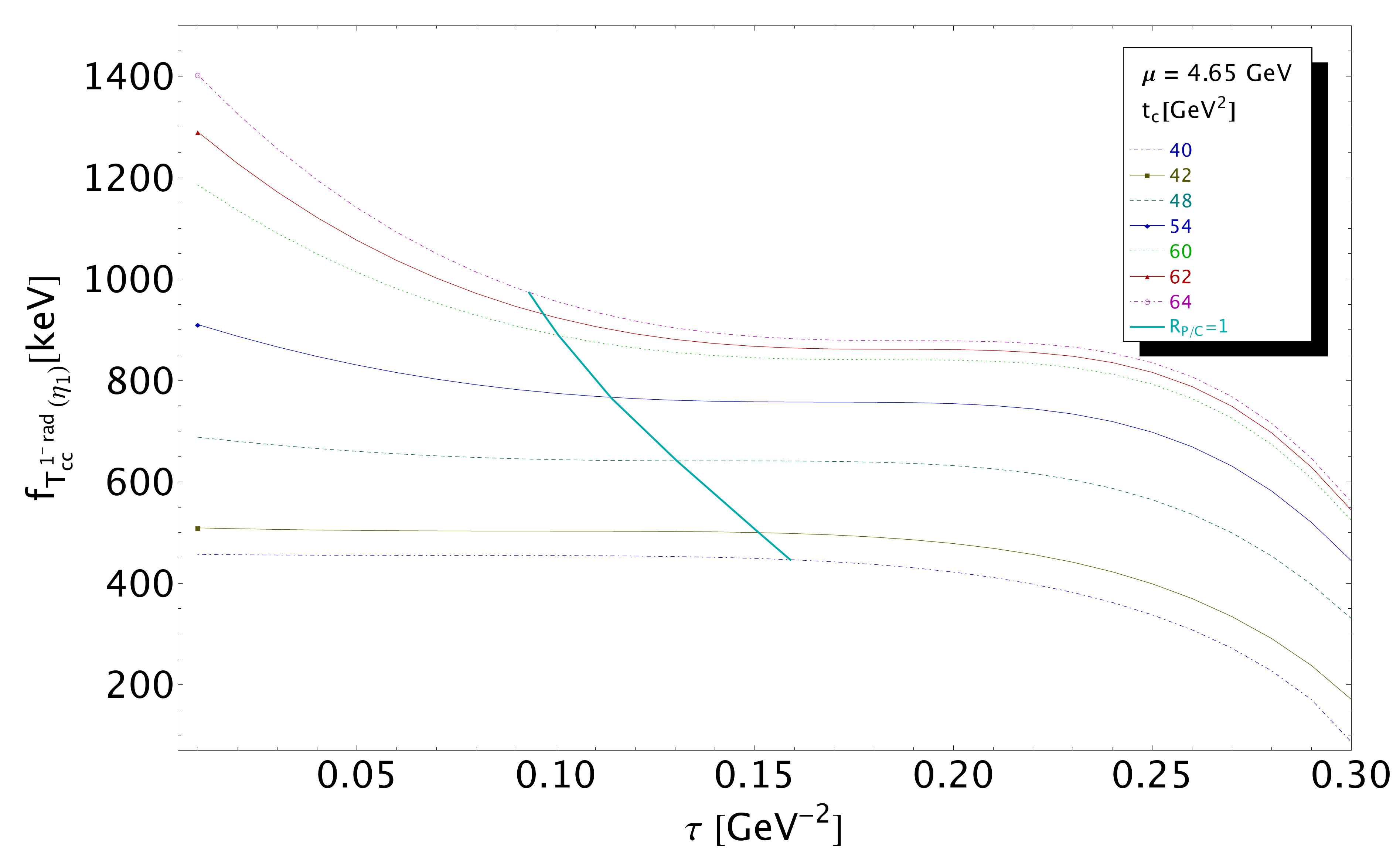}
\includegraphics[width=8cm]{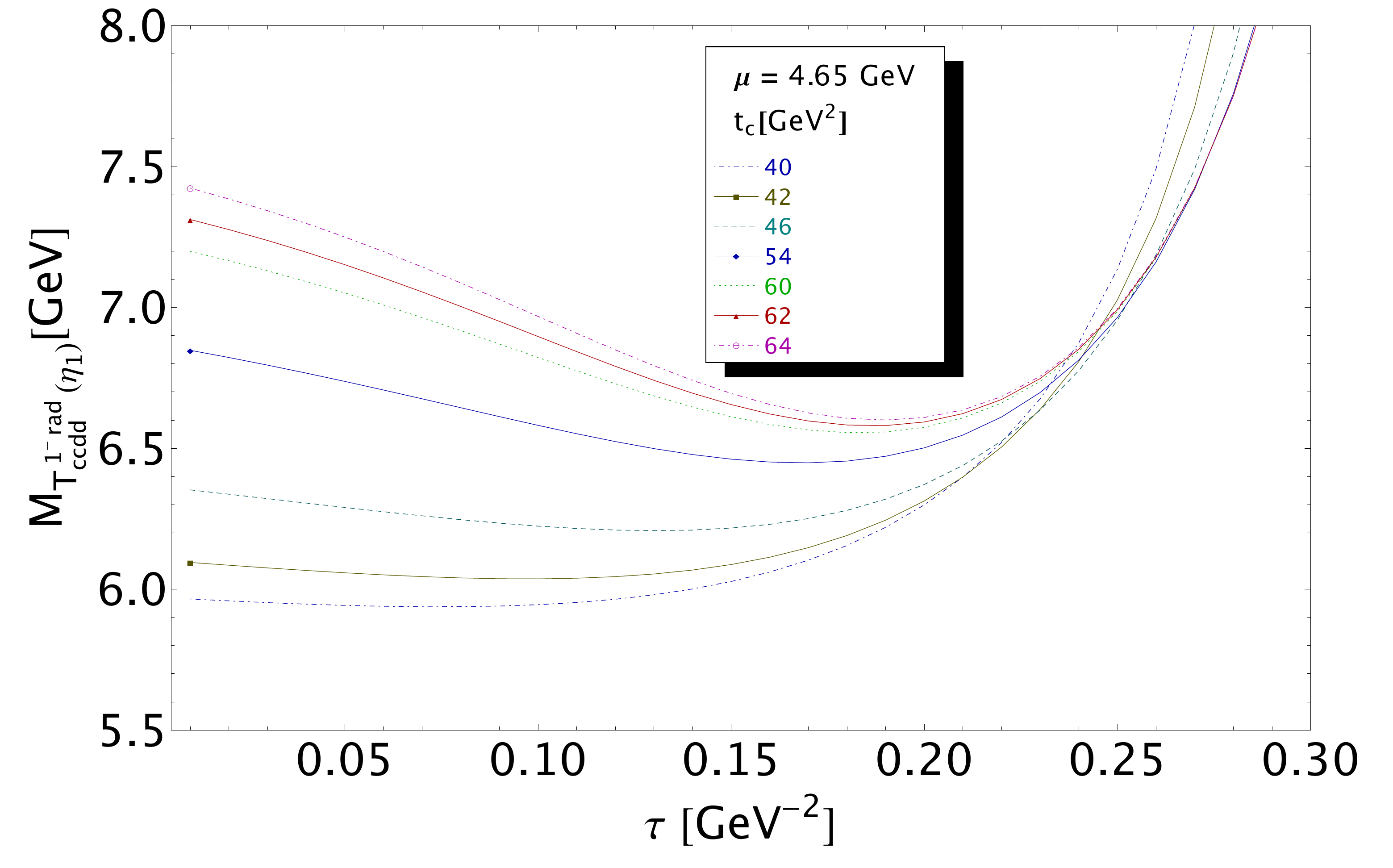}
\vspace*{-0.5cm}
\caption{\footnotesize  $f_{T^{1^-}_{cc\bar d\bar d}}$ and $M_{T^{1^-}_{cc\bar d\bar d}}$ radial excitation coupling and mass for the $\eta_1$ vector  current as a function of $\tau$ at NLO and for different values of $t_c$.   } 
\label{fig:rad-eta1-L}
\end{center}
\vspace*{-0.5cm}
\end{figure}

\begin{figure}[hbt]
\begin{center}
\centerline {\hspace*{-7.5cm} \bf a)\hspace{8cm} b)}
\includegraphics[width=8cm]{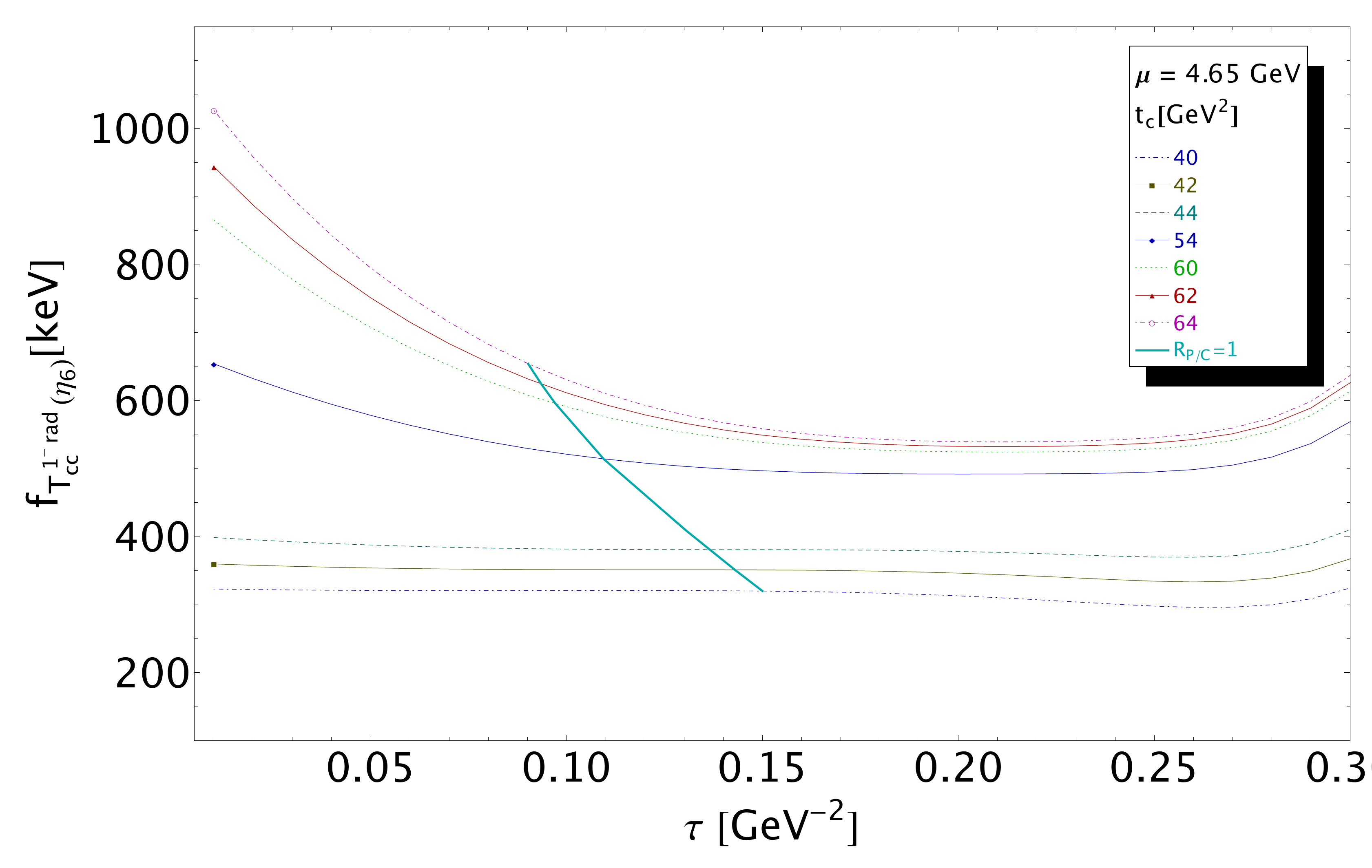}
\includegraphics[width=8cm]{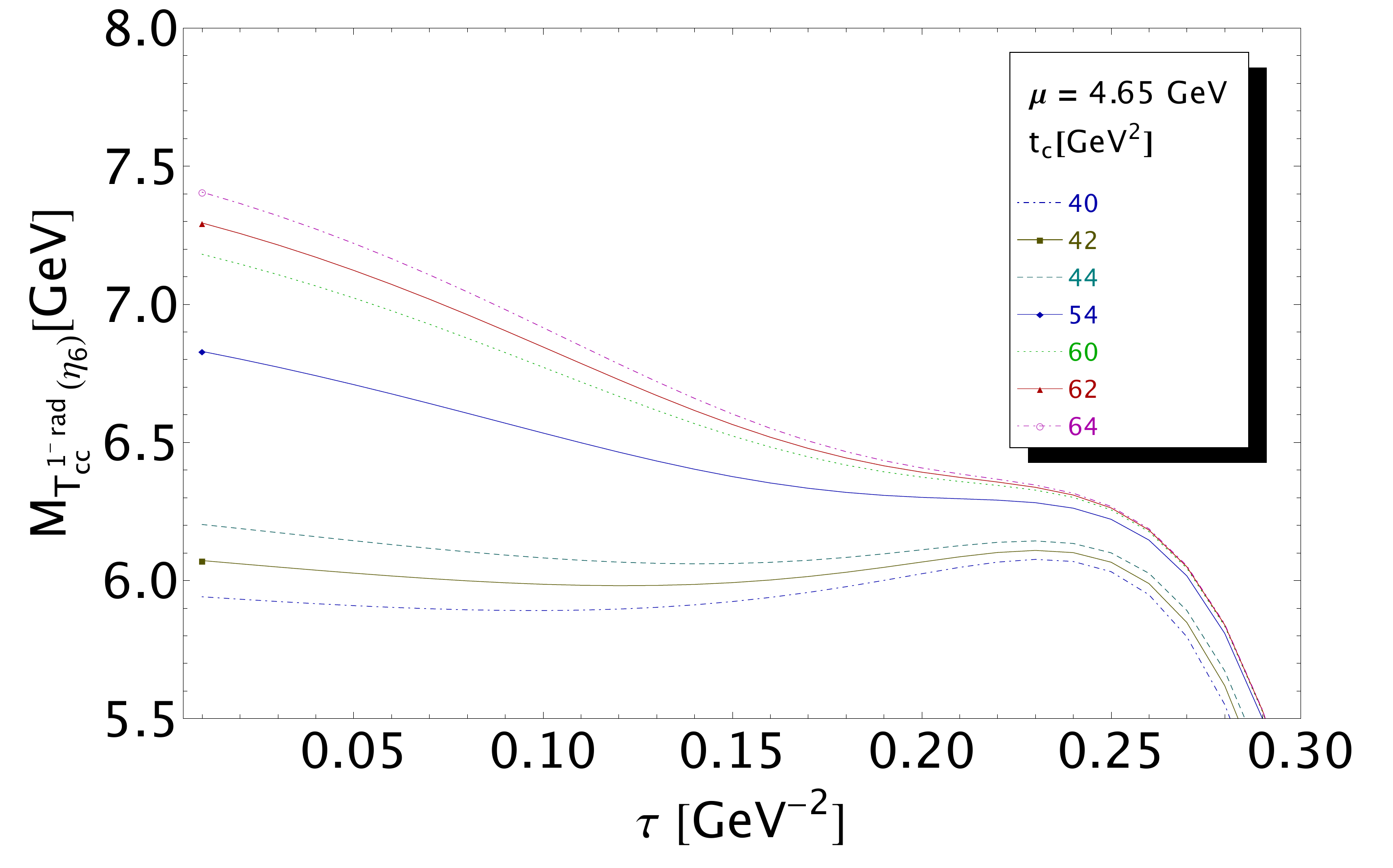}
\vspace*{-0.5cm}
\caption{\footnotesize  $f_{T^{1^-}_{cc\bar d\bar d}}$ and $M_{T^{1^-}_{cc\bar d\bar d}}$ radial excitation coupling and mass for the $\eta_6$ vector current as a function of $\tau$ at NLO and for different values of $t_c$.   } 
\label{fig:rad-eta6-L}
\end{center}
\vspace*{-0.5cm}
\end{figure}

The shape of the curves versus $\tau$ for different values of $t_c$ is similar for the  $\eta_1$ pseudoscalar and $\eta_6$ vector currents. Similar observation occurs for the $\eta_5$ pseudoscalar and $\eta_1$ vector currents.
We illustrate the two cases in Fig.\,\ref{fig:rad-eta1-L} for the $\eta_1$ and in Fig.\,\ref{fig:rad-eta6-L} for the $\eta_6$ vector currents.  
\begin{table}[H]
\setlength{\tabcolsep}{0.35pc}
{\scriptsize{
\begin{tabular}{ll ll  ll  ll ll ll ll ll l c}
\hline
\hline
                States & Currents
                    &\multicolumn{1}{c}{$\Delta t_c$}
					&\multicolumn{1}{c}{$\Delta \tau$}
					&\multicolumn{1}{c}{$\Delta \mu$}
					&\multicolumn{1}{c}{$\Delta \alpha_s$}
					&\multicolumn{1}{c}{$\Delta m_c$}
					&\multicolumn{1}{c}{$\Delta \overline{\psi}\psi$}
					&\multicolumn{1}{c}{$\Delta G^2$}
					&\multicolumn{1}{c}{$\Delta G^3$}
					&\multicolumn{1}{c}{$\Delta \overline{\psi}\psi^2$}
					
					&\multicolumn{1}{c}{$\Delta OPE$}
					&\multicolumn{1}{c}{$\Delta (M_{G},f_G)$}
					&\multicolumn{1}{c}{$\Delta M_{G_1}$}
					&\multicolumn{1}{r}{Total}
\\
					
\hline
{\bf Class H} &&&&&&&&&&\\
\it Masses [MeV]\\
$0^-$ &&&&&&&&&&\\
$T_{cc\bar{u}\bar{d}}$ & ${\cal O}^{0^-}_{T}$, $\eta_4$ &119&42.0&6&18&13&0.0&9&9&15&7&$170$&$\cdots$&214 \\
$T_{cc\bar{d}\bar{d}}$ & $\eta_2$ &204&122&15&23&46&0.0&2&0.0&55&12&250&$\cdots$&353 \\
$1^-$ &&&&&&&&&&\\
$T_{cc\bar{u}\bar{d}}$ & ${\cal O}^{1^-}_{T}$, $\eta_5$ &177&68&8&15&23&0.0&0.0&0.0&20&4&261&$\cdots$&325\\

$T_{cc\bar{d}\bar{d}}$ & $\eta_2$ &139&85&19&31&60&0.0&1&1&81&17&249&$\cdots$&316 \\
\it Couplings [keV]\\

$0^-$ &&&&&&&&&&\\

$T_{cc\bar{u}\bar{d}}$ & ${\cal O}^{0^-}_{T}$, $\eta_4$ &120&55&4&8&11&0.0&6&6&12&2&53&197&212\\
$T_{cc\bar{d}\bar{d}}$ & $\eta_2$ &368&37&5&12&15&0.0&1&0.0&34&0.0&42&358&514 \\
$1^-$ &&&&&&&&&&\\
$T_{cc\bar{u}\bar{d}}$ & ${\cal O}^{1^-}_{T}$, $\eta_5$ &176&7&1.&1&5&0.0&0.0&0.0&3&1&40&133&224\\

$T_{cc\bar{d}\bar{d}}$ & $\eta_2$ &206&13&7&12&22&0.0&0.5&0.0&20&5&88&356&423\\
{\bf Class L} &&&&&&&&&&\\
\it Masses [MeV]\\
$0^-$ &&&&&&&&&&\\
$T_{cc\bar{u}\bar{d}}$ & $\eta_5$&98&4&16&49&46&0.0&1&1&59&26&163&$\cdots$&213\\
$T_{cc\bar{d}\bar{d}}$ & $\eta_1$&157&25&13&66&39&0.0&3&1&120&52&122&$\cdots$&251 \\
$1^-$ &&&&&&&&&&\\
$T_{cc\bar{u}\bar{d}}$ & $\eta_6$&157&25&10&36&31&0.0&1&1&71&30&297&$\cdots$&349 \\
$T_{cc\bar{d}\bar{d}}$ & $\eta_1$&207&33&6&32&20&0.0&2&1&65&26&142&$\cdots$&265\\
\it Couplings [keV]\\
$0^-$ &&&&&&&&&&\\
$T_{cc\bar{u}\bar{d}}$ & $\eta_5$&85&0.1&3&7&5&0.0&0.0&0.5&10&5&39&97&135\\
$T_{cc\bar{d}\bar{d}}$ & $\eta_1$&153&3&0&11&3&0.0&0.0&0.9&21&1&53&41&169 \\
$1^-$ &&&&&&&&&&\\
$T_{cc\bar{u}\bar{d}}$ & $\eta_6$&81&3&2&4&7&0.0&2&1&9&12&10&55&100 \\
$T_{cc\bar{d}\bar{d}}$ & $\eta_1$&133&4&2&2&4&0.0&1&1&5&5&6&20&135\\
\hline
\hline
\end{tabular}
}}
 \caption{Sources of errors of the $T_{cc\bar u\bar d}$ and $T_{cc\bar d\bar d}$ 1st radial excitations masses and couplings. We take $\ve \Delta \mu\ve=0.05$ GeV and $\ve \Delta \tau\ve =0.02$ GeV$^{-2}$. 
 For $\Delta OPE$, we assume that the high-dimension condensates contribute as $m_c^2(\tau/3)\times d=6$ contributions.}
\label{tab:error-rad-c}
\end{table}
-- Like in the case of the Class H states, the splittings between the lowest ground states and the 1st radial excitations are about 2 GeV while their couplings to the currents are about 2 times the ones of the corresponding ground states which signal the new dynamics of these $T_{cc\bar q\bar q'}$ states. 

-- The 1st radial excitations of the Class L vector states are difficult to disentangle from the ground states of the corresponding Class H currents as they are almost degenerated.  However, the ones 
 of the Class L pseudoscalar states are about ($300\sim 500$) MeV below the corresponding ground states Class H currents.

-- The eventual finding of the $T_{cc\bar u\bar d}$ (6.3) Class H pseudoscalar ground state which may not be obscured by the 1st radial excitation $T'_{cc\bar u\bar d}$ (5.8) of the corresponding Class L one can serve as an  alternative test of the violation of four-quark condensate factorization hypothesis.

\subsubsection*{\b  $T_{bb\bar u\bar d}$ and $T_{bb\bar d\bar d}$ states}
The results of the analysis are summarized in Table\,\ref{tab:res-rad-b}.  The different sources of the errors are compiled in Table\,\ref{tab:error-rad-b}.
\subsubsection*{\d Class H states}
\begin{figure}[hbt]
\begin{center}
\centerline {\hspace*{-7.5cm} \bf a)\hspace{8cm} b)}
\includegraphics[width=8cm]{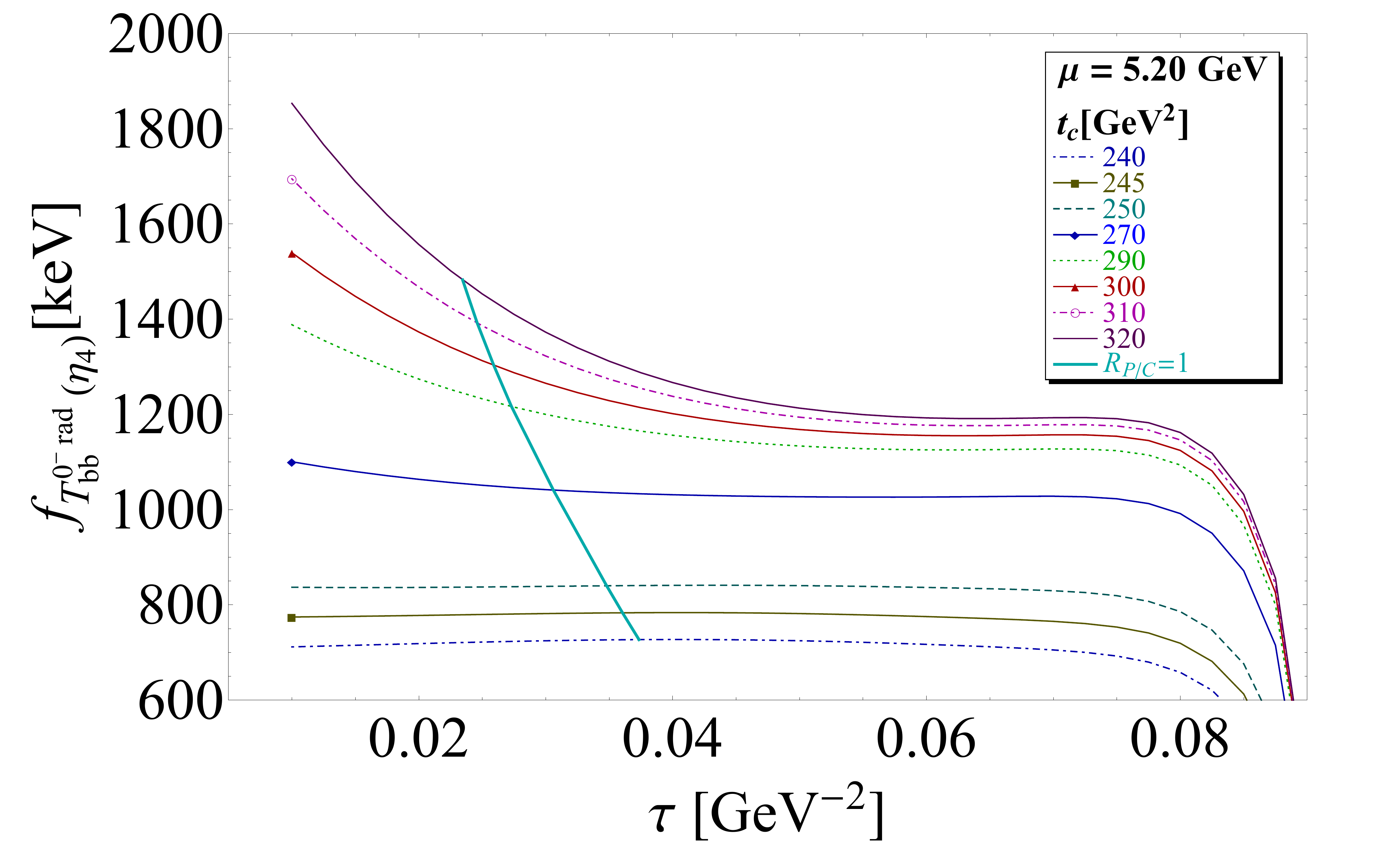}
\includegraphics[width=8cm]{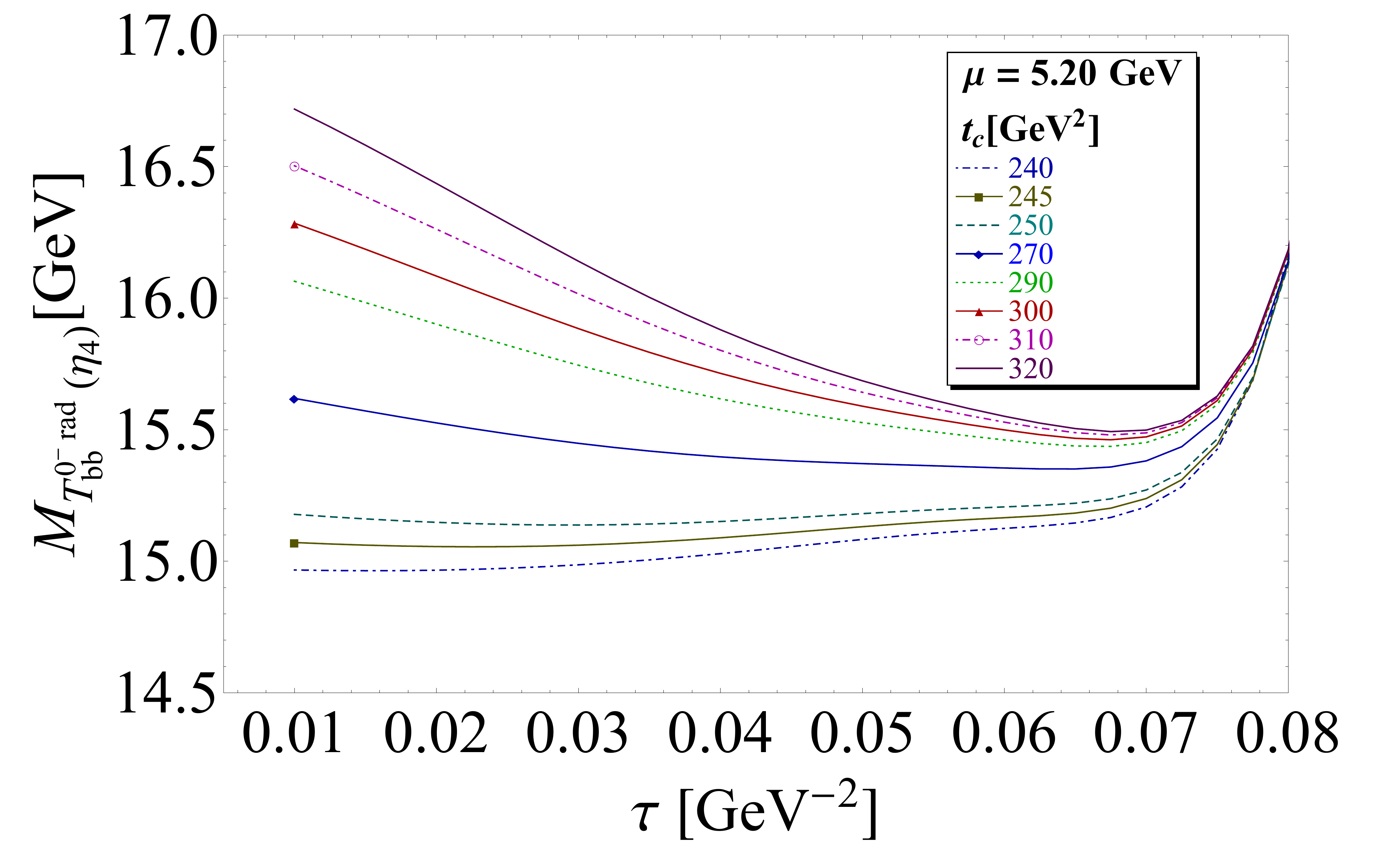}
\vspace*{-0.5cm}
\caption{\footnotesize  $f_{T^{0^-}_{bb\bar u\bar d}}$ and $M_{T^{0^-}_{bb\bar u\bar d}}$ 1st radial excitation coupling and mass for the $\eta_4$ current as a function of $\tau$ and for different values of $t_c$.   } 
\label{fig:eta4-rad-b0}
\end{center}
\vspace*{-0.5cm}
\end{figure}

-- Like the case of the charm, the Class H $\eta_5$ vector and $\eta_2$ pseudoscalar and vector currents can be illustrated by the $\eta_4$ $0^-$ pseudoscalar states which present the same shape of $\tau$ and $t_c$ behaviours as in Fig.\,\ref{fig:eta4-rad-b0}. 

\begin{figure}[H]
\begin{center}
\centerline {\hspace*{-7.5cm} \bf a)\hspace{8cm} b)}
\includegraphics[width=8cm]{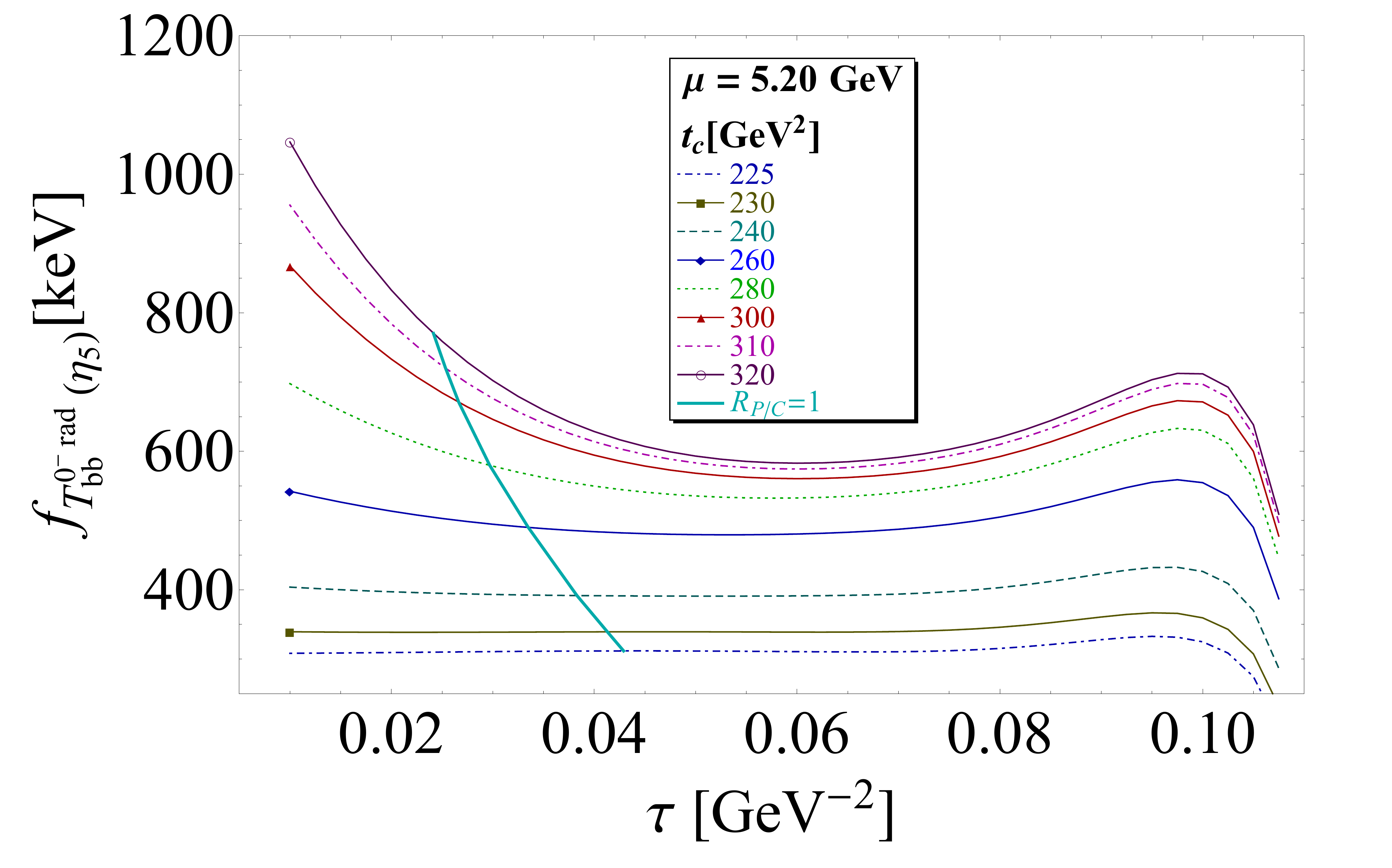}
\includegraphics[width=8cm]{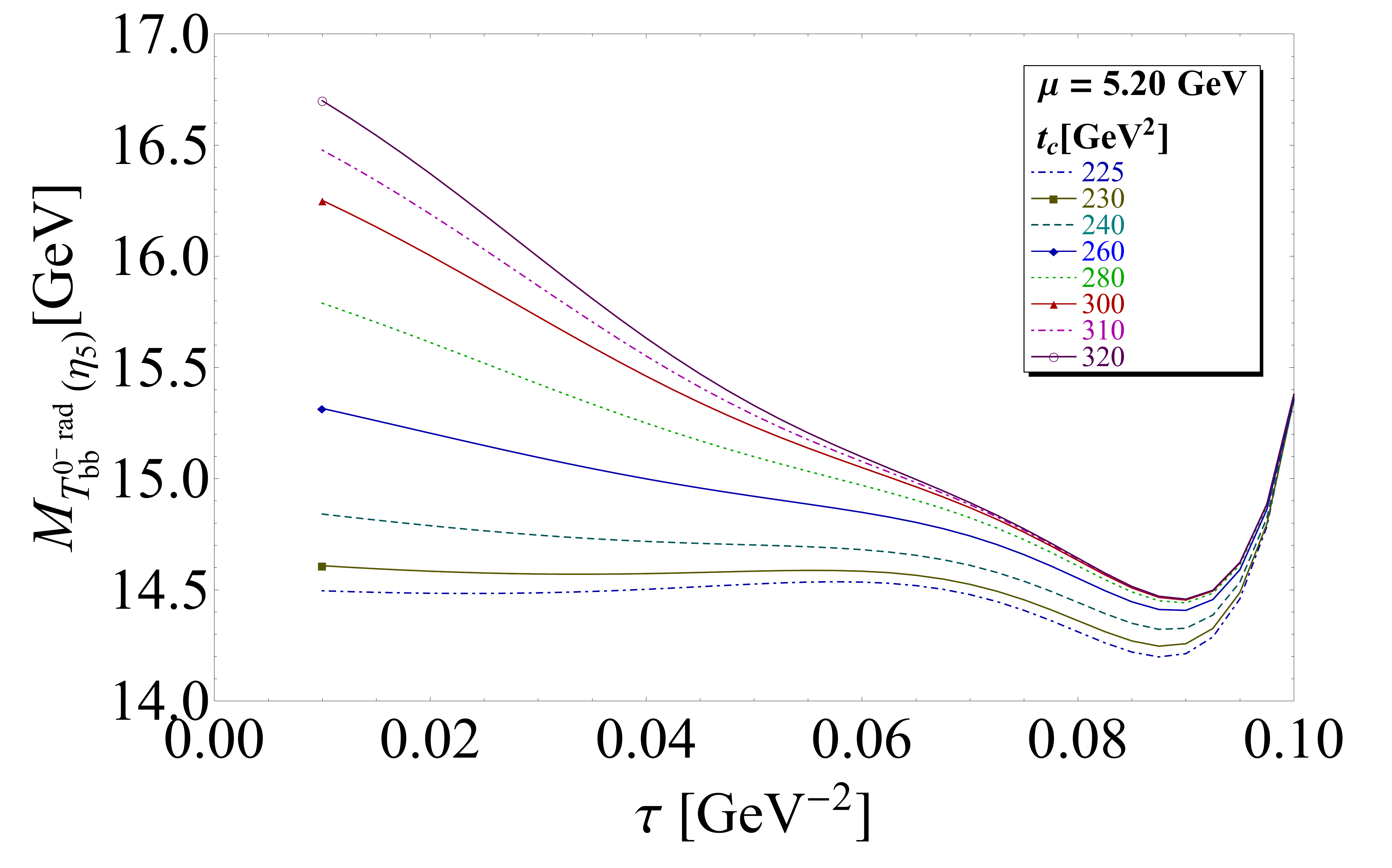}
\vspace*{-0.5cm}
\caption{\footnotesize  $f_{T^{0^-}_{bb\bar u\bar d}}$ and $M_{T^{0^-}_{bb\bar u\bar d}}$ radial excitation coupling and mass for the $\eta_5$ current as a function of $\tau$ and for different values of $t_c$.   } 
\label{fig:eta5-rad-b0}
\end{center}
\vspace*{-0.5cm}
\end{figure}

-- The $\eta_5$ pseudoscalar  current which becomes a Class H for the $b$-quark channel is shown in Fig.\,\ref{fig:eta5-rad-b0} as the corresponding curves have a different shape. One can notice like in the charm channel that the minimum of the mass at 0.9 GeV$^{-2}$ is misleading where the OPE breaks down. Instead, we consider as optimal values, the ones from the inflexion points below 0.07 GeV$^{-2}$. 

-- The masses of the $T_{bb\bar q\bar q'}$ 1st radial excitations are around 15 GeV (Table\,\ref{tab:res-rad-b}) which are  about 2 GeV above the ground state ones (Table\,\ref{tab:resb}) like in the case of the $T_{cc\bar q\bar q'}$ states. 

-- The couplings of the $T_{bb\bar q\bar q'}$ 1st radial excitations to the currents are about $(2.5\sim 3)$ times the ones of the ground states. Here, these are about 3 times larger than the one of the ground states indicating the new dynamics of these four-quark states. 

-- One can notice that the sum rule scale for the first radial excitations and for the ground states of the $T_{bb\bar q\bar q'}$ are about : $\tau\simeq (0.04 \sim 0.07)$ GeV$^{-2}$ which is lower than the one of the ground states : $\tau\simeq (0.07 \sim 0.09)$ GeV$^{-2}$ like in the case of the  $T_{cc\bar q\bar q'}$.

\subsubsection*{\d Class L states}
\begin{figure}[hbt]
\begin{center}
\centerline {\hspace*{-7.5cm} \bf a)\hspace{8cm} b)}
\includegraphics[width=8cm]{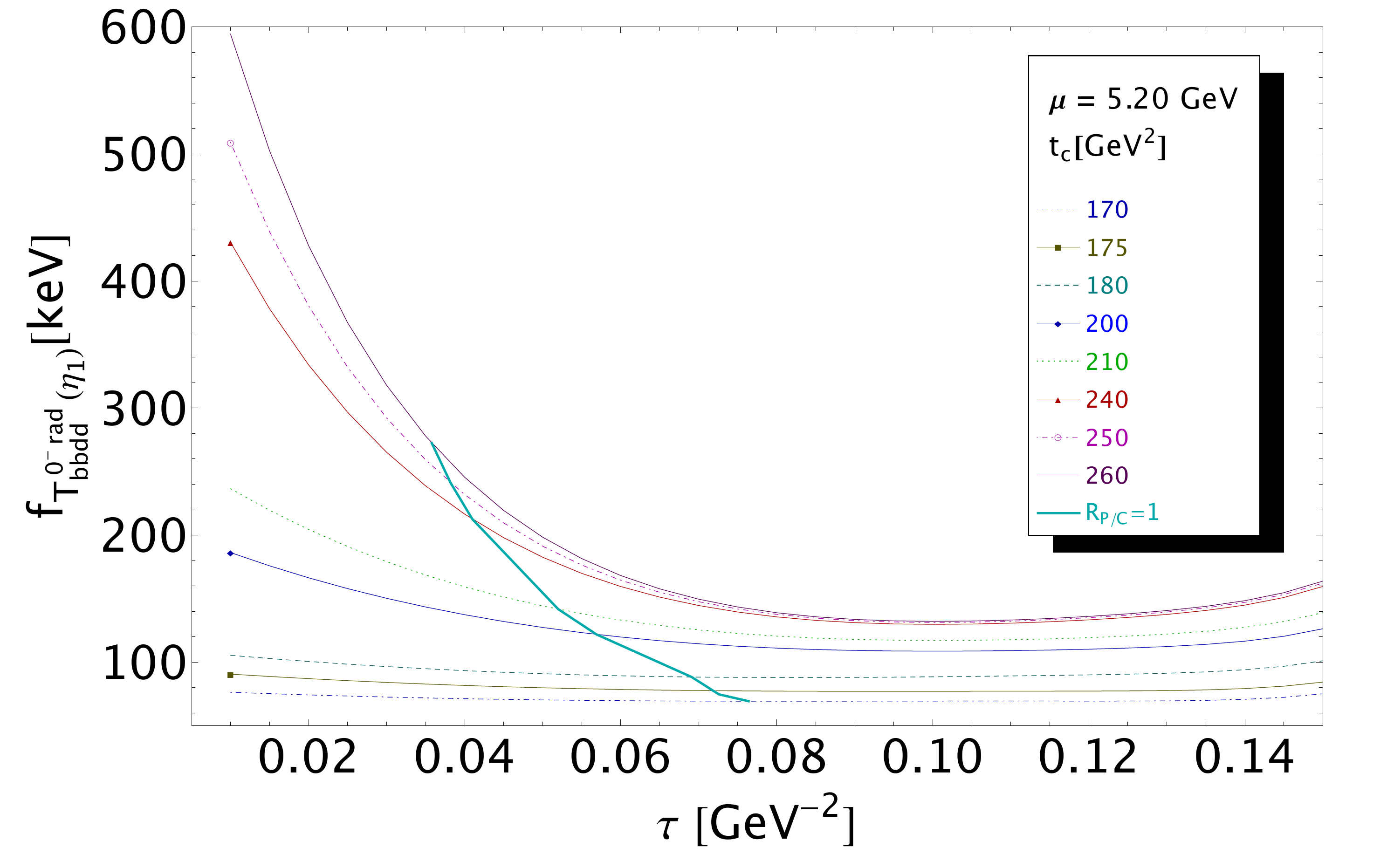}
\includegraphics[width=8cm]{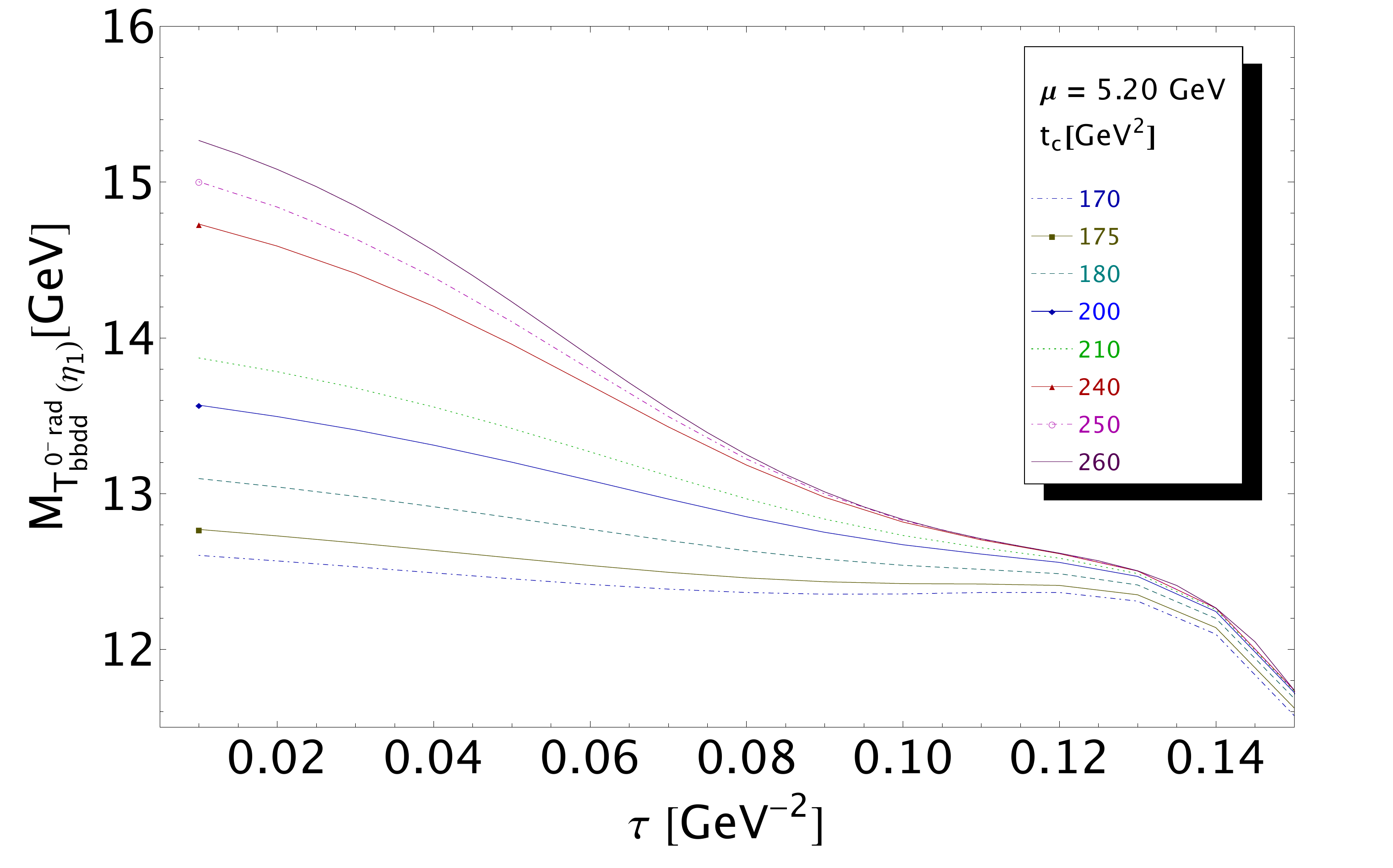}
\vspace*{-0.5cm}
\caption{\footnotesize  $f_{T^{0^-}_{bb\bar d\bar d}}$ and $M_{T^{0^-}_{bb\bar d\bar d}}$ radial excitation coupling and mass for the $\eta_1$ $0^-$ current as a function of $\tau$ and for different values of $t_c$.   } 
\label{fig:eta1-rad-b0}
\end{center}
\vspace*{-0.5cm}
\end{figure}

\begin{figure}[hbt]
\begin{center}
\centerline {\hspace*{-7.5cm} \bf a)\hspace{8cm} b)}
\includegraphics[width=8cm]{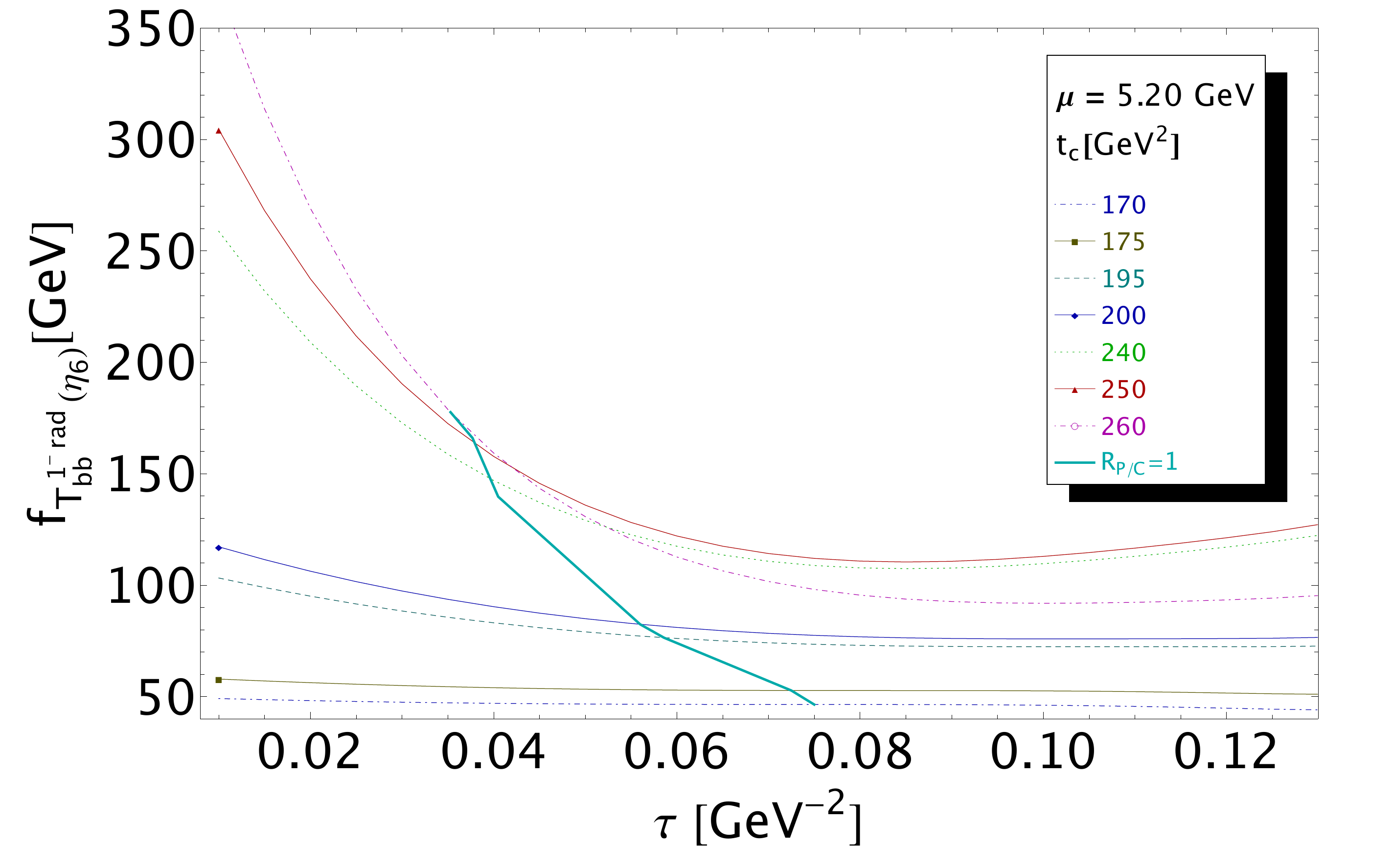}
\includegraphics[width=8cm]{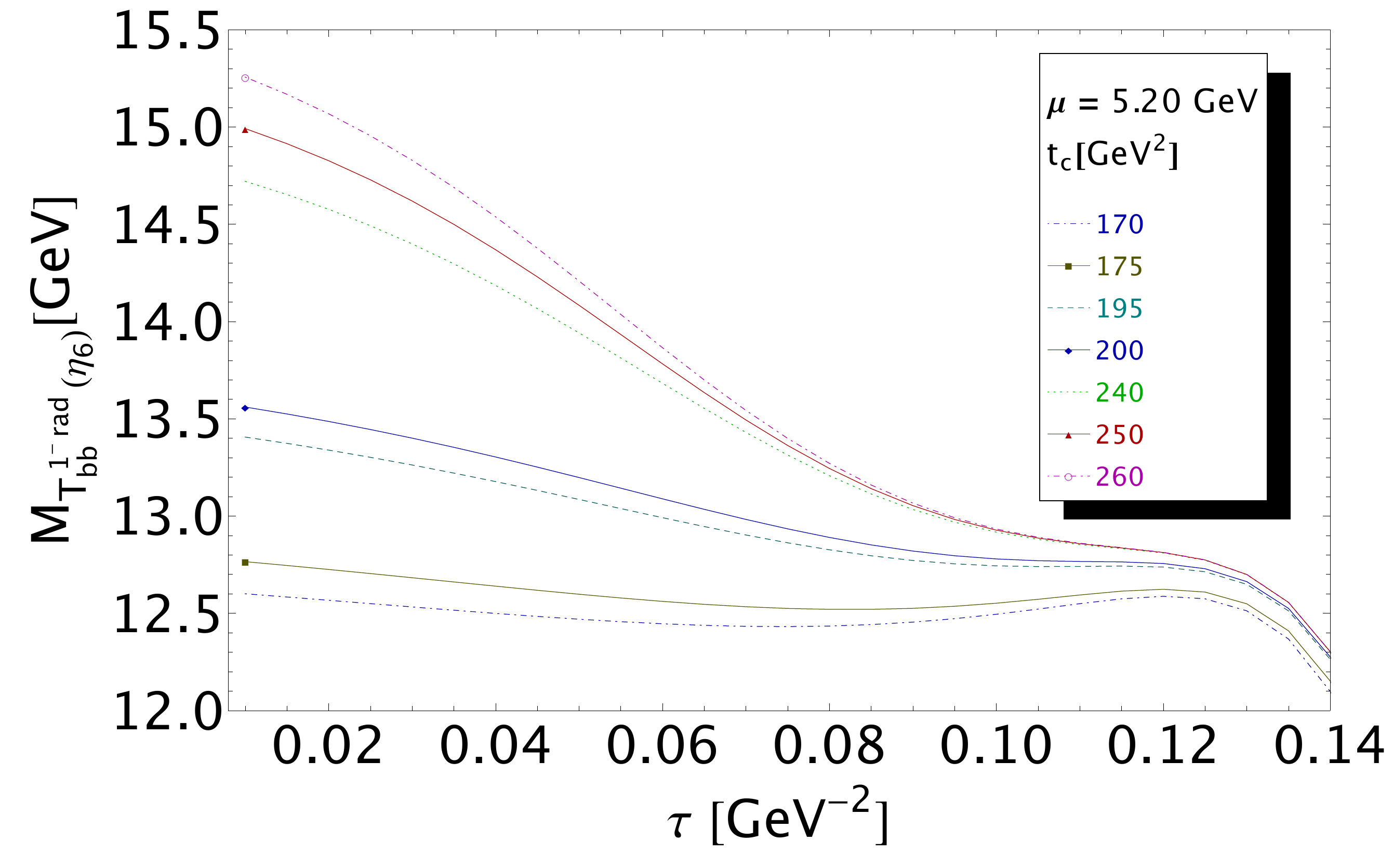}
\vspace*{-0.5cm}
\caption{\footnotesize  $f_{T^{1^-}_{bb\bar u\bar d}}$ and $M_{T^{1^-}_{bb\bar u\bar d}}$ radial excitation coupling and mass for the $\eta_6$ current as a function of $\tau$ and for different values of $t_c$.   } 
\label{fig:eta6-rad-b1}
\end{center}
\vspace*{-0.5cm}
\end{figure}

\begin{figure}[hbt]
\begin{center}
\centerline {\hspace*{-7.5cm} \bf a)\hspace{8cm} b)}
\includegraphics[width=8cm]{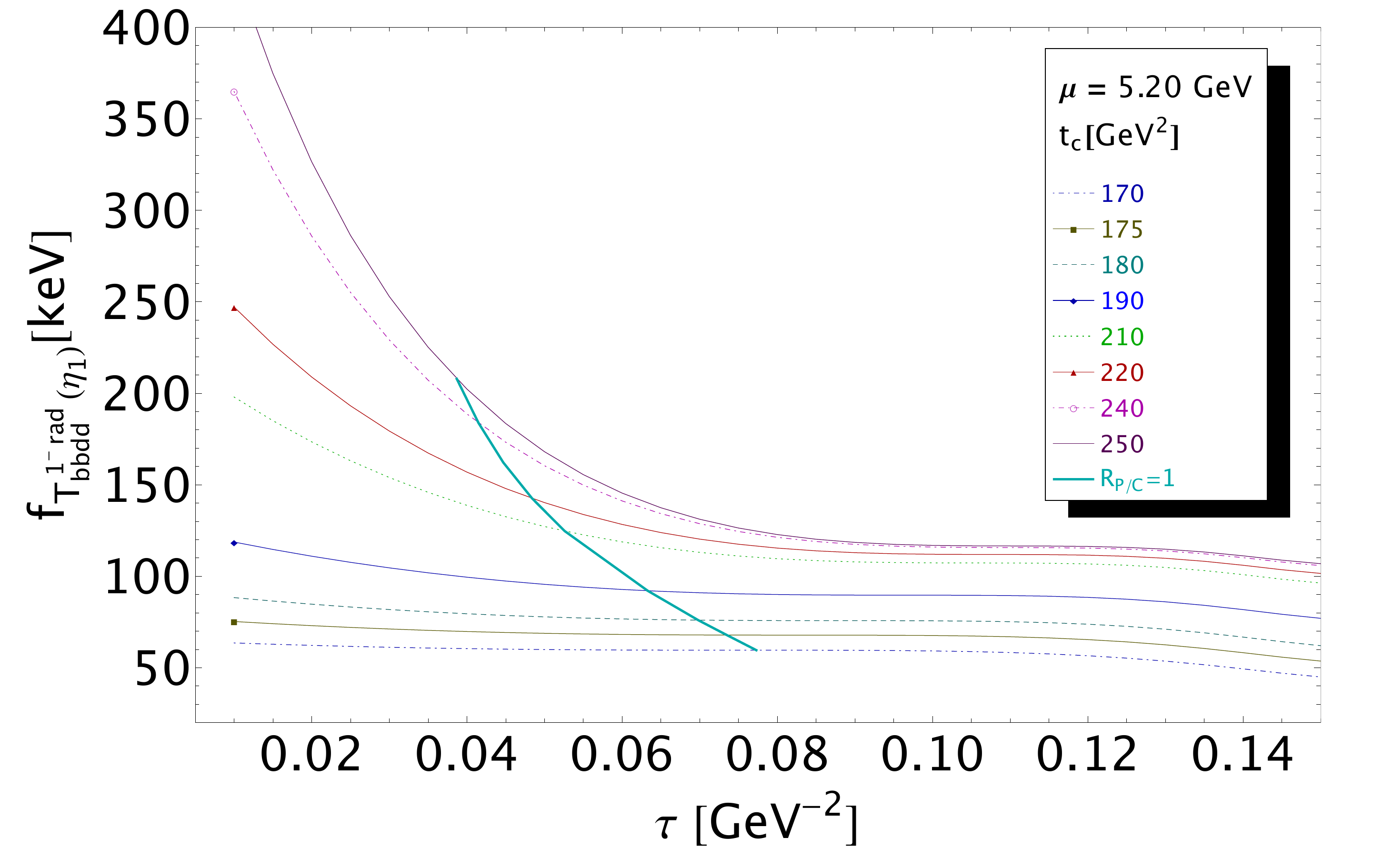}
\includegraphics[width=8cm]{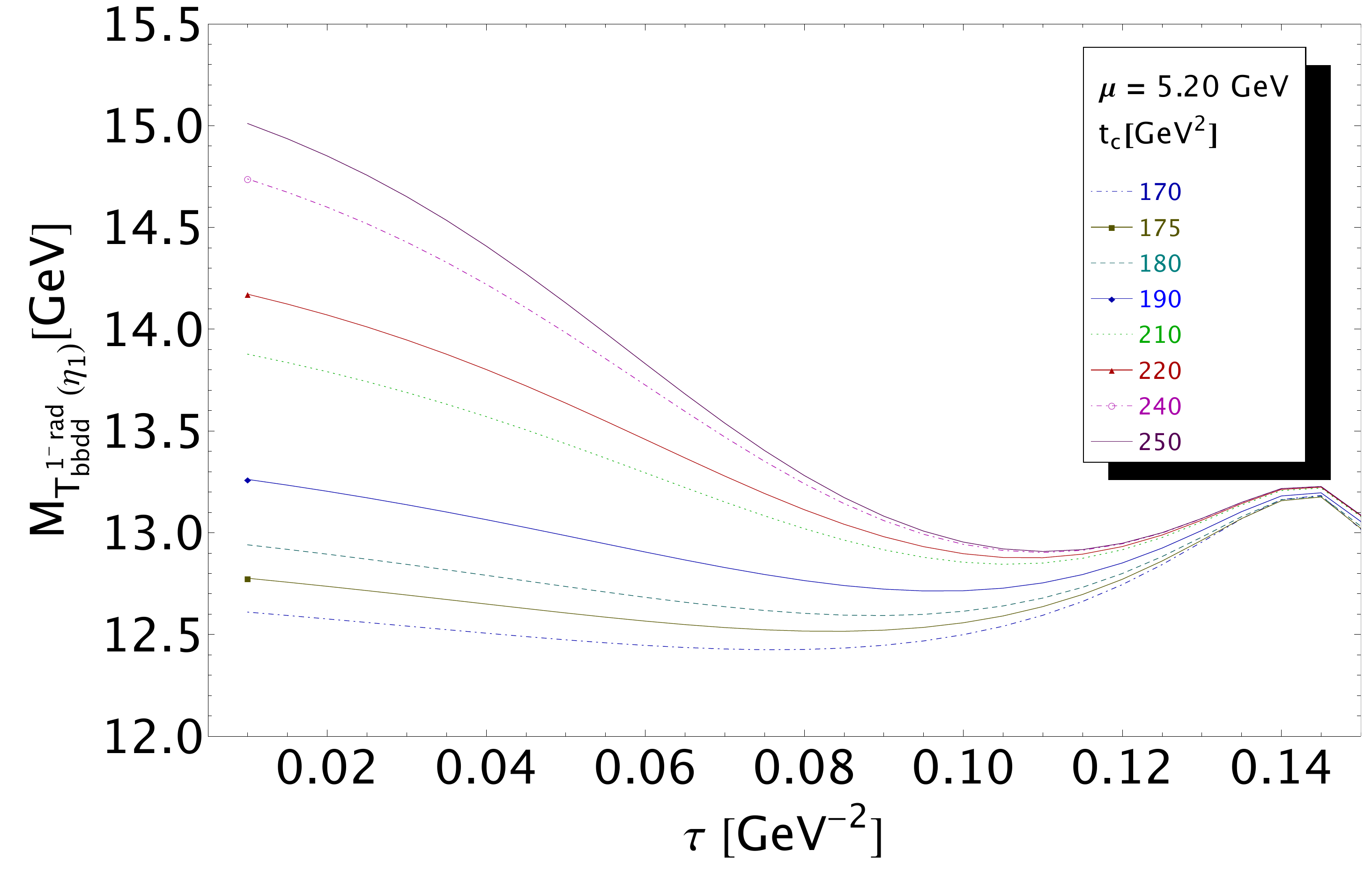}
\vspace*{-0.5cm}
\caption{\footnotesize  $f_{T^{1^-}_{bb\bar d\bar d}}$ and $M_{T^{1^-}_{bb\bar d\bar d}}$ radial excitation coupling and mass for the $\eta_1$ $1^-$ current as a function of $\tau$ and for different values of $t_c$.   } 
\label{fig:eta1-rad-b1}
\end{center}
\vspace*{-0.5cm}
\end{figure}

The curves for each current have different shapes. They are shown in Fig.\,\ref{fig:eta1-rad-b0} for the $\eta_1$ $0^-$pseudoscalar and in Figs. \,\ref{fig:eta6-rad-b1}, \,\ref{fig:eta1-rad-b1}  for the $\eta_6$ and $\eta_1$ $ 1^-$ vector currents. 

-- The Class L radial excitations spectra are in the same region as the Class H ground state ones which may obscure the search for the latter (compare Table\,\ref{tab:res-rad-b} and \ref{tab:resb}).


\begin{center}
   {\scriptsize
\begin{table}[hbt]
\setlength{\tabcolsep}{0.7pc}
    {\small
  \begin{tabular}{lllllll }
&\\
\hline
\hline
States&Current  &$t_c$ [GeV$^2$] &$\tau$  [GeV$^{-2}$]  & $f^{NLO} _{T_{ccqq'}}$ [keV]& $\tau$  [GeV$^{-2}$] & $M^{NLO}_{T_{ccqq'}}$ [MeV]\\ 
 \hline 
\bf Class H&&&&&&\\
  { \bma  $0^-$} &&&&&&\\
  $T_{bb\bar u\bar d}$&$ {\cal O}_T^{0^-},\eta_4$&$250\to 310$&$0.0425\to 0.0675$&1009(314)&$0.0425\to 0.0675$&15319(645)\\
  $T_{bb\bar u\bar d}$&$ \eta_5$&$230\to 310$&$0.0450\to 0.0600$&457(151)&$0.0475\to 0.0600$&14829(350) \\
$T_{bb\bar d\bar d}$ & $\eta_2$&$250\to 310$&$0.0400\to 0.0700$&905(308)&$0.0425\to 0.0700$&15266(509) \\
 { \bma  $1^-$} &&&&&&\\
 $T_{bb\bar u\bar d}$& $ {\cal O}_T^{1^-},\eta_5 $&$250\to 310$&$0.0425\to 0.0700$&587(208)&$0.0425\to 0.0700$&15275(628)\\
  $T_{bb\bar d\bar d}$& $\eta_2 $&$250\to 310$&$0.0400\to 0.0700$&816(303)&$0.0400\to 0.0700$&15373(576)\\

  \bf Class L&&&&&&\\
  { \bma  $0^-$} &&&&&&\\
   $T_{bb\bar d\bar d}$& $ \eta_1 $&$ 170\to 250$& $0.08 \to 0.10$&100(37)&$ 0.08\to 0.10$&12596(405) \\
  { \bma  $1^-$} &&&&&&\\
$T_{bb\bar u\bar d}$&  $\eta_6 $&$ 170\to 250$&$ 0.07\to 0.11$&69(29)&$ 0.075\to 0.10$&12679(437)\\
$T_{bb\bar d\bar d}$& $\eta_1$&$ 170\to 250$&$ 0.08\to 0.10$&88(35)& $0.075 \to 0.11$&12667(408) \\
   \hline\hline
  \vspace*{-0.5cm}
\end{tabular}}
 \caption{Masses and couplings of the $T_{bb\bar u\bar d}$ and $T_{bb\bar d\bar d}$1st radial excitations. The different sources of the errors are in Table\,\ref{tab:error-rad-b}. }  

\label{tab:res-rad-b}
\end{table}
} 
\end{center}
\vspace*{-0.5cm} 
\begin{table}[hbt]
\setlength{\tabcolsep}{0.35pc}
{\scriptsize{
\begin{tabular}{ll ll  ll  ll ll ll ll ll l c}
\hline
\hline
                States & Currents
                    &\multicolumn{1}{c}{$\Delta t_c$}
					&\multicolumn{1}{c}{$\Delta \tau$}
					&\multicolumn{1}{c}{$\Delta \mu$}
					&\multicolumn{1}{c}{$\Delta \alpha_s$}
					&\multicolumn{1}{c}{$\Delta m_b$}
					&\multicolumn{1}{c}{$\Delta \overline{\psi}\psi$}
					&\multicolumn{1}{c}{$\Delta G^2$}
					&\multicolumn{1}{c}{$\Delta G^3$}
					&\multicolumn{1}{c}{$\Delta \overline{\psi}\psi^2$}
					
					&\multicolumn{1}{c}{$\Delta OPE$}
					&\multicolumn{1}{c}{$\Delta (M_{G},f_G)$}
					&\multicolumn{1}{c}{$\Delta M_{G_1}$}
					&\multicolumn{1}{r}{Total}
\\
					
\hline
{\bf Class H} &&&&&&&&&&\\
\it Masses [MeV]\\
$0^-$ &&&&&&&&&&\\
$T_{bb\bar{u}\bar{d}}$ & ${\cal O}^{0^-}_{T}$, $\eta_4$ &161&105& 19&55&59&0.0&0.7&0.2&60&97&599&$\cdots$&645 \\
$T_{bb\bar{u}\bar{d}}$ & $\eta_5$ & 249&97& 9& 59& 56&0.0&0.2&0.0& 27& 38& 205&$\cdots$& 350 \\
$T_{bb\bar{d}\bar{d}}$ & $\eta_2$ &145&91& 19&54&54&0.0&1.7&0.0&64&111&455&$\cdots$&509 \\
$1^-$ &&&&&&&&&&\\
$T_{bb\bar{u}\bar{d}}$ & ${\cal O}^{1^-}_{T}$, $\eta_5$ &143&121& 19&52&52&0.0&2&0.6&45&112&582&$\cdots$&628\\
$T_{bb\bar{d}\bar{d}}$ & $\eta_2$ &163&170& 22&67&67&0.0&0.1&0.0&78&95&502&$\cdots$& 576 \\
\it Couplings [keV]\\
$0^-$ &&&&&&&&&&\\
$T_{bb\bar{u}\bar{d}}$ & ${\cal O}^{0^-}_{T}$, $\eta_4$ &168&3&11&24&24&0.0&0.0&0.0& 8& 5&170&201& 314 \\
$T_{bb\bar{u}\bar{d}}$ & $\eta_5$ &118&4& 11& 18& 8&0.0&0.2&0.2& 3& 1& 45& 79& 151 \\
$T_{bb\bar{d}\bar{d}}$ & $\eta_2$ &137&2&9&18&19&0.0&0.3&0.1&6&7&222&161&308 \\
$1^-$ &&&&&&&&&&\\
$T_{bb\bar{u}\bar{d}}$ & ${\cal O}^{1^-}_{T}$, $\eta_5$ &89&2&11&12&12&0.0&0.4&0.4&5&5&141&123&208\\
$T_{bb\bar{d}\bar{d}}$ & $\eta_2$ &137&6&10&15&15&0.0&0.6&0.6&7& 3&173& 206& 303 \\
{\bf Class L} &&&&&&&&&&\\
\it Masses [MeV]\\
$0^-$ &&&&&&&&&&\\
 $T_{bb\bar d\bar d}$& $ \eta_1 $& 231& 81&21& 78& 73&0.0&0.4&0.1& 138& 253& 95&$\cdots$& 405 \\
$1^-$ &&&&&&&&&&\\
$T_{bb\bar u\bar d}$& $ \eta_6 $& 248& 53& 19& 92& 70&0.0& 2.5& 0.2& 107& 279& 154&$\cdots$& 437 \\
$T_{bb\bar d\bar d}$& $ \eta_1 $& 242& 26&17& 94& 58&0.0&1.0&0.3& 144& 247& 112&$\cdots$& 408 \\
\it Couplings [keV]\\
$0^-$ &&&&&&&&&&\\
 $T_{bb\bar d\bar d}$& $ \eta_1 $& 31& 1.0&1.4&3&3&0.0&0.0&0.0&2&7&2& 18& 37 \\
$1^-$ &&&&&&&&&&\\
$T_{bb\bar u\bar d}$& $ \eta_6 $&18&0.7&1.0&3&2&0.0&0.6&0.5&2&5&3& 22& 29 \\
$T_{bb\bar d\bar d}$& $ \eta_1 $& 28& 1.2&1.0&3&2&0.0&0.0&0.0&3&4&4& 19& 35 \\
\hline
\hline
\end{tabular}
}}
 \caption{Sources of errors of $T_{bb\bar u\bar d}$ and $T_{bb\bar d\bar d}$ 1st radial excitations masses and couplings. We take $\ve \Delta \mu\ve=0.05$ GeV and $\ve \Delta \tau\ve =0.01$ GeV$^{-2}$. $\Delta OPE$ has been estimated assuming that the high-dimension condensates are given as $m_c^2(\tau/3)\times d=6$ contributions.}
\label{tab:error-rad-b}
\end{table}

-- The couplings of these Class L 1st radial excitation states to the current are about one order of magnitude of the one of the ground states which may suggest that these states can be narrow using a Golberger-Treiman-like relation : $\Gamma\to \bar BB\sim 1/f^2_{T^{1^-}_{bb\bar u\bar d}}$. These large  couplings are unexpected compared to ordinary hadrons.  We do not fully understand the origin of this numerical enhancement.

\section{Quark mass behaviour of the $T_{QQ\bar q\bar q'}$  couplings and masses }
In this section, we discuss the quark mass behaviour of the couplings and masses of the $T_{QQ\bar q\bar q'}$
states deduced from an empirical observation of the numerical results in Tables\,\ref{tab:resc}, \ref{tab:resb}
and \ref{tab:res-rad-c}, \ref{tab:res-rad-b}.  We may expect that  these empirical observation can help for a future attempt to built an effective theory for these exotics. 

\subsubsection*{\b  Masses}
\d On can notice that the $T_{cc\bar q\bar q'}$ ground state masses are in the range of 6 (resp. 13) GeV for the Class L (resp. H) while for the $T_{bb\bar q\bar q'}$ ones they are about 13 (resp. 15) GeV.  This result suggests that the masses increase like $m_Q$ (running mass evaluated at the scale $\mu$) as expected from the leading behaviour of the ratio of moments.

\d The mass-splittings between the lowest ground states and their corresponding 1st radial excitations are about 2 GeV which suggest that they are quark masses  independent. These splittings are large compared to the ones of ordinary hadrons which may indicate the new dynamics of these exotic states. 

\subsubsection*{\b  Couplings}
The quark mass behaviours of the ratios $R_{c/b}$ of couplings of $T_{cc\bar q\bar q'}$ over $T_{bb\bar q\bar q'}$ are more subtle.  

\d For the Class H ground state, $R_{c/b}\simeq (3.5-4)$ which suggests that the couplings behave as $1/m_Q$ while, for the Class L states, it is about 24 indicating an approximate  $1/m_Q^{5/2}$ behaviour.

\d For the Class H radial excitations,  $R_{c/b}\simeq 2$ which corresponds to a $1/m_Q^{1/2}$ behaviour.  For the Class L
states $R_{c/b}\simeq (4-6)$ which suggests an approximate $1/m_Q$ behaviour.

\section{Summary and Comments}

\b We have systematically analyzed different interpolating pseudoscalar and vector low-dimension interpolating currents having the quantum numbers of the $T_{QQ\bar q\bar q'}$ states. This work completes our previous one on scalar and axial-vector states in Ref.\,\cite{Tcc}. 

\b We have especially compared our results with the previous LO ones of Ref.\,\cite{ZHUT}. We disagree on the size of the mixed condensate contributions for the $d=6,8$ dimensions and on some numerical analysis as explained in details in the text, where, in the latter, the favoured result is taken at the lowest value of $t_c$ where some results are not yet $t_c$-stable and in some cases not $\tau$-stable while some others do not satisfy the $R_{P/C}$ constraint. 

\b The results for the ground states are summarized in Tables\,\ref{tab:resc} and \,\ref{tab:resb} while the different sources of the errors can be found in Tables\,\ref{tab:error-fc},  \ref{tab:error-mc}, \ref{tab:error-fb} and \ref{tab:error-mb}:

\d One can notice that the mass-splittings due to SU3 breakings $(m_s,\la\bar ss\ra)$ are tiny ($\leq$ 50 MeV).

\d We found that the pseudoscalar and vector states can be divided into two classes : 

\hspace*{0.5cm} -- Class H (Heavy) states where the masses of the $T_{cc\bar q\bar q'}$ (resp. $T_{bb\bar q\bar q'}$) states are around 6 (resp. 13) GeV which are 2 GeV higher than the ones of their axial partners $(1^+,0^+)$\,\cite{Tcc}. This feature is in line with our previous findings for XYZ states\,\cite{MOLE12,MOLE16,MOLE16X,SU3}. 

\hspace*{0.5cm} -- Class L (Light) states  
around (3.8-4.4) GeV, where the pseudoscalar $T_{cc\bar q\bar q}$ (resp. all
vector states $T_{cc\bar q\bar q'}$) are below the  $\bar D D_0, \bar D_sD_{s0}$ (resp. $\bar D D_1, \bar D_sD_{s1}$) open charm thresholds. 

\hspace*{0.5cm}-- Class L  $T_{bb\bar q\bar q'}$ pseudoscalar and vector  states have masses around 10.4 GeV where all of them are below the open beauty thresholds. 

\b The results for the 1st radial excitations of the $T_{QQ\bar u\bar d}$ and  $T_{QQ\bar u\bar d}$ are summarized in Tables\,\ref{tab:res-rad-c} and \,\ref{tab:res-rad-b} while the sources of different errors are given in Tables\,\ref{tab:error-rad-c} and \ref{tab:error-rad-b} :

\d The radial excitations of the Class L  states are expected to be in the region ($ 5.8\sim 6.4$) GeV (resp.  ($12.7\sim 13.1$)  GeV) for the charm (resp. bottom) channels . As these radial excitations couple strongly to the corresponding interpolating currents, their presence in these regions may mask the searches for some Class H ground states having similar masses.

\d  The Class H $T_{cc\bar u\bar d}$ $0^-$ pseudoscalar ground state is exceptionnally (300-500) MeV above the radial excitation of the corresponding Class L state. Its eventual detection can be an alternative test of the violation of the factorization of the four-quark condensates. 

\b One may finally assume that some physical states emerge from some non-trivial mixing of the different interpolating operators discussed here. However, looking at these eventual mixings is beyond the aim of the present paper. 

 \newpage
\appendix
\section{Pseudoscalar Currents $0^-$ }
One can evaluate the two-point correlation
function using the hadronic currents in Tables\,\ref{tab:current} and \,\ref{tab:zhu}:

The spectral densities can be evaluated using sum rules 
techniques\,\cite{SNB1,SNB2,RAPHAEL} and the expressions are given below. \\
\vspace{0.5cm}
{\bma $ {\cal O}_{T_{ud}^{0^-}}, \eta_4$ \bf current}
\vspace{-0.5cm}
\begin{eqnarray*}
  \rho_{ud}^{pert}(s) &=& 
  \frac{m_c^8}{5\cdot 3 \cdot 2^{8} \,\pi^6} 
  \bigg[ v \big( 1080 + 5400/x + 306/x^2 - 28/x^3 
  + 1/x^4 \big) + \\
  && + 120 {\cal L}_v \big( 18x + 15 - 6(3 + 1/x)\log(x) 
  - 32/x \big) 
  - 1440{\cal L}_+ \big( 3 + 1/x \big) \bigg] \\ 
  \rho_{ud}^{\langle \bar{q}q \rangle}(s) &=& 0 \\ 
  \rho_{ud}^{\langle G^2 \rangle}(s) &=& \frac{
  m_c^4 \langle G^2 \rangle} {3 \cdot 2^{8} \,\pi^6} 
  \bigg[ v \big( 6 + 17/x + 1/x^2 \big) 
  + 12 {\cal L}_v \big( x - \log(x) - 1/x \big)
  - 24{\cal L}_+ \bigg] \\ 
  \rho_{ud}^{\langle \bar{q}Gq \rangle}(s) &=& 0 \\ 
  \rho_{ud}^{\langle \bar{q}q \rangle^2}(s) &=& 
  - \frac{m_c^2 \langle \bar{q}q \rangle^2}{3 \,\pi^2} 
  v \big( 2 + 1/x \big) \\ 
  \rho_{ud}^{\langle G^3 \rangle}(s) &=& 
  \frac{m_c^2 \langle G^3 \rangle}
  {5 \cdot 3^4 \cdot 2^{8} \,\pi^6} \bigg[ 
  v \bigg( 1200x^2 - 280x + 770 + 1145/x - 54/x^2
  + 2m_c^2 \tau \big( 300x + \\
  && - 190 + 30/x + 67/x^2 + 9/x^3 \big) \bigg)
  + 12 {\cal L}_v \bigg( 200x^3 - 80x^2 
  + 135x + 21 + \\ 
  && - 15(4 + 1/x)\log(x) - 60/x + m_c^2 \tau 
  \big( 100x^2 - 80x + 20 - 9/x^2 \big) \bigg) 
  - 360 {\cal L}_+ \big( 4 + 1/x) \bigg] \\
   \rho_{4}^{\langle 8 \rangle}(s) &=& 
  \frac{\langle \bar{u} u \rangle 
  \langle \bar{q} G q \rangle + \langle \bar{q} q \rangle 
  \langle \bar{u} G u \rangle}{3 \cdot 2^{4} \,\pi^2} ~
  v \bigg( 8x + 5 + 3m_c^2 \tau \big( 4 - 1/x \big) + 
  12 m_c^4 \tau^2  /x \bigg) + \\
  && + \frac{\langle G^2 \rangle^2}{3^4 \cdot 2^{13} \,\pi^6} \bigg[ 
  v \bigg( 351 + 108/x - 36m_c^2 \tau \big( 2/x + 1/x^2 \big) + 
  2 m_c^4 \tau^2 \big( 24/x + 2/x^2 + 1/x^3 \big) \bigg) + \\
  && + 24 {\cal L}_v \bigg( 18x - 18 - 6m_c^2 \tau \big( 1 - 1/x \big) 
  + m_c^4 \tau^2 \big( 4 - 3/x \big) \bigg) \bigg]
\end{eqnarray*}
{\bma $ {\cal O}_{T^{0^-}_{us}},\eta_4$ \bf current}
\vspace{-0.25cm}
\begin{eqnarray*}
\rho_{us}^{pert}(s) &=& 
  \frac{m_c^8}{5\cdot 3 \cdot 2^{9} \,\pi^6} 
  \bigg[ v \big( 1080 + 5400/x + 306/x^2 - 28/x^3 
  + 1/x^4 \big) + \\
  && + 120 {\cal L}_v \big( 18x + 15 - 6(3 + 1/x)\log(x) 
  - 32/x \big) - 1440{\cal L}_+ 
  \big(3 + 1/x \big) \bigg] \\ 
  \rho_{us}^{\langle \bar{q}q \rangle}(s) &=& 
  \frac{m_s m_c^4}{3 \cdot 2^4 \,\pi^4} \bigg[ 
  v \bigg( \langle \bar{q}q \rangle (24 + 2/x + 1/x^2)
  + \langle \bar{s}s \rangle (6 - 7/x + 1/x^2) \bigg) + \\
  && + 12 {\cal L}_v \bigg( \langle \bar{q}q \rangle 
  (4x-3) + \langle \bar{s}s \rangle x \bigg) \bigg]
  \\ 
   \rho_{us}^{\langle G^2 \rangle}(s) &=& 
  \frac{m_c^4 \langle G^2 \rangle}{3 \cdot 2^{9} \,\pi^6} 
  \bigg[ v \big( 6 + 17/x + 1/x^2 \big) 
  + 12 {\cal L}_v \big( x - \log(x) - 1/x \big)
  - 24{\cal L}_+ \bigg] \\ 
  \rho_{us}^{\langle \bar{q}Gq \rangle}(s) &=&
  -\frac{m_s m_c^2}{3 \cdot 2^6 \,\pi^4} \bigg[ 
  v \bigg( 3\langle \bar{q}Gq \rangle
  + \langle \bar{s}Gs \rangle \bigg) \big( 8 + 1/x \big) + 
  6{\cal L}_v \bigg( \langle \bar{q}Gq \rangle 
  (4x-1) + 2\langle \bar{s}Gs \rangle x \bigg) \bigg]
  \\ 
  \rho_{us}^{\langle \bar{q}q \rangle^2}(s) &=& 
  - \frac{m_c^2 \langle \bar{q}q \rangle 
  \langle \bar{s}s \rangle}{6 \,\pi^2} 
  v \big( 2 + 1/x \big) \\ 
  \end{eqnarray*}
  \begin{eqnarray*}
  \rho_{us}^{\langle G^3 \rangle}(s) &=& 
  \frac{m_c^2 \langle G^3 \rangle}
  {5 \cdot 3^4 \cdot 2^{9} \,\pi^6} \bigg[ 
  v \bigg( 1200x^2 - 280x + 770 + 1145/x - 54/x^2
  + 2m_c^2 \tau \big( 300x + \\
  && - 190 + 30/x + 67/x^2 + 9/x^3 \big) \bigg)
  + 12 {\cal L}_v \bigg( 200x^3 - 80x^2 
  + 135x + 21 + \\ 
  && - 15(4 + 1/x)\log(x) - 60/x + m_c^2 \tau 
  \big( 100x^2 - 80x + 20 - 9/x^2 \big) \bigg) 
  - 360 {\cal L}_+ \big( 4 + 1/x) \bigg]
\end{eqnarray*}

We note that the spectral function associated to the current $ {\cal O}_{T^{0^-}_{ss} }$ is identically zero. 

\vspace{0.6cm}
{\bma $ \eta_1$ \bf current}
\begin{eqnarray*}
  \rho_{1}^{pert}(s) &=& 
  -\frac{m_c^8}{5 \cdot 3 \cdot 2^{9} \,\pi^6} 
  \bigg[ v \big( 720 + 6420/x + 1434/x^2 + 58/x^3 
  - 1/x^4 \big) + \\
  && + 120 {\cal L}_v \big( 12x + 33 - 6(3 + 2/x)\log(x) 
  - 40/x - 6/x^2 \big) - 1440{\cal L}_+ 
  \big(3 + 2/x \big) \bigg] \\ 
  \rho_{1}^{\langle \bar{q}q \rangle}(s) &=& 
  \frac{m_q m_c^4 \langle \bar{q}q \rangle}{3 \cdot 2^4 \,\pi^4} 
  \bigg[ v \big( 12 + 16/x - 1/x^2 \big) 
  + 12 {\cal L}_v \big( 2x-3 \big) \bigg]
  \\ 
  \rho_{1}^{\langle G^2 \rangle}(s) &=& 
  -\frac{m_c^4 \langle G^2 \rangle}{3 \cdot 2^{9} \,\pi^6} 
  \bigg[ v \big( 30 + 31/x + 5/x^2 \big) 
  + 6 {\cal L}_v \big( 10x - 6 - (5-1/x)\log(x) - 4/x \big)
  - 12{\cal L}_+ (5-1/x) \bigg] \\ 
  \rho_{1}^{\langle \bar{q}Gq \rangle}(s) &=&
  -\frac{5 m_q m_c^2 \langle \bar{q}Gq \rangle}
  {3 \cdot 2^5 \,\pi^4} \,v \big( 4 - 1/x \big)
  \\ 
  \rho_{1}^{\langle \bar{q}q \rangle^2}(s) &=& 
  - \frac{m_c^2 \langle \bar{q}q \rangle^2}{3 \,\pi^2} 
  v \big( 4 - 1/x \big) \\ 
  \rho_{1}^{\langle G^3 \rangle}(s) &=& 
  -\frac{m_c^2 \langle G^3 \rangle}
  {5 \cdot 3^2 \cdot 2^{8} \,\pi^6} \bigg[ 
  v \bigg( 10 - 35/x - 2/x^2 + m_c^2 \tau \big( 
  26/x^2 + 1/x^3 \big) \bigg) +  \\
  && + {\cal L}_v \bigg( 20x + 56 + 5/x - 45\log(x) 
  - 12 m_c^2 \tau \big( 1/x + 1/x^2 \big) \bigg) 
  - 90 {\cal L}_+ \bigg]
\end{eqnarray*}
{\bma $ \eta_2$ \bf current}
\begin{eqnarray*}
  \rho_{2}^{pert}(s) &=& 
  \frac{m_c^8}{5 \cdot 3 \cdot 2^{9} \,\pi^6} 
  \bigg[ v \big( 1680 + 9340/x + 886/x^2 - 18/x^3 
  + 1/x^4 \big) + \\
  && + 120 {\cal L}_v \big( 28x + 31 - 6(5 + 2/x)\log(x) 
  - 56/x - 2/x^2 \big) - 1440{\cal L}_+ 
  \big(5 + 2/x \big) \bigg] \\ 
  \rho_{2}^{\langle \bar{q}q \rangle}(s) &=& 
  \frac{m_q m_c^4 \langle \bar{q}q \rangle}{8 \,\pi^4} 
  \bigg[ v \big( 12 - 4/x + 1/x^2 \big) 
  + 12 {\cal L}_v \big( 2x-1 \big) \bigg]
  \\ 
  \rho_{2}^{\langle G^2 \rangle}(s) &=& 
  \frac{m_c^4 \langle G^2 \rangle}{3^2 \cdot 2^{9} \,\pi^6} 
  \bigg[ v \big( 102 + 23/x + 1/x^2 \big) + \\
  && + 6 {\cal L}_v \big( 34x - 34 - 3(3-1/x)\log(x) 
  + 2/x \big) - 36{\cal L}_+ (3-1/x) \bigg] \\ 
  \rho_{2}^{\langle \bar{q}Gq \rangle}(s) &=&
  -\frac{19 m_q m_c^2 \langle \bar{q}Gq \rangle}
  {3 \cdot 2^5 \,\pi^4} \,v/x
  \\ 
  \rho_{2}^{\langle \bar{q}q \rangle^2}(s) &=& 
  - \frac{m_c^2 \langle \bar{q}q \rangle^2}{3 \cdot 2^5 \,\pi^2} 
  \,v/x \\ 
  \rho_{2}^{\langle G^3 \rangle}(s) &=& 
  \frac{m_c^2 \langle G^3 \rangle}
  {5 \cdot 3^3 \cdot 2^{8} \,\pi^6} \bigg[ 
  v \bigg( 60 - 10/x - 14/x^2 + m_c^2 \tau \big( 
  58/x^2 + 5/x^3 \big) \bigg) + \\
  && + 3{\cal L}_v \bigg( 40x + 4 + 5/x - 45\log(x) 
  - 4 m_c^2 \tau \big( 1/x + 3/x^2 \big) \bigg) 
  - 270 {\cal L}_+ \bigg]
\end{eqnarray*}
{\bma $ \eta_5$ \bf current}
\begin{eqnarray*}
  \rho_{5}^{pert}(s) &=& 
  -\frac{m_c^8}{5\cdot 3 \cdot 2^{9} \,\pi^6} 
  \bigg[ v \big( 120 + 2480/x + 854/x^2 + 48/x^3 
  - 1/x^4 \big) + \\
  && + 120 {\cal L}_v \big( 2x + 17 - 6(1 + 1/x)\log(x) 
  - 16/x - 4/x^2 \big) - 1440{\cal L}_+ 
  \big(1 + 1/x \big) \bigg] \\ 
  \rho_{5}^{\langle \bar{q}q \rangle}(s) &=& 
  \frac{m_q m_c^4}{3 \cdot 2^4 \,\pi^4} \bigg[ 
  v \bigg( \langle \bar{u}u \rangle (24 + 22/x - 1/x^2)
  - \langle \bar{q} q \rangle (6 + 13/x - 1/x^2) \bigg) + \\
  && + 12 {\cal L}_v \bigg( \langle \bar{u}u \rangle 
  (4x - 5) - \langle \bar{q}q \rangle (x - 2) \bigg) \bigg]
  \\ 
  \rho_{5}^{\langle G^2 \rangle}(s) &=& 
  -\frac{m_c^4 \langle G^2 \rangle}{3^2 \cdot 2^{9} \,\pi^6} 
  \bigg[ v \big( 6 + 61/x + 5/x^2 \big) 
  + 12 {\cal L}_v \big( x + 2 - 3\log(x) - 4/x \big)
  - 72{\cal L}_+ \bigg] \\ 
  \rho_{5}^{\langle \bar{q}Gq \rangle}(s) &=&
  -\frac{m_q m_c^2}{3 \cdot 2^6 \,\pi^4} \bigg[ 
  v \bigg( 3\langle \bar{u}Gu \rangle (16 - 1/x)
  - \langle \bar{q}Gq \rangle (24 - 1/x) \bigg) + \\ 
  && + 6{\cal L}_v \bigg( \langle \bar{u}Gu \rangle 
  (4x-3) - 2\langle \bar{q}Gq \rangle (x - 1) \bigg) \bigg]
  \\ 
  \rho_{5}^{\langle \bar{q}q \rangle^2}(s) &=& 
  - \frac{m_c^2 \langle \bar{u}u \rangle 
  \langle \bar{q}q \rangle}{6 \,\pi^2} 
  \,v \big( 6 - 1/x \big) \\
  \rho_{5}^{\langle G^3 \rangle}(s) &=& 
  \frac{m_c^2 \langle G^3 \rangle}
  {5 \cdot 3^4 \cdot 2^{9} \,\pi^6} \bigg[ 
  v \bigg( 1200x^2 - 280x + 230 + 1895/x + 6/x^2
  + 2m_c^2 \tau \big( 300x + \\
  && - 190 + 30/x - 137/x^2 - 3/x^3 \big) \bigg)
  + 12 {\cal L}_v \bigg( 200x^3 - 80x^2 
  + 45x + 21 + \\ 
  && - 15(4 + 1/x)\log(x) - 90/x + m_c^2 \tau 
  \big( 100x^2 - 80x + 20 + 12/x + 9/x^2 \big) \bigg) 
  - 360 {\cal L}_+ \big( 4 + 1/x) \bigg] \\
    \rho_{5}^{\langle 8 \rangle}(s) &=& 
  - \frac{\langle \bar{q} q \rangle 
  \langle \bar{s} G s \rangle + \langle \bar{s} s \rangle 
  \langle \bar{q} G q \rangle}{3 \cdot 2^{4} \,\pi^2} ~
  v \bigg( 8x + 1 + m_c^2 \tau \big( 12 + 1/x \big) 
  - 4 m_c^4 \tau^2  /x \bigg) + \\
  && - \frac{\langle G^2 \rangle^2}{3^4 \cdot 2^{13} \,\pi^6} \bigg[ 
  v \bigg( 81 + 108/x - 36m_c^2 \tau \big( 2/x + 1/x^2 \big) - 
  2 m_c^4 \tau^2 \big( 24/x + 22/x^2 - 1/x^3 \big) \bigg) + \\
  && + 24 {\cal L}_v \bigg( 18x - 18 - 6m_c^2 \tau \big( 1 - 1/x \big) 
  - m_c^4 \tau^2 \big( 4 - 5/x \big) \bigg) \bigg]
\end{eqnarray*}

\section{ Vector Currents $1^-$}
\vspace{-0.25cm}
{\bma ${\cal O}_{T_{ud}^{1^-}},\eta_5$ \bf current}
\vspace{-0.25cm}
\begin{eqnarray*}
  \rho_{ud}^{pert}(s) &=& 
  \frac{m_c^8}{5\cdot 3^2 \cdot 2^{11} \,\pi^6} 
  \bigg[ v \big( 840x + 7340 + 52528/x + 5796/x^2 - 62/x^3 
  + 5/x^4 \big) + \\
  && + 120 {\cal L}_v \big( 14x^2 + 120x + 207 - 18(9 + 4/x)\log(x) 
  - 320/x -15/x^2 \big) - 4320{\cal L}_+ \big( 9 + 4/x \big) 
  \bigg] \\ 
  \rho_{ud}^{\langle \bar{q}q \rangle}(s) &=& 0 \\ 
  \rho_{ud}^{\langle G^2 \rangle}(s) &=& \frac{
  m_c^4 \langle G^2 \rangle} {3^3 \cdot 2^{10} \,\pi^6} 
  \bigg[ v \big( 102x + 557 + 538/x + 18/x^2 \big) + \\ 
  && + 6 {\cal L}_v \big( 34x^2 + 180x - 123 - 63\log(x) 
  - 44/x \big) - 756{\cal L}_+ \bigg] \\ 
   \end{eqnarray*}
\begin{eqnarray*}
  \rho_{ud}^{\langle \bar{q}Gq \rangle}(s) &=& 0 \\ 
  \rho_{ud}^{\langle \bar{q}q \rangle^2}(s) &=& 
  - \frac{m_c^2 \langle \bar{q}q \rangle^2}{9 \,\pi^2} 
  \,v \big( 2 + 1/x \big) \\ 
  \rho_{ud}^{\langle G^3 \rangle}(s) &=& 
  -\frac{m_c^2 \langle G^3 \rangle}
  {5 \cdot 3^4 \cdot 2^{11} \,\pi^6} \bigg[ 
  v \bigg( 420x + 1750 + 1490/x + 165/x^2
  - 12m_c^2 \tau \big( 58/x^2 + 5/x^3 \big) \bigg) + \\
  && + 24 {\cal L}_v \bigg( 35x^2 + 140x - 135 - 30/x 
  + 6 m_c^2 \tau \big( 1/x + 3/x^2 \big) \bigg) \bigg]
\end{eqnarray*}
{\bma ${\cal O}_{T^{1^-}_{us}},\eta_5$ \bf current}
\vspace*{-0.25cm}
\begin{eqnarray*}
  \rho_{us}^{pert}(s) &=& 
  \frac{m_c^8}{5\cdot 3^2 \cdot 2^{12} \,\pi^6} 
  \bigg[ v \big( 840x + 7340 + 52528/x + 5796/x^2 - 62/x^3 
  + 5/x^4 \big) + \\
  && + 120 {\cal L}_v \big( 14x^2 + 120x + 207 - 18(9 + 4/x)\log(x) 
  - 320/x -15/x^2 \big) - 4320{\cal L}_+ \big( 9 + 4/x \big) 
  \bigg] \\ 
  \rho_{us}^{\langle \bar{q}q \rangle}(s) &=& 
  \frac{m_s m_c^4}{3 \cdot 2^8 \,\pi^4} \big( 
  2\langle \bar{q}q \rangle + \langle \bar{s}s \rangle \big)
  \bigg[ v (12x + 50 - 2/x + 3/x^2) + 24 {\cal L}_v 
  (x^2 + 4x -3) \bigg]
  \\ 
  \rho_{us}^{\langle G^2 \rangle}(s) &=& 
  \frac{m_c^4 \langle G^2 \rangle}{3^3 \cdot 2^{11} \,\pi^6} 
  \bigg[ v \big( 102x + 557 + 538/x + 18/x^2 \big) + \\ 
  && + 6 {\cal L}_v \big( 34x^2 + 180x - 123 - 63\log(x) 
  - 44/x \big) - 756{\cal L}_+ \bigg] \\ 
  \rho_{us}^{\langle \bar{q}Gq \rangle}(s) &=&
  -\frac{m_s m_c^2}{3^2 \cdot 2^6 \,\pi^4} \big( 
  6 \langle \bar{q}Gq \rangle - \langle \bar{s}Gs \rangle \big)
  \,v (2 + 1/x) \\ 
  \rho_{us}^{\langle \bar{q}q \rangle^2}(s) &=& 
  -\frac{m_c^2 \langle \bar{q}q \rangle 
  \langle \bar{s}s \rangle}{18 \,\pi^2} 
  \,v \big( 2 + 1/x \big) \\ 
  \rho_{us}^{\langle G^3 \rangle}(s) &=& 
  -\frac{m_c^2 \langle G^3 \rangle}
  {5 \cdot 3^4 \cdot 2^{12} \,\pi^6} \bigg[ 
  v \bigg( 420x + 1750 + 1490/x + 165/x^2
  - 12m_c^2 \tau \big( 58/x^2 + 5/x^3 \big) \bigg) + \\
  && + 24 {\cal L}_v \bigg( 35x^2 + 140x - 135 - 30/x 
  + 6 m_c^2 \tau \big( 1/x + 3/x^2 \big) \bigg) \bigg]
\end{eqnarray*}
{\bma $ \eta_1$ \bf current} 
where $q = u ~\mbox{or}~ s$ and $m_u = 0$. 
\begin{eqnarray*}
  \rho_{1}^{pert}(s) &=& 
  \frac{m_c^8}{5 \cdot 3^2 \cdot 2^{10} \,\pi^6} 
  \bigg[ v \big( 840x - 7060 - 42032/x - 8124/x^2 - 302/x^3 
  + 5/x^4 \big) + \\
  && + 120 {\cal L}_v \big( 14x^2 - 120x - 177 + 18(7 + 4/x)\log(x) 
  + 256/x + 33/x^2 \big) + 4320{\cal L}_+ 
  \big( 7 + 4/x \big) \bigg] \\ 
  \rho_{1}^{\langle \bar{q}q \rangle}(s) &=& 
  -\frac{m_q m_c^4 \langle \bar{q}q \rangle}{3 \cdot 2^5 \,\pi^4} 
  \bigg[ v \big( 12x - 46 - 50/x + 3/x^2 \big) 
  + 24 {\cal L}_v \big( x^2 - 4x + 5 \big) \bigg]
  \\ 
  \rho_{1}^{\langle G^2 \rangle}(s) &=& 
  \frac{m_c^4 \langle G^2 \rangle}{3^3 \cdot 2^{10} \,\pi^6} 
  \bigg[ v \big( 78x - 815 - 1192/x - 69/x^2 \big) + \\
  && + 6 {\cal L}_v \big( 26x^2 - 276x + 111 + 135\log(x) 
  + 128/x \big) + 1620{\cal L}_+ \bigg] \\ 
  \rho_{1}^{\langle \bar{q}Gq \rangle}(s) &=&
  -\frac{5 m_q m_c^2 \langle \bar{q}Gq \rangle}
  {3^2 \cdot 2^4 \,\pi^4} \,v \big( 4 - 1/x \big)
  \\ 
  \rho_{1}^{\langle \bar{q}q \rangle^2}(s) &=& 
  - \frac{2 m_c^2 \langle \bar{q}q \rangle^2}{9 \,\pi^2} 
  \,v \big( 4 - 1/x \big) \\ 
  \rho_{1}^{\langle G^3 \rangle}(s) &=& 
  -\frac{m_c^2 \langle G^3 \rangle}
  {5 \cdot 3^4 \cdot 2^{10} \,\pi^6} \bigg[ 
  v \bigg( 60x - 1910 - 2350/x - 75/x^2 + 36m_c^2 \tau \big( 
  26/x^2 + 1/x^3 \big) \bigg) + \\
  && + 24 {\cal L}_v \bigg( 5x^2 - 160x + 180 + 30/x
  - 18 m_c^2 \tau \big( 1/x + 1/x^2 \big) \bigg) \bigg]
\end{eqnarray*}
{\bma $ \eta_2$ \bf current}
\begin{eqnarray*}
\vspace*{-0.25cm}
  \rho_{2}^{pert}(s) &=& 
  \frac{m_c^8}{5 \cdot 3^2 \cdot 2^{10} \,\pi^6} 
  \bigg[ v \big( 2520x + 420 + 32664/x + 3888/x^2 - 86/x^3 
  + 5/x^4 \big) + \\
  && + 120 {\cal L}_v \big( 42x^2 + 183 - 6(15 + 8/x)\log(x) 
  - 208/x - 9/x^2 \big) - 1440{\cal L}_+ 
  \big(15 + 8/x \big) \bigg] \\ 
  \rho_{2}^{\langle \bar{q}q \rangle}(s) &=& 
  \frac{m_q m_c^4 \langle \bar{q}q \rangle}{8 \,\pi^4} 
  \bigg[ v \big( 12 - 4/x + 1/x^2 \big) 
  + 12 {\cal L}_v \big( 2x-1 \big) \bigg]
  \\ 
  \rho_{2}^{\langle G^2 \rangle}(s) &=& 
  \frac{m_c^4 \langle G^2 \rangle}{3^3 \cdot 2^{10} \,\pi^6} 
  \bigg[ v \big( 390x + 137 + 70/x + 15/x^2 \big) + \\
  && + 6 {\cal L}_v \big( 130x^2 + 24x - 93 - 3(9-4/x)\log(x) + 
  4/x \big) - 36{\cal L}_+ (9-4/x) \bigg] \\ 
  \rho_{2}^{\langle \bar{q}Gq \rangle}(s) &=&
  -\frac{5 m_q m_c^2 \langle \bar{q}Gq \rangle}
  {3^2 \cdot 2^5 \,\pi^4} \,\bigg[ v( 6x + 1 + 8/x) + 
  12x^2 {\cal L}_v \bigg] \\ 
  \rho_{2}^{\langle \bar{q}q \rangle^2}(s) &=& 
  - \frac{m_c^2 \langle \bar{q}q \rangle^2}{3 \,\pi^2} 
  \,v/x \\ 
  \rho_{2}^{\langle G^3 \rangle}(s) &=& 
  \frac{m_c^2 \langle G^3 \rangle}
  {5 \cdot 3^6 \cdot 2^{10} \,\pi^6} \bigg[ 
  v \bigg( 870x - 7955 + 4160/x - 1215/x^2 + 432m_c^2 \tau \big( 
  8/x^2 + 1/x^3 \big) \bigg) + \\
  && + 6 {\cal L}_v \bigg( 290x^2 - 2700x + 3015 - 370/x - 
  15(81 + 10/x)\log(x) - 432 m_c^2 \tau / x^2 \bigg) + \\
  && - 180 {\cal L}_+ (81 + 10/x) \bigg]
\end{eqnarray*}
{\bma $ \eta_6$ \bf current}
\begin{eqnarray*}
  \rho_{6}^{pert}(s) &=& 
  -\frac{m_c^8}{5 \cdot 3^2 \cdot 2^{11} \,\pi^6} 
  \bigg[ v \big( 840x + 140 + 22168/x + 6216/x^2 + 278/x^3 
  - 5/x^4 \big) + \\
  && + 120 {\cal L}_v \big( 14x^2 + 153  - 6(9 + 8/x)\log(x) 
  - 144/x -27/x^2 \big) - 1440{\cal L}_+ \big( 9 + 8/x \big) 
  \bigg] \\ 
  \rho_{6}^{\langle \bar{q}q \rangle}(s) &=& 
  \frac{m_q m_c^4}{3 \cdot 2^7 \,\pi^4} \bigg[ 
  \langle \bar{u}u \rangle \bigg( v (96 + 128/x - 8/x^2) + 
  96 {\cal L}_v (2x - 3) \bigg) + \\
  && - \langle \bar{q}q \rangle \bigg( v (36x + 6 + 42/x - 
  3/x^2) + 72 {\cal L}_v (x^2 - 1)  \bigg) \bigg]
  \\ 
  \rho_{6}^{\langle G^2 \rangle}(s) &=& 
  -\frac{m_c^4 \langle G^2 \rangle}{3^2 \cdot 2^{9} \,\pi^6} 
  \bigg[ v \big( 6 + 61/x + 5/x^2 \big) + 12 {\cal L}_v 
  \big( x + 2 - 3\log(x) - 4/x \big) - 72{\cal L}_+ \bigg] 
  \\ 
  \rho_{6}^{\langle \bar{q}Gq \rangle}(s) &=&
  -\frac{m_q m_c^2}{3^2 \cdot 2^7 \,\pi^4} \bigg[
  36 \langle \bar{u}Gu \rangle \,v(4 - 1/x) - 
  \langle \bar{q}Gq \rangle \bigg( v(30x + 1 - 4/x) + 
  60x^2 {\cal L}_v \bigg) \bigg]
  \\ 
  \rho_{6}^{\langle \bar{q}q \rangle^2}(s) &=& 
  -\frac{m_c^2 \langle \bar{q}q \rangle 
  \langle \bar{s}s \rangle}{6 \,\pi^2} 
  \,v \big( 4 - 1/x \big) \\ 
  \rho_{6}^{\langle G^3 \rangle}(s) &=& 
  \frac{m_c^2 \langle G^3 \rangle}
  {5 \cdot 3^6 \cdot 2^{11} \,\pi^6} \bigg[ 
  v \bigg( 11130x - 6245 + 12440/x + 405/x^2
  - 216m_c^2 \tau \big( 26/x^2 + 1/x^3 \big) \bigg) + \\
  && + 6 {\cal L}_v \bigg( 3710x^2 - 2700x - 1035 +
  15(81-10/x) \log(x) - 730/x + 432 m_c^2 \tau \big( 
  1/x + 1/x^2 \big) \bigg) + \\
  && + 180 {\cal L}_+ (81-10/x) \bigg]
\end{eqnarray*}

\vfill\eject

\end{document}